\documentclass[11pt,a4paper]{article}
\pdfoutput=1
\usepackage{jheppub}

\usepackage[numbers,sort&compress]{natbib}
\usepackage{subcaption}
\renewcommand*{\le}{\left}
\newcommand*{\ri}{\right}

\newcommand*{\beq}{\begin{equation}}
\newcommand*{\eeq}{\end{equation}}
\newcommand*{\bea}{\begin{eqnarray}}
\newcommand*{\eea}{\end{eqnarray}}

\renewcommand*{\a}{\alpha}
\renewcommand*{\b}{\beta}
\newcommand*{\g}{\gamma}
\renewcommand*{\d}{\delta}
\newcommand*{\e}{\epsilon}
\newcommand*{\ve}{\varepsilon}
\newcommand*{\h}{\eta}
\newcommand*{\q}{\theta}
\renewcommand*{\l}{\lambda}
\newcommand*{\m}{\mu}
\newcommand*{\n}{\nu}
\renewcommand*{\r}{\rho}
\newcommand*{\s}{\sigma}

\newcommand*{\w}{\omega}
\newcommand*{\f}{\phi}
\newcommand*{\y}{\psi}

\newcommand*{\cF}{\mathcal{F}}
\newcommand*{\cL}{\mathcal{L}}
\newcommand*{\cN}{\mathcal{N}}
\newcommand*{\N}{\mathcal{N}}
\newcommand*{\cO}{\mathcal{O}}
\newcommand*{\Om}{\mathcal{O}_m}
\newcommand*{\Omv}{\langle \mathcal{O}_m \rangle}
\newcommand*{\Ophi}{\mathcal{O}_{\phi}}
\newcommand*{\Ophiv}{\langle \mathcal{O}_{\phi} \rangle}
\newcommand*{\T}{\mathcal{T}}
\newcommand*{\Pa}{\mathcal{P}}
\newcommand{\Jx}{\langle J_x \rangle}
\newcommand{\Jy}{\langle J_y \rangle}

\newcommand*{\p}{\partial}

\title{A Weyl Semimetal from AdS/CFT with Flavour}
\author[1]{Kazem Bitaghsir Fadafan,}
\author[2]{Andy O'Bannon,}
\author[3]{Ronnie Rodgers,}
\author[2]{Matthew Russell}

\affiliation[1]{Faculty of Physics, Shahrood University of Technology, P.O.Box 3619995161 Shahrood, Iran}
\affiliation[2]{STAG Research Centre, Physics and Astronomy, University of Southampton,\\Southampton, SO17 1BJ, United Kingdom}
\affiliation[3]{Institute for Theoretical Physics, Utrecht University, \\Princetonplein 5, 3584 CE Utrecht, the Netherlands}

\emailAdd{bitaghsir@shahroodut.ac.ir}
\emailAdd{a.obannon@soton.ac.uk}
\emailAdd{r.j.rodgers@uu.nl}
\emailAdd{m.j.russell@soton.ac.uk}

\abstract{We construct a top-down holographic model of Weyl semimetal states using $(3+1)$-dimensional $\N=4$ supersymmetric $SU(N_c)$ Yang-Mills theory, at large $N_c$ and strong coupling, coupled to a number $N_f \ll N_c$ of $\N=2$ hypermultiplets with mass $m$. A $U(1)$ subgroup of the R-symmetry acts on the hypermultiplet fermions as an axial symmetry. In the presence of a constant external axial gauge field in a spatial direction, $b$, we find the defining characteristic of a Weyl semi-metal: a quantum phase transition as $m/b$ increases, from a topological state with non-zero anomalous Hall conductivity to a trivial insulator. The transition is first order. Remarkably, the anomalous Hall conductivity is independent of the hypermultiplet mass, taking the value dictated by the axial anomaly. At non-zero temperature the transition remains first order, and the anomalous Hall conductivity acquires non-trivial dependence on the hypermultiplet mass and temperature.}

\begin{document}
    
\maketitle

\section{Introduction}
\label{sec:intro}

Weyl semimetals (WSMs) are a class of recently-discovered materials in which two electronic bands touch at isolated points in momentum space at or near the Fermi surface, such that the low energy excitations near these nodal points are $(3+1)$-dimensional relativistic Weyl fermions, with the Fermi velocity playing the role of the speed of light~\cite{Yan_2017,Hasan_2017,Burkov_2018,Armitage_2018,Gao_2019}.

If both parity (or inversion), $\Pa$, and time reversal, $\T$, are preserved then left- and right-handed Weyl fermions must appear in degenerate pairs. (For $\T$ this is the Kramers theorem.) Each such pair forms a Dirac fermion. To split the Dirac fermion into separate left- and right-handed Weyl fermions, either $\T$ or $\Pa$ must be broken. WSMs breaking either $\Pa$ or $\T$ have been experimentally discovered: TaAs~\cite{2015Sci...349..613X,2015Sci...349..622L,2015PhRvX...5c1013L,2015NatPh..11..724L} and its cousins TaP~\cite{liu2016a}, NbAs~\cite{Xu_2015}, and NbP~\cite{liu2016a,Belopolski_2016} break $\Pa$ and preserve $\T$, whereas magnetic WSMs like Co$_3$Sn$_2$S$_2$~\cite{Morali1286,Liu1282} and Co$_2$MnGa~\cite{Belopolski1278} preserve $\Pa$ and break $\T$.

WSMs are ``topological'' materials in the following sense. Each Weyl point has an associated topological invariant: the integral of the Berry curvature over a surface enclosing the Weyl point is a Chern number $\pm 1$, depending on the point's chirality~\cite{PhysRevB.83.205101}. The Weyl fermions are thus topologically protected, meaning they cannot be destroyed by any continuous deformation that leaves the discrete symmetries unchanged.

In lattice systems, the Nielsen-Ninomiya theorem~\cite{NIELSEN1981173} guarantees zero net chirality in the Brillouin zone, or equivalently zero net Chern number. In lattice realisations of WSMs, Weyl fermions will thus always appear in positive and negative chirality pairs, and a Weyl point can disappear only by annihilating against another Weyl point of opposite chirality. In particular, a lattice system can never support a single isolated Weyl fermion.

The presence of Weyl points has (at least) two major phenomenological consequences. The first is Fermi arcs at the material's surface, meaning lines at or near the Fermi energy in the \textit{surface} Brillouin zone, connecting the projections of the Weyl points~\cite{PhysRevB.83.205101}. Given that the bulk Weyl points are topologically protected, the existence of these Fermi arcs is as well. Fermi arc states can give rise to phenomena such as quantum oscillations~\cite{Armitage_2018}. The second consequence is exotic transport in the WSM's bulk, including the chiral magnetic effect, negative magneto-resistance, and the anomalous Hall effect~\cite{2004PhRvL..93t6602H}. These exotic effects arise in whole or in part from the axial anomaly of the Weyl fermions.

Perhaps the simplest field theory exhibiting the physics of WSMs is a free Dirac fermion \(\y\), of mass \(m\), with a non-dynamical background axial vector field $A^5_j$, where $j=x,y,z$ labels spatial coordinates. In units with $\hbar \equiv 1$ and the Fermi velocity \(v_f \equiv 1\), the Lagrangian $\cL$ of such a Dirac fermion is~\cite{Colladay:1998fq,Grushin:2012mt}
\beq
\label{eq:wsm_toy_model}
    \cL = \bar{\y} \le(i \g^\m \p_\m - m + A^5_j \g^j \g^5\ri) \y,
\eeq
where $\mu = t,x,y,z$ labels spacetime coordinates, $\g^\m$ are the Dirac matrices, and $\g^5 \equiv i \g^t\g^x\g^y\g^z$. In eq.~\eqref{eq:wsm_toy_model}, on the right-hand-side the kinetic term and mass term each preserve $\Pa$ and $\T$, while the coupling to $A^5_j$ preserves $\Pa$ but breaks $\T$. If we choose $A_j^5$ to have constant magnitude $b/2$ (the factor of $1/2$ is for later convenience), and use rotational symmetry to orient it in the $z$ direction, $A^5_j = b/2 \, \delta_{jz}$, then the energy $\ve$ of the Dirac fermion is, for spatial momentum $k_j$,
\beq
\label{eq:wsm_toy_model_spectrum}
    \ve = \pm \sqrt{
        k_x^2 + k_y^2 + \le(\frac{b^2}{4} \pm \sqrt{k_z^2 + m^2} \ri)^2,
    }
\eeq
where the $\pm$ signs are uncorrelated, so that eq.~\eqref{eq:wsm_toy_model_spectrum} describes four energy levels.

The qualitative form of the spectrum in eq.~\eqref{eq:wsm_toy_model_spectrum} depends on the dimensionless ratio \(|m/b|\). If \(|m/b| < 1/2\), then two of the four energy levels meet at two points in momentum space, \((k_x,k_y,k_z) = (0, 0, \pm \sqrt{(b/2)^2- m^2})\). At these points, \(\ve = 0\). The effective theory governing the low energy excitations near these two nodal points is then a pair of Weyl fermions, and thus for \(|m/b| < 1/2\) the system is a WSM. On the other hand, if \(|m/b| > 1/2\) then an energy gap appears, and the system is a trivial insulator. At the critical point \(|m/b|=1/2\), a single node at \((k_x,k_y,k_z) = 0\) appears.

The $\cL$ in eq.~\eqref{eq:wsm_toy_model} is invariant under \(U(1)_V\) vector transformations \(\y \to e^{i \a} \y\) with constant $\a$, and when $m=0$ and $b=0$ also under \(U(1)_A\) axial transformations \(\y \to e^{i \b \g^5} \y\), with constant $\b$. The corresponding $U(1)_V$ and $U(1)_A$ currents are, respectively,
\beq
    J^\m = i \bar{\y} \g^\m \y,
    \qquad
    J^\m_5 = i \bar{\y} \g^\m \g^5 \y.
\eeq
However, $U(1)_A$ is anomalous: in the presence of a background $U(1)_V$ field strength $F_{\mu\nu}$, the axial current \(J^\m_5\) is not conserved. Moreover, non-zero $m$ explicitly breaks $U(1)_A$. To be specific, if $m$ and $F_{\m\n}$ are both non-zero, then the divergence of the $U(1)_A$ current is
\beq
\label{eq:axialcons}
    \p_\m J^\m_5 = \frac{1}{16 \pi^2} \,\e^{\m\n\r\s} F_{\m\n} F_{\r\s} - 2 m \,\bar{\y} \g^5 \y.
\eeq

As mentioned above, the $U(1)_A$ anomaly gives rise to exotic transport. Our focus will be the anomalous Hall effect: if we introduce $A^5_j = b/2 \, \delta_{jz}$, then a constant, external $U(1)_V$ electric field, $E$, in a perpendicular direction, say $x$, induces a $U(1)_V$ Hall current~\cite{Grushin:2012mt,Goswami:2012db},
\beq
\label{eq:wsm_toy_model_hall_current}
J^y = \sigma_{yx} E, \qquad \sigma_{yx} = - \sigma_{xy} = \frac{1}{4\pi^2} \, \sqrt{b^2 - 4 m^2}\, \Theta\le(|b| - 2|m|\ri),
\eeq
with $J^x=0$ and $J^z=0$. In eq.~\eqref{eq:wsm_toy_model_hall_current}, the Heaviside step function \(\Theta\le(|b| - 2|m|\ri)\) makes manifest that the anomalous Hall effect occurs only when $|m/b|<1/2$, in the WSM phase.

The free Dirac fermion theory of eq.~\eqref{eq:wsm_toy_model} thus has a quantum phase transition as $|m/b|$ increases. As mentioned above, when $|m/b| < 1/2$ the system is a WSM, with $\Pa$ preserved and $\T$ broken at low energy, and an AHE with the $\sigma_{xy}\neq0$ in eq.~\eqref{eq:wsm_toy_model_hall_current}. When $|m/b|>1/2$ the system is a trivial insulator, with both $\Pa$ and $\T$ preserved at low energy and $\sigma_{xy}=0$. This quantum phase transition is second order, with a quantum critical point described by a scale-invariant field theory.

A crucial question is: what if the low-energy excitations of a material cannot be described by the free Dirac fermion theory of eq.~\eqref{eq:wsm_toy_model}? In fact, what if the low-energy excitations cannot be described by band theory at all? What if the low-energy excitations are not weakly-interacting, long-lived quasi-particles? Are WSM states possible in strongly-correlated materials, and if so, then what are their properties?

Some phenomena of the free Dirac fermion theory are independent of interactions, as long as those interactions do not change the discrete symmetries---specifically phenomena that are topological and/or determined by the $U(1)_A$ anomaly in eq.~\eqref{eq:axialcons}. Examples include the presence of Weyl points in the Brilloin zone and the corresponding Fermi arcs, both of which are topological, and the anomalous Hall conductivity $\sigma_{xy}$ in eq.~\eqref{eq:wsm_toy_model_hall_current} when $m=0$, which is completely determined by the $U(1)_A$ anomaly.

However, practically any other property will be affected by interactions, including the exact shape and energy dispersion of Fermi arcs, the value of the anomalous Hall conductivity when $m \neq 0$, and the thermodynamic equation of state, which in turn determines the order of any (quantum) phase transition and the critical value of $|m/b|$ at which it occurs. Indeed, effective field theory techniques have shown that with sufficiently strong \textit{short-range} interactions a WSM will experience either a first-order transition to a band insulator or a continuous transition to a broken symmetry phase~\cite{Roy_2017}.

An alternative approach to strongly-interacting WSMs is the Anti-de Sitter/CFT (AdS/CFT) correspondence, or simply holography~\cite{Maldacena:1997re,Gubser:1998bc,Witten:1998qj}. In holography, a strongly-interacting CFT, typically a non-abelian gauge theory in the 't Hooft limit of a large number of colors $N_c$, is equivalent to weakly-coupled gravity in one higher dimension, typically Einstein-Hilbert gravity coupled to matter fields in asymptotically AdS space. The CFT ``lives'' at the AdS boundary. A CFT describes massless fields and hence \textit{long-range} interactions. In holography, single-trace CFT operators of sufficiently low dimension are dual to sufficiently light fields in the gravity theory. For example, a conserved $U(1)$ current is dual to a massless $U(1)$ gauge field in AdS. Furthermore, CFT states with order $N_c^2$ entropy are dual to black holes, whose Bekenstein-Hawking entropy is order $N_c^2$~\cite{Gubser:1996de,Witten:1998zw}.

Current holographic models of strongly-interacting WSMs fall into three classes. The first consists of fermions with strong interactions mediated by a holographic CFT~\cite{Gursoy:2012ie,Jacobs:2015fiv}. Such a mix of non-holographic fermions with a holographic CFT is called ``semi-holographic''~\cite{Faulkner:2010tq,Gursoy:2011gz}. These models behave as undoped WSMs exhibiting quantum criticality generically with non-integer scaling of the conductivity in frequency and $T$~\cite{Gursoy:2012ie,Jacobs:2015fiv}.

The second class of models is fully holographic~\cite{Landsteiner:2015lsa,Landsteiner:2015pdh,Copetti:2016ewq,Liu:2018spp,Landsteiner:2019kxb,Juricic:2020sgg}, and consists of Einstein-Hilbert gravity in $(4+1)$-dimensional asymptotically AdS space ($AdS_5$) coupled to a complex scalar field and two $U(1)$ gauge fields, one of which has a five-dimensional Chern-Simons term. These matter fields are dual to the complex Dirac mass operator, the $U(1)_V$ current, and the $U(1)_A$ current, respectively, and roughly speaking the Chern-Simons term is dual to the $U(1)_A$ anomaly. Upon introducing non-zero $m$ and $b$, most models in this class exhibit a quantum phase transition as $|m/b|$ increases, from a WSM with $\Pa$ preserved and $\T$ broken at low energy, to a trivial semimetal with $\Pa$ and $\T$ preserved at low energy, with a Lifshitz critical point in between. However, some models in this class have a first-order quantum phase transition from a WSM to a Chern insulator, for suitable choices of scalar field couplings~\cite{Liu:2018spp}. A key feature of these models is an anomalous Hall conductivity completely determined by the product of the Chern-Simons coefficient and the $U(1)_A$ gauge field's value at the black hole horizon~\cite{Landsteiner:2015lsa,Landsteiner:2015pdh}. Models in this class have realised edge currents indicating Fermi arcs~\cite{Ammon:2016mwa}, odd viscosity~\cite{Landsteiner:2016stv}, chaos~\cite{Baggioli:2018afg}, and much more~\cite{Grignani:2016wyz,Ammon:2018wzb,Liu:2018djq,Baggioli:2020cld}.

Crucially, most models in these two classes are ``bottom-up,'' meaning they are \textit{ad hoc}, and may or may not be realised in a genuine string or supergravity (SUGRA) theory. Whether a dual CFT actually exists is thus unclear. However, one model in the second class was ``top-down,'' being a consistent truncation of 10-dimensional SUGRA, and thus having a dual CFT with known Lagrangian~\cite{Copetti:2016ewq}. Bottom-up models have the advantage of revealing generic phenomena independent of any specific model's details, while top-down models have the advantage of a known CFT dual. The latter offers the possibility of a non-holographic, purely CFT approach, for example using weak-coupling perturbation theory, and comparing to holography to reveal how observables evolve with coupling strength.

The third class is fully holographic and top-down, namely 10-dimensional type IIB SUGRA in $AdS_5 \times S^5$, with $N_c$ units of five-form flux on the five-phere $S^5$, plus a number $N_f$ of coincident probe D7-branes along $AdS_5 \times S^3$. The SUGRA is dual to a $(3+1)$-dimensional CFT, namely $\N=4$ supersymmetric (SUSY) $SU(N_c)$ Yang-Mills (SYM) theory, with large $N_c$ and large coupling~\cite{Maldacena:1997re}. The D7-branes are dual to a number $N_f$ of $\N=2$ hypermultiplets in the fundamental representation of $SU(N_c)$, i.e. flavour fields, in the probe limit $N_f \ll N_c$~\cite{Karch:2002sh}. The $\N=4$ SYM theory has $SO(6)$ R-symmetry, which the hypermultiplets break to $SO(4) \times U(1)_A$, where as indicated the $U(1)$ factor acts on the hypermultiplet fermions, i.e. the quarks, as the axial symmetry. When all $N_f$ flavours have the same mass $m$, the system has a global $U(N_f)$ symmetry whose diagonal $U(1)_V$ acts on the quarks as the vector symmetry. Using the holographic description, refs.~\cite{Hashimoto:2016ize,Kinoshita:2017uch} showed that a $U(1)_V$ electric field rotating in space produced non-equilibrium steady states that were in fact WSMs, among other remarkable properties, such as an effective temperature and fluctuation-dissipation relation~\cite{Hashimoto:2016ize}.

In this paper we initiate the analysis of a fully holographic, top-down model unlike those above, thought it borrows from the latter two classes. In particular, we consider $\N=4$ SYM coupled to probe hypermultiplets, similar to the third class of models mentioned above. However, instead of a rotating $U(1)_V$ electric field we introduce a $U(1)_A$ field $A_j^5 = b/2 \, \delta_{jz}$, similar to the second class of models mentioned above.

We find that this model exhibits several remarkable phenomena distinct from all previous models. For example, when $T=0$ we find a \textit{first order} transition from a WSM to a trivial insulator as $|m/b|$ increases, in contrast to the second order transitions of most previous models, and to the first order transition to a Chern insulator of ref.~\cite{Liu:2018spp}.

At $T=0$ our small $|m/b|$ phase is a WSM with $\sigma_{xx}=0$ and $\sigma_{xy}\neq0$, signaling broken $\mathcal{T}$ at low energy, while the large $|m/b|$ phase is a trivial insulator with $\sigma_{xx}=0$ and $\sigma_{xy}=0$, signaling restored $\mathcal{T}$ at low energy. Most remarkably, at $T=0$ in the WSM phase our $\sigma_{xy}$ is \textit{independent of $m$}, and in particular retains its $m=0$ value, dictated by the $U(1)_A$ anomaly, for all $|m/b|$ in the WSM phase. To our knowledge such behaviour does not occur in any other model. We also find that at $T=0$ in the WSM phase the low energy effective theory is a CFT, namely $\N=4$ SYM coupled to \textit{massless} probe hypermultiplets. In other words, the non-zero $m$ in the ultraviolet (UV) is renormalised to zero in the infra-red (IR).

For any $T>0$ we again find a first order transition, now from a WSM with $\sigma_{xx}\neq0$ and $\sigma_{xy}\neq0$ to a trivial insulator with $\sigma_{xx}=0$ and $\sigma_{xy}=0$. In other words, when $T>0$ in the WSM phase $\sigma_{xx}\neq0$ and $\sigma_{xy}\neq0$ both acquire non-trivial dependence on $m$ and $T$. We also explore our model's thermodynamics by computing our model's entropy density, heat capacity, and speed of sound. We find various curious features. For example, at sufficiently low $T$, in the WSM phase near the transition we find a rapid increase in the entropy density, presumably arising from the emergent IR CFT degrees of freedom.

Broadly speaking, individual holographic models can reveal what is possible with strong interactions, while families of holographic models can reveal what is universal with strong interactions. Our model shows that first order transitions from WSMs to trivial insulators are possible with strong interactions, accompanied by remarkable behaviour of thermodynamics and transport, and our model provides further evidence that transport properties controlled by anomalies are universal in the presence of strong interactions.

In section~\ref{sec:holographic_model} we describe our model in detail. In section~\ref{sec:thermo} we present our solutions for the D7-brane worldvolume fields, and use them to study the thermodynamics of our model. In section~\ref{sec:conductivities} we holographically compute the longitudinal and Hall conductivities of our model. We conclude in section~\ref{sec:summary} with a summary and outlook for future research using our model. We collect many technical results in three appendices.

\section{Holographic Model}
\label{sec:holographic_model}

In type IIB SUGRA in flat space with coordinates $x_0,x_1,\ldots,x_9$, we consider the following SUSY intersection of $N_c$ coincident D3-branes with $N_f$ coincident D7-branes (table~\ref{tab:embedding}).

\begin{table}
\begin{center}
\begin{tabular}{c | c c c c c c c c c c}
    & \(x_0\) & \(x_1\) & \(x_2\) & \(x_3\) & \(x_4\) & \(x_5\) & \(x_6\) & \(x_7\) & \(x_8\) & \(x_9\)
    \\
    \hline
    D3 & \(\times\) & \(\times\) & \(\times\) & \(\times\)
    \\
    D7 & \(\times\) & \(\times\) & \(\times\) & \(\times\)  & \(\times\) & \(\times\) & \(\times\) & \(\times\)
\end{tabular}
\end{center}
\caption{The D-brane intersection we consider. The $N_c$ D3-branes giving rise to SU($N_c$) $\mathcal{N}=4$ SYM are placed along the directions labelled $(x_0,\ldots,x_3)$. The $N_f$ D7-branes placed along $(x_0,\ldots,x_7)$ give rise to $\mathcal{N}=2$ hypermultiplets transforming in the fundamental representation of SU($N_c$).}
\label{tab:embedding}
\end{table}

On the D3-brane worldvolume, open strings with both ends on the D3-branes give rise at low energy to $\N=4$ SYM with gauge group $U(N_c)$, where the overall $U(1) \in U(N_c)$ will play no role in what follows so we will ignore it. The YM coupling constant given by $g_\mathrm{YM}^2=4\pi g_s$ with $g_s$ the closed string coupling.  The field content of $\N=4$ SYM is a vector field, four Weyl fermions, and six real scalar fields, all in the adjoint representation of $SU(N_c)$. The theory has an $SO(6)_R$ R-symmetry, realised in the D-brane intersection as the rotational symmetry in the directions $(x_4, \ldots, x_9)$ normal to the D3-branes.

Open strings with one end on the D3-branes and one end on the D7-branes give rise at low energy on the D3-brane worldvolume to $N_f$ $\N=2$ hypermultiplets in the fundamental representation of $SU(N_c)$, i.e. flavour fields. A hypermultiplet's field content is a Dirac fermion $\psi$ and a pair of complex scalar fields $q$ and $\tilde{q}$. The Dirac fermion $\psi$ will be our ``quark'' or ``electron'', with the other fields serving to mediate their interactions.

In the D-brane intersection the D7-branes clearly break the $SO(6)_R$ rotational symmetry in the directions $(x_4, \ldots, x_9)$ down to $SO(4) \simeq SU(2) \times SU(2)$ rotations in the directions $(x_4,x_5,x_6,x_7)$ normal to the D3-branes but parallel to the D7-branes, and $SO(2)\simeq U(1)$ rotations in the directions $(x_8,x_9)$ normal to both sets of D-branes. Correspondingly, in the D3-brane worldvolume theory the couplings of massless flavour fields break $SO(6)_R \to SU(2) \times SU(2)_R \times U(1)_A$ where as indicated $SU(2)_R$ acts as an R-symmetry and $U(1)_A$ is an R-symmetry that acts on the quarks as the axial symmetry.

Separating the D7-branes from the D3-branes in $(x_8,x_9)$ by a distance $R$ preserves $\N=2$ SUSY and gives the open strings stretched between them a minimum mass $R/(2 \pi \alpha')$, with $\alpha'$ the string length squared. We thus identify the hypermultiplet mass as $m = R/(2 \pi \alpha')$. Such separation preserves the rotational symmetry in $(x_4,x_5,x_6,x_7)$ but breaks that in $(x_8,x_9)$, corresponding to $m$ preserving $SU(2) \times SU(2)_R$ but explicitly breaking $U(1)_A$. The angle of separation in $(x_8,x_9)$ is the phase of the hypermultiplet mass, $\phi$.

We will consider only a number $N_f$ of coincident D7-branes in $(x_8,x_9)$, so the hypermultiplets have a global $U(N_f)$ symmetry whose diagonal acts on the quarks as $U(1)_V$.

Being top-down, our model has the attractive feature that the Lagrangian is known: see for example refs.~\cite{Chesler:2006gr,Erdmenger:2007cm} for explicit expressions. However, its full form is lengthy, so we will write only the terms we need, namely terms in the potential $V$ that involve the complex hypermultiplet mass $m \, e^{i \phi}$,
\beq
\label{eq:mass_terms}
    V \supset m \bar{\y} e^{i \f \g^5} \y - m q^\dagger \le(e^{i\f} \Gamma^\dagger + e^{-i\f} \Gamma \ri) q  - m \tilde{q}^\dagger \le(e^{i\f} \Gamma^\dagger + e^{-i\f} \Gamma \ri) \tilde{q} + m^2 \le( q^\dagger q + \tilde{q}^\dagger \tilde{q} \ri),
\eeq
with $\Gamma$ a complex scalar field formed from two of the six real scalar fields of $\N=4$ SYM.

If we transform $\psi \to e^{-i \f \g^5/2} \psi$, then the first term in eq.~\eqref{eq:mass_terms} becomes $m \, \bar{\psi} \psi$. Moreover, if $\phi$ depends on the field theory spacetime coordinates $(x_0,x_1,x_2,x_3)=(t,x,y,z)$ then the derivative in $\psi$'s kinetic term will act on $\phi$, producing a new term that we may include in the potential. The terms in $V$ that depend only on $\y$ then become
\beq
\label{eq:axialpot}
    V \supset  \bar{\y} \le(m - \frac{\p_\m \f}{2} \g^\m \g^5 \ri)\y.
\eeq
Comparing to the Dirac Lagrangian in eq.~\eqref{eq:wsm_toy_model}, in eq.~\eqref{eq:axialpot} the second term on the right-hand side clearly represents a coupling to an external, non-dynamical $U(1)_A$ gauge field, $A^5_{\mu} = \partial_{\mu} \phi/2$. As in the effective theory of eq.~\eqref{eq:wsm_toy_model}, to produce a WSM we will choose $\phi = b \, z$.\footnote{In this model the effect of a $U(1)_A$ chemical potential $\mu_5$, introduced as $\phi = 2 \mu_5 t$, was studied holographically for example in refs.~\cite{Das:2010yw,Hoyos:2011us}.} In the D-brane intersection, $\phi = b\,z$ corresponds to D7-branes spiraling around the D3-branes in the $(x_8,x_9)$ plane as they extend along $z$.

The operators sourced by $m$ and $\phi$ are, respectively,
\begin{align}
\label{eq:field_theory_operators}
    \cO_m &\equiv \frac{\partial V}{\partial m} = \bar{\y} e^{i \f \g^5} \y - q^\dagger \le( e^{i\f} \Gamma^\dagger + e^{- i \f} \Gamma \ri) q 
    - \tilde{q}^\dagger  \le( e^{i\f} \Gamma^\dagger + e^{- i \f} \Gamma \ri)  \tilde{q}
    + 2 m \le(q^\dagger q + \tilde{q}^\dagger \tilde{q} \ri),
    \nonumber \\
    \cO_\f &\equiv  \frac{\partial V}{\partial \f} = i m \bar{\y} e^{i \f \g^5} \g^5 \y - i m q^\dagger \le( e^{i\f} \Gamma^\dagger -  e^{- i \f} \Gamma \ri) q
    - i m \tilde{q}^\dagger  \le( e^{i\f} \Gamma^\dagger - e^{- i \f} \Gamma \ri)  \tilde{q}.
\end{align}
The operator $\Om$ is dimension 3, and when $\phi$ is constant is just the SUSY completion of the Dirac mass operator. The operator $\Ophi$ is dimension 4, and obeys $\Ophi \propto m$, so that if $m=0$ then $\Ophi=0$. The conserved \(U(1)_V\) current is
\begin{equation}
    J^\mu = i \bar{\psi} \gamma^\mu \psi - i \left[q^\dagger (D^\mu q) - (D^\mu q)^\dagger q \right]- i \left[\tilde{q} (D^\mu \tilde{q})^\dagger - (D^\mu \tilde{q}) \tilde{q}^\dagger \right],
\end{equation}
where \(D^\mu\) denotes the gauge-covariant derivative.

We will take the 't Hooft limit $g_\mathrm{YM}^2 \to 0$ and $N_c \to \infty$ with fixed 't Hooft coupling $\l \equiv g_\mathrm{YM}^2 N_c$, followed by the strong coupling limit $\l \to \infty$. In these limits $\N=4$ SYM is holographically dual to type IIB SUGRA in the near-horizon geometry of the D3-branes, $AdS_5 \times S^5$~\cite{Maldacena:1997re,Gubser:1998bc,Witten:1998qj}. We will also take the probe limit, $N_f \ll N_c$, in which case the $N_f$ hypermultiplets are holographically dual to $N_f$ probe D7-branes along $AdS_5 \times S^3$~\cite{Karch:2002sh}.

To study our system with non-zero temperature $T$, we generalise the $AdS_5$ factor to an $AdS_5$-Schwarzschild black brane~\cite{Gubser:1996de,Witten:1998zw}. The type IIB SUGRA solution then has all fields trivial except for the metric and the four-form, $C_4$, which are given by
\begin{subequations}
 \label{eq:background_solution}
\begin{align}
    d s^2 &= \frac{\r^2}{L^2} \le(
        - \frac{g^2(\r)}{h(\r)} d t^2 + h(\r) d \vec{x}^2
    \ri)
    + \frac{L^2}{\r^2} \le(
        d r^2 + r^2 d s_{S^3}^2 + d R^2 + R^2 d \f^2
    \ri),
    \nonumber \\
    C_4 &= \frac{\r^4}{L^4} h^2(\r) d t \wedge d x \wedge d y \wedge d z - \frac{L^4 r^4}{\r^4} d \f \wedge \w(S^3),
\end{align}
\beq
    \rho^2 = r^2 + R^2, \qquad g(\r) \equiv 1 - \frac{\r_H^4}{\r^4},
    \qquad
    h(\r) \equiv 1 + \frac{\r_H^4}{\r^4},
\eeq
\end{subequations}
where $\rho \in [\rho_H , \infty)$ with the black brane horizon at $\rho_H$ and the asymptotic $AdS_5$ boundary at $\rho \to \infty$, the $AdS_5$ radius $L$ is given by $L^4/\a'^2 = \lambda$, and $d s_{S^3}^2$ and $\w(S^3)$ denote the round metric and volume form on a unit-radius $S^3$, respectively. The black brane's Hawking temperature is
\beq
\label{eq:hawkingT}
    T = \frac{\sqrt{2}}{\pi} \frac{\r_H}{L^2},
\eeq
which is also the dual field theory's temperature~\cite{Gubser:1996de,Witten:1998zw}. If $T=0$ then $\rho_H=0$ and so $g(\r)=1$ and $h(\r) = 1$, in which case the solution in eq.~\eqref{eq:background_solution} becomes exactly $AdS_5 \times S^5$.

The part of the D7-brane action that we will need is a sum of abelian Dirac-Born-Infeld (DBI) and Wess-Zumino (WZ) terms,
\beq
\label{eq:d7_action}
    S_\mathrm{D7} = - N_f T_\mathrm{D7} \int d^8 \xi \sqrt{-\det (P[G] + F)}
    + \frac{1}{2} N_f T_\mathrm{D7} \int P[C_4] \wedge F \wedge F,
\eeq
where the D7-brane tension is $T_\mathrm{D7} = (2\pi)^{-7} g_s^{-1} \a'^{-4}$, $\xi^a$ with $a=1,2,\ldots,8$ are the worldvolume coordinates, $P[G]$ and $P[C_4]$ denote pullbacks of the bulk metric and four-form to the worldvolume, respectively, and $F = d A$ is the field strength of the $U(1)$ worldvolume gauge field $A$. Compared to $F$'s textbook definition~\cite{Polchinski:1998rr}, we have absorbed a factor of $(2\pi\alpha')$ into our $F$, which is thus dimensionless. The D7-branes' worldvolume $U(N_f)$ gauge invariance is dual to the $U(N_f)$ flavour symmetry, and in particular the $U(1)$ gauge field $A$ is dual to the $U(1)_V$ current $J^{\mu}$.

The D7-branes are extended along $AdS_5 \times S^3$. We parametrise the D7-branes' worldvolume coordinates as $\xi^a = (t,x,y,z,r)$ plus the $S^3$ coordinates. The two worldvolume scalars are then $R$ and $\phi$, where $R$ is holographically dual to $\Om$ and $\phi$ is dual to $\Ophi$ in eq.~\eqref{eq:field_theory_operators}. More specifically, in the near-boundary region on the worldvolume, $r \to \infty$, the leading asymptotic values of $R$ and $\phi$ determine the sources for $\Om$ and $\Ophi$, i.e. the modulus $m$ and phase $\phi$ of the hypermultiplet mass, respectively, while the sub-leading behaviours determine the expectation values $\Omv$ and $\Ophiv$.

We will use the simplest ansatz for the worldvolume scalars that introduces the phase $bz$ in the hypermutiplets' mass and allows for non-zero $\Omv$ and $\Ophiv$, namely $R(r)$ and $\phi(z,r) = bz + \Phi(r)$. A similar ansatz for \(\phi\), in the presence of a magnetic field, has appeared in refs.~\cite{Kharzeev:2011rw,Bu:2018trt}. Crucially, because of the background's Killing symmetry that shifts $\phi$, the D7-branes' action will depend only on derivatives of $\phi$, and hence with our ansatz will not depend explicitly on $z$. As a result, our ansatz can otherwise self-consistently depend on $r$ alone. In practical terms, our goal will be to solve for $R(r)$ and $\Phi(r)$, and from these extract the dual flavour fields' thermodynamic and transport properties.

Our ansatz has been chosen to reduce the D7-brane equations of motion to ordinary differential equations, greatly simplifying calculations. Of course, there is a risk that the solution that globally minimises the action for given boundary conditions is not captured by this ansatz, and so a complete and thorough analysis should eventually consider general \(\f(z,r)\). That the solutions we find are at least \textit{local} minima can be tested by computing the spectrum of linear perturbations of the D7-brane: a necessary, but not sufficient, condition is that there are no unstable modes, i.e. perturbations that grow with time. We intend to report on this in future work.

With our ansatz, the pullback of $C_4$ to the worldvolume becomes
\beq
\label{eq:C4_pullback}
    P[C_4] = \frac{\r^4}{L^4} d t \wedge d x \wedge d y \wedge d z - \frac{L^4 r^4}{\r^4} \le(b \, d z +  \frac{\p \f}{\p r} d r \ri) \wedge \w(S^3),
\eeq
and hence the WZ term in the action eq.~\eqref{eq:d7_action} includes a term $\propto \int_{AdS_5} \frac{r^4}{\r^4} b \, dz \wedge F \wedge F$, which holographically encodes the $U(1)_A$ anomaly.

Plugging our ansatz into the D7-branes' action eq.~\eqref{eq:d7_action} gives
\begin{subequations}
\beq
    S_\mathrm{D7} = - \cN \mathrm{vol}(\mathbb{R}^{1,3}) \int d r \, r^3 \, g \, h
        \sqrt{ \le(1 + \frac{L^4 b^2 R^2}{h\left(r^2+R^2\right)^2} \ri)\le(1 + R'^2\ri) +  R^2 \f'^2},
        \label{eq:d7_action_on_ansatz}
\eeq
\beq
\label{eq:cNdef}
\cN \equiv  2 \pi^2 N_f T_\mathrm{D7} = \frac{\l N_f N_c}{16 \pi^4}\frac{1}{L^8},
\eeq
\end{subequations}
where $R' \equiv \partial R/\partial r$ and similarly for $\phi'$, and the factor $\mathrm{vol}(\mathbb{R}^{1,3})$ denotes the infinite volume of Minkowski space, arising from integration over the field theory directions $(t,x,y,z)$. Starting now we will divide both sides of eq.~\eqref{eq:d7_action_on_ansatz} by $\mathrm{vol}(\mathbb{R}^{1,3})$, so that $S_\mathrm{D7}$ will be an action \textit{density}. Correspondingly, quantities derived from $S_\mathrm{D7}$ will be densities.

For our ansatz, the canonical momentum $P_\f$ conjugate to $\f$ is
\beq
\label{eq:phi_momentum}
    P_\f \equiv \frac{\d S_\mathrm{D7}}{\d \f'} = - \cN r^3 \, g \, h \, \frac{R^2 \f'}{\sqrt{ \le(1 + \frac{L^4 b^2 R^2}{h\left(r^2+R^2\right)^2} \ri)\le(1 + R'^2\ri) +  R^2 \f'^2}}.
\eeq
The equation of motion for $\f$ is then $\partial_r P_\f=0$, so that $P_\f$ is a constant of motion in the worldvolume holographic direction, $r$. We thus write the solution as $P_\f = \cN p_\f$ where the factor of $\cN$ is a convenient normalisation, and $p_\f$ is a constant.

Plugging $P_\f = \N p_\f$ into eq.~\eqref{eq:phi_momentum} and solving for $\phi'^2$ gives
\beq
\label{eq:phisol}
\phi'^2 = p_\f \, \frac{\left[ \left(r^2+R^2\right)^2 + L^4 b^2 R^2\right]\left( 1+ R'^2\right)}{\left(r^2+R^2\right)^2 R^2\left(R^2 r^6-p_\f^2\right)}.
\eeq
Subsequently plugging $\phi'^2$ in eq.~\eqref{eq:phisol} into the action eq.~\eqref{eq:d7_action_on_ansatz} and Legendre transforming with respect to $\phi$ then gives an effective action for $R(r)$ alone,
\begin{align}
\label{eq:d7_action_LT}
    \tilde{S}_\mathrm{D7} &\equiv S_\mathrm{D7} - \int dr \, P_\f \f'
    \nonumber \\
    &= - \cN \int dr \, r^3 \, g \, h \sqrt{1 + R'^2} \sqrt{1 + \frac{L^4 b^2 R^2}{h\left(r^2+R^2\right)^2}} \sqrt{1 - \frac{p_\f^2}{r^6 g^2 h^2 R^2}},
\end{align}
whose variation gives $R(r)$'s equation of motion.

When $T>0$ two classes of solutions for $R(r)$ are possible~\cite{Mateos:2006nu}. In both classes, at the asymptotic $AdS_5$ boundary $r \to \infty$ the D7-branes wrap an equatorial $S^3 \in S^5$, and as the D7-branes extend into the bulk, towards smaller $r$, the $S^3$ radius $r^2/(r^2+R(r)^2)$ shrinks. In the first class of solutions, the D7-branes intersect  the horizon at some value $r_H>0$ of $r$ such that $\rho_H^2 = r_H^2 + R(r_H)^2$. These are called ``black hole embeddings.'' In the second class, the D7-branes do not intersect the horizon, and instead extend all the way to $r=0$. These are called ``Minkowski embeddings.'' These two classes are distinguished by topology: in black hole embeddings the $S^3$ maintains non-zero radius for all $r$ down to the horizon, whereas in Minkowski embeddings the $S^3$ radius shrinks to zero at $r=0$. In the latter case the D7-branes ``end'' at $\rho = R(r=0)$~\cite{Karch:2002sh}. Black hole embeddings describe a gapless and continuous spectrum of excitations, i.e. conducting states, while Minkowski embeddings describe a gapped and discrete spectrum, i.e. insulating states~\cite{Hoyos:2006gb,Karch:2007pd}

The integrand of $\tilde{S}_\mathrm{D7}$ in eq.~\eqref{eq:d7_action_LT} includes a product of three square roots. For both black hole and Minkowski embeddings, the arguments of the first and second square roots are positive for all $r$, hence both of these square roots are real-valued for all $r$.

However, if $p_\f\neq0$ then the third square root is never real-valued for all $r$. At the asymptotic $AdS_5$ boundary $r \to \infty$ the argument of the third square root is positive and hence the third square root is real-valued. For black hole embeddings, the argument of the third square root diverges to negative infinity at $r_H$ because $g(\rho_H)=0$, while for Minkowski embeddings it diverges to negative infinity at $r=0$. In each case the argument of the third square root must change sign at some $r$ between the asymptotic $AdS_5$ boundary and the horizon or brane endpoint, so for some values of $r$ the third square root always acquires a non-zero imaginary part. As a result, $\tilde{S}_{\mathrm{D7}}$ acquires a non-zero imaginary part, which signals a tachyonic instability with decay rate $\propto |\textrm{Im} \,\tilde{S}_{\mathrm{D7}}|$~\cite{Hashimoto:2013mua,Hashimoto:2014dza,Hashimoto:2014yya}.

Similar tachyons appear in other probe brane systems, when a square root factor acquires a non-zero imaginary part: see for example refs.~\cite{Karch:2007pd,OBannon:2007cex,Das:2010yw,Hoyos:2011us}. In those cases we can ``fix the problem,'' i.e. prevent the instability, by adding to our ansatz non-zero components of the worldvolume gauge field $A$. These come with their own integration constants, and typically produce additional factors under the square root that can be arranged such that the action remains real. Indeed, we will do precisely this in section~\ref{sec:conductivities}, where we will introduce a constant, non-dynamical, external $U(1)_V$ electric field $E$, and to avoid a tachyonic instability we introduce components of $A$. In field theory terms, we will introduce $E$ which will in turn induce $U(1)_V$ currents.

However, that strategy does not work when $E=0$ and $p_\f\neq0$. In that case, even if we introduce all components of $A$ in field theory directions, $(A_t(r),A_x(r),A_y(r),A_z(r))$, then the corresponding integration constants cannot be arranged to keep the square root real for all $r$. In particular, these integration constants appear in the Legendre-transformed action under the third square root as terms added to those in eq.~\eqref{eq:d7_action_LT}, but with powers of $r$ sub-leading compared to the $p_\f$ term at small $r$. As a result, these integration constants cannot be adjusted to keep the square root real for all $r$.

The upshot is that we will set $p_\f=0$ in all that follows, to guarantee that $\tilde{S}_{\mathrm{D7}}$ in eq.~\eqref{eq:d7_action_LT} is always real, and hence no tachyonic instability appears. In appendix~\ref{sec:holo_rg} we perform the holographic renormalisation of our model and find $\Ophiv  = P_\f = \N p_\f$, so our choice $p_\f=0$ means $\Ophiv=0$.

With our choice $p_\f=0$, the near-boundary asymptotic expansion of $R(r)$ is
\beq
\label{eq:R_near_boundary}
    R(r) =  M L^2 \le(1 - \frac{L^4 b^2}{2 \, r^2} \log (r/L)\ri) + C \, \frac{L^6}{r^2} + \mathcal{O}\left(\frac{\log (r/L)}{r^4}\right),
\eeq
where the constants $M$ and $C$ determine all subsequent coefficients in the large-$r$ expansion, and hence determine the entire solution $R(r)$. Consequently, $M$ and $C$ must map to $m$ and $\Omv$. Indeed, as mentioned above, the asymptotic separation $\lim_{r\to \infty} R(r) = M L^2$ determines $m = M L^2/(2 \pi \a') = M \sqrt{\l}/(2\pi)$. In appendix~\ref{sec:holo_rg} we show that $M$ and the sub-leading asymptotic coefficient $C$ together determine $\Omv$ as 
\beq
\label{eq:condensate_formula}
    \langle \cO_m \rangle = \frac{\sqrt{\l} N_f N_c}{8 \pi^3} \le[- 2 C + \frac{b^2 M}{2} + b^2 M \log\le(M L\ri)\ri].
\eeq

\section{Thermodynamics}
\label{sec:thermo}

In this section we will explore our model's thermodynamics. Specifically, for different classes of solutions of $R(r)$, characterised by boundary conditions, we will compute the hypermultiplets' contribution to the (Helmholtz) free energy density, $f$. Given $f$ we can also compute the thermal expectation value $\Omv = \partial f/\partial m$, which in terms of the near-boundary asymptotic coefficients $M$ and $C$ is given by eq.~\eqref{eq:condensate_formula}, and the hypermultiplets' contribution to the entropy density, $s$, and heat capacity density, $c_V$,
\beq
\label{eq:entropy}
s = -\frac{\partial f}{\partial T}, \qquad c_V = T \frac{\partial s}{\partial T}.
\eeq
In our case, where all chemical potentials vanish, we can also compute the speed of sound, $v$, from these thermodynamic quantities, as follows. The entropy density of the $\N=4$ SYM fields is $s_\mathrm{YM} = \frac{\pi^2}{2} N_c^2 T^3$ and their heat capacity density is $c_V^\mathrm{YM} = 3 \, s_\mathrm{YM}$~\cite{Gubser:1996de}. The total entropy density and heat capacity density are then $s_\mathrm{tot} = s_\mathrm{YM} + s$ and $c_V^\mathrm{tot}=c_V^\mathrm{YM}+c_V$, respectively. The speed of sound is then given by
\beq
\label{eq:soundspeed}
v^2 = \frac{s_\mathrm{tot}}{T} \frac{\partial T}{\partial s_\mathrm{tot}} = \frac{s_\mathrm{tot}}{c_V^\mathrm{tot}} = \frac{s_\mathrm{YM} + s}{c_V^\mathrm{YM}+c_V} = v^2_\mathrm{YM} + \delta v^2 + \mathcal{O}\left(N_f^2/N_c^2\right),
\eeq
where in the final equality we expanded in the probe limit $N_f \ll N_c$, with leading term $v^2_\mathrm{YM}=s_\mathrm{YM}/c_V^\mathrm{YM}$, which takes the value required for a $(3+1)$-dimensional CFT, $v^2_\mathrm{YM}=1/3$, and the $\mathcal{O}\left(N_f/N_c\right)$ correction due to the hypermultiplets is
\beq
\label{eq:soundcorr}
\delta v^2 = \frac{s_\mathrm{YM}}{c_V^\mathrm{YM}}\left(\frac{s}{s_\mathrm{YM}}-\frac{c_V}{c_V^\mathrm{YM}}\right).
\eeq
Given $f$ we can thus compute $s$ and $c_V$, and hence $\delta v^2$.

In holography, $f$ is simply minus the on-shell D7-brane action $S_{\mathrm{D7}}$ in eq.~\eqref{eq:d7_action} (not $\tilde{S}_{\mathrm{D7}}$ in eq.~\eqref{eq:d7_action_LT}) in Euclidean signature~\cite{Witten:1998zw}. In appendix~\ref{sec:holo_rg} we show that
\begin{align}
\label{eq:free_energy_formula}
    f = \frac{\l N_f N_c}{16 \pi^4 L^8} \lim_{r_c\to\infty}\biggl[&
        \int^{r_c} dr \, r^3 \,g\, h\, \sqrt{1 + \frac{L^4 b^2 R^2}{h\left(r^2+R^2\right)^2}} \sqrt{1 + R'^2}
    \\  &\hspace{1.5cm}
        - \frac{r_c^4}{4} - \frac{L^8 b^2 M^2}{2} \log \le( r_c / L\ri) + \frac{L^8 b^2 M^2}{4} \le(1 + 2 \log \le(M L \ri)\ri)
    \biggr],   \nonumber
\end{align}
where $r_c$ is a large-$r$ cutoff, and the lower endpoint of integration is $r_H$ for black hole embeddings and $r=0$ for Minkowski embeddings.

In the field theory the free parameters are $m$, $b$, and $T$, all with dimensions of mass. We will plot most physical quantities in units of $b$, and specifically as functions of the dimensionless ratios  $T/b$ and $m/(b\sqrt{\l})=M/(2\pi b)$. Most of our results will be numerical, although we will obtain closed-form results in certain limits.

As mentioned in section~\ref{sec:intro}, our main result is that for all $T/b$ we find a first-order transition as $m/(b\sqrt{\l})$ increases. In holographic terms, the transition is from black hole to Minkowski embeddings. In CFT terms, we find that $f$ is of course continuous, but has a discontinuous first derivative $\Omv = \partial f/\partial m$ at the transition. Our results are summarised in the phase diagram of figure~\ref{fig:phase_diagram}. In section~\ref{sec:conductivities}, by computing the conductivity we show that the transition is in fact from a WSM to a trivial insulator.

\subsection{Phase Transition at Zero Temperature}
\label{sec:T0_solutions}

We start with $T/b=0$, in which case the only scale in the field theory is $m/(b\sqrt{\l})$. When $T=0$ in eq.~\eqref{eq:hawkingT} the horizon disappears, $\rho_H=0$, and in eq.~\eqref{eq:background_solution} $g= 1$ and $h= 1$. Taking also $p_\f=0$, the equation of motion for $R(r)$ following from the Legendre transformed action $\tilde{S}_{\mathrm{D7}}$ in eq.~\eqref{eq:d7_action_LT} is
\begin{multline}
\label{eq:T0_EOM}
    R''+ \le(\frac{3}{r} - \frac{2 L^4 b^2 r R^2}{(r^2 + R^2) \le[(r^2+ R^2)^2 + L^4 b^2 R^2 \ri]} \ri) R' (1 + R'^2)
    \\
    + \frac{L^4 b^2 R (R^2 - r^2)}{(r^2 + R^2) \le[(r^2+ R^2)^2 + L^4 b^2 R^2 \ri]} (1 + R'^2) = 0.
\end{multline}

Without a horizon, all embeddings reach $r=0$. We can divide the embeddings into two classes, distinguished by whether $R_0 \equiv R(r=0)$ vanishes. In the first class of embeddings, $R_0\neq0$. Specifically, by expanding $R(r)$ around $r=0$ in eq.~\eqref{eq:T0_EOM} we find
\begin{subequations}
\label{eq:T0_boundary_conditions}
\beq \label{eq:T0_boundary_condition_analytic}
    R(r) = R_0 - \frac{L^4 b^2 r^2}{8 R_0\le(L^4 b^2 + R_0^2 \ri)} + \cO(r^4).
\eeq
These are Minkowski embeddings: at $r=0$ we have $\rho  = R_0 \neq 0$, so the $S^3 \in S^5$ collapses to zero size outside the $AdS_5$ Poincar\'e horizon $\rho=0$. In the second class of embeddings $R_0=0$, and in fact from eq.~\eqref{eq:T0_EOM} we find $R(r)$ vanishes exponentially quickly as $r\to0$,
\beq
\label{eq:T0_boundary_condition_non_analytic}
    R(r) = \eta \frac{e^{-L^2 b/r}}{\sqrt{r}} \le[1 + \cO(r^2) \ri],
\eeq
where $\eta$ is a constant. These are analogous to $T>0$ black hole embeddings: since $R_0=0$ we find that $\rho^2 = r^2 + R(r)^2$ vanishes at $r=0$, so that the D7-branes intersect the $AdS_5$ Poincar\'e horizon. These two classes are separated by a critical embedding, $R_{\mathrm{c}}(r)$, which from eq.~\eqref{eq:T0_EOM} we find approaches $R_0=0$ linearly in $r$,
\beq
\label{eq:T0_boundary_condition_critical}
    R_{\mathrm{c}}(r) = \frac{r}{\sqrt{3}} - \frac{32 r^3}{27 \sqrt{3} L^4 b^2} + \cO(r^5).
\eeq
\end{subequations}

For any value of $b$, eq.~\eqref{eq:T0_EOM} admits a trivial solution, $R(r)=0$, which has $R_0=0$ and in eq.~\eqref{eq:R_near_boundary} also $M=0$ and $C=0$. As a result, this solution describes $m=0$ and $\Omv=0$, and a straightforward calculation shows that also $f=0$.

We can obtain approximate solutions with non-zero $m$ in two limits, large $m$ and small $m$. More precisely, large mass means $m/(b \sqrt{\lambda}) \gg 1$. In that limit, following ref.~\cite{Filev:2007gb} we take $R(r) = ML^2 + \d R(r)$ and linearise the equation of motion eq.~\eqref{eq:T0_EOM} in $\d R$, also keeping only leading-order terms in $1/\le(L^4M^2 + r^2\ri)$, with the result
\beq
\label{eq:zeroTlargeMsol}
    \d R'' + \frac{3}{r} \d R' + \frac{L^6 b^2 M \le( M^2 - r^2 \ri)}{\le(L^4 M^2 + r^2\ri)^3} = 0, \qquad \frac{m}{b\sqrt{\l}} \gg 1.
\eeq
The solution of eq.~\eqref{eq:zeroTlargeMsol} regular as $r\to0$ and with the large-$r$ asymptotics of eq.~\eqref{eq:R_near_boundary} is
\beq
\label{eq:T0_solution_large_mass}
    R(r) \approx L^2 M + \frac{L^6 b^2 M}{4} \le[\frac{1}{L^4 M^2 + r^2} - \frac{1}{r^2} \log\le(1 + \frac{r^2}{L^4 M^2} \ri)\ri], \qquad \frac{m}{b\sqrt{\l}} \gg 1.
\eeq
This solution has $R_0=L^2 M \neq 0$, as in eq.~\eqref{eq:T0_boundary_condition_analytic}, and is therefore a Minkowski embedding. This solution has the large-$r$ asymptotics of eq.~\eqref{eq:R_near_boundary}, with
\beq
\label{eq:largemC}
C = \frac{1}{4} M b^2 \le[1 + 2 \log (M L) \ri], \qquad \frac{m}{b\sqrt{\l}} \gg 1.
\eeq
Substituting this into eq.~\eqref{eq:condensate_formula} then gives $\Omv = 0$. Integrating $\Omv$ over $m$ then trivially gives a free energy independent of \(m\), as expected in the limit $\frac{m}{b\sqrt{\l}} \gg 1$ where the hypermultiplets decouple. Concretely, we find \(f \to - b^4 N_f N_c / (512 \pi^4)\) as \(m/(b\sqrt{\l}) \to \infty\).

Small mass means $m/(b \sqrt{\lambda}) \ll 1$, where we may linearise the equation of motion eq.~\eqref{eq:T0_EOM} in $R(r)$, finding
\beq
\label{eq:zeroTsmallMsol}
    R'' + \frac{3}{r} R' - \frac{L^4 b^2}{r^4} R = 0, \qquad
    \frac{m}{b\sqrt{\l}} \ll 1.
\eeq
The solution of eq.~\eqref{eq:zeroTsmallMsol} regular as $r \to 0$ and with large-$r$ asymptotics as in eq.~\eqref{eq:R_near_boundary} is
\beq
 \label{eq:T0_solution_small_mass}
    R(r) \approx \frac{L^4 b M}{r} K_1(L^2 b/r), \qquad \frac{m}{b\sqrt{\l}} \ll 1,
\eeq
with modified Bessel function $K_1$. This solution vanishes exponentially as $r\to 0$, as in eq.~\eqref{eq:T0_boundary_condition_non_analytic}, with $\eta = L^3 M \sqrt{\pi b/2}$, and hence is analogous to a $T>0$ black hole embedding. This solution has the large-$r$ asymptotics of eq.~\eqref{eq:R_near_boundary}, with
\beq
C = \frac{1}{4} b^2 M\le[2 \log \le(bL/2\ri)2 -1 + 2\g_\mathrm{E} \ri], \qquad \frac{m}{b\sqrt{\l}} \ll 1,
\eeq
with Euler-Mascheroni constant $\g_\mathrm{E} \approx 0.577$. Using $m = M\sqrt{\l}/2\pi$, eq.~\eqref{eq:condensate_formula} then gives
\beq
    \langle \cO_m \rangle \approx \frac{N_f N_c}{4 \pi^2} \, m \, b^2 \le[
        \log \le( \frac{4 \pi m}{b \sqrt{\l}} \ri) + 1 - \g_\mathrm{E}
    \ri],
    \qquad
    \frac{m}{b\sqrt{\l}} \ll 1.
    \label{eq:small_m_condensate_approximate}
\eeq
We then obtain $f$ by integrating eq.~\eqref{eq:small_m_condensate_approximate} with respect to $m$, fixing the integration constant using the fact that the trivial solution $R(r)=0$ has $f=0$, with the result
\beq
\label{eq:small_m_free_energy_approximate}
    f \approx \frac{N_f N_c}{16 \pi^2} \, m^2 \, b^2 \le[
        2 \log\le(\frac{4 \pi m}{b \sqrt{\l}}\ri) + 1 - 2 \g_\mathrm{E}
    \ri],
    \qquad
    \frac{m}{b\sqrt{\l}} \ll 1.
\eeq

We will obtain more general solutions with non-zero $m$ numerically, by shooting from $r=0$, with the boundary conditions in eq.~\eqref{eq:T0_boundary_conditions}, towards the asymptotic $AdS_5$ boundary $r \to \infty$. For solutions obeying eq.~\eqref{eq:T0_boundary_condition_analytic} we impose $R'(r=0)=0$ and choose the free parameter $R_0\neq0$. For solutions obeying eq.~\eqref{eq:T0_boundary_condition_non_analytic} we impose \(R(r) =  \eta L (2 b/\pi)^{1/2} r^{-1} K_1(L^2 b/r)\) at small \(r\), with free parameter \(\eta\).\footnote{This is the correct small-\(r\) behaviour of solutions obeying eq.~\eqref{eq:T0_boundary_condition_non_analytic}, up to corrections of order \(e^{-3 L^2 b/r}\).} In each case, for a given value of $R_0$ or $\eta$, we numerically integrate to large $r$, and then perform a numerical fit to the large-$r$ asymptotic form in eq.~\eqref{eq:R_near_boundary}, and extract $M$ and $C$. Since every solution is determined by a single parameter, $R_0$ or  $\eta$, the asymptotic coefficient $C$ will always implicitly depend on $M$. For given values of $M$ and $C$, we compute $\Omv$ from eq.~\eqref{eq:condensate_formula}, and for a given numerical solution for $R(r)$ we compute $f$ by performing the integral in eq.~\eqref{eq:free_energy_formula} numerically. For the unique, critical solution $R_{\mathrm{c}}(r)$ obeying eq.~\eqref{eq:T0_boundary_condition_critical} we have $R_0=0$ and $R'_c(r=0) = 1/\sqrt{3}$, which map to the unique values 
\begin{subequations}
\beq
M_{\mathrm{c}} =  0.4875, \qquad C_{\mathrm{c}} = -0.07804,
\eeq
\beq
\Omv_{\mathrm{c}}= 0.04965 \, b^3 \, \frac{\sqrt{\l}\,N_f N_c}{8\pi^3}, \qquad f_{\mathrm{c}} = -0.02671 \, b^4 \, \frac{\l\,N_f N_c }{16 \pi^4}.
\eeq
\end{subequations}

\begin{figure}
\begin{subfigure}{0.5\textwidth}
    \includegraphics[width=\textwidth]{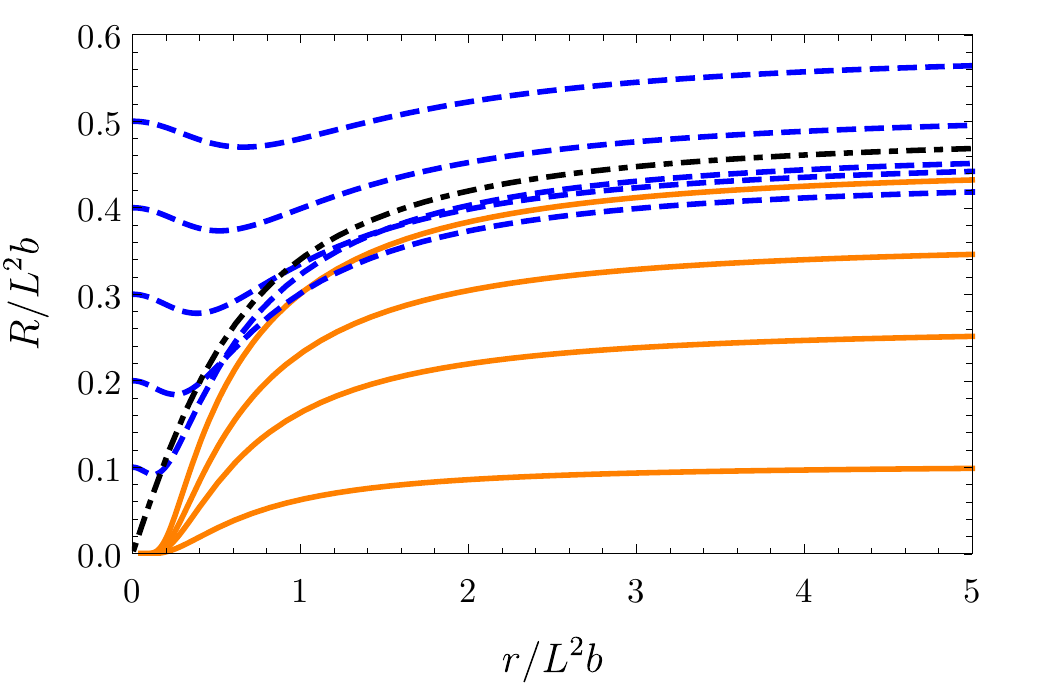}
    \caption{Embeddings at $T/b=0$}
    \label{fig:T0_solutions}
\end{subfigure}
\begin{subfigure}{0.5\textwidth}
    \includegraphics[width=\textwidth]{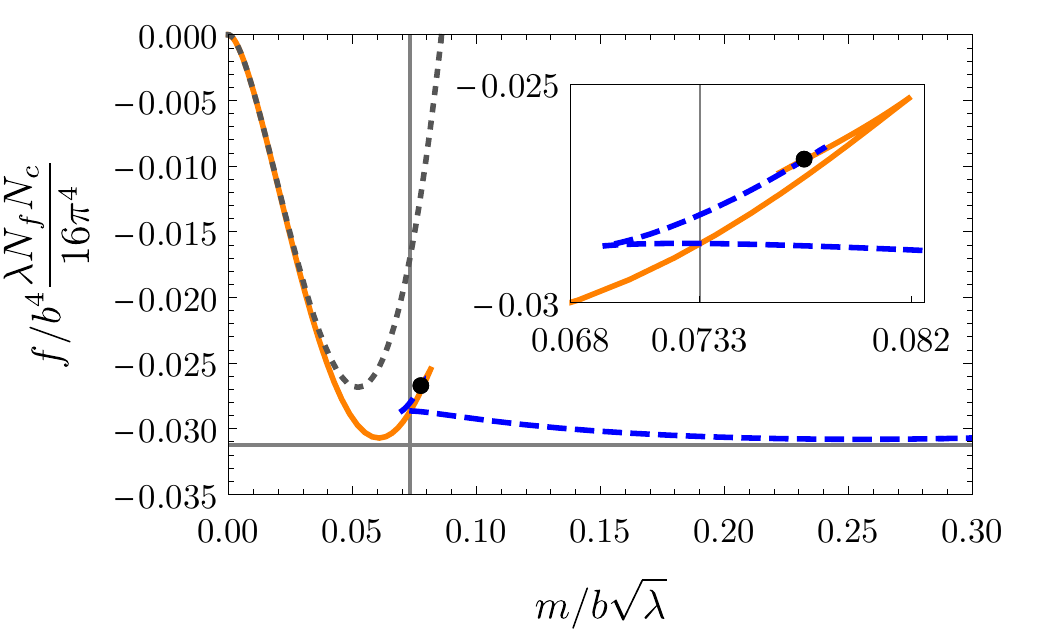}
    \caption{Free energy density at $T/b=0$}
    \label{fig:T0_free_energy}
\end{subfigure}
\begin{center}
\begin{subfigure}{0.5\textwidth}
    \includegraphics[width=\textwidth]{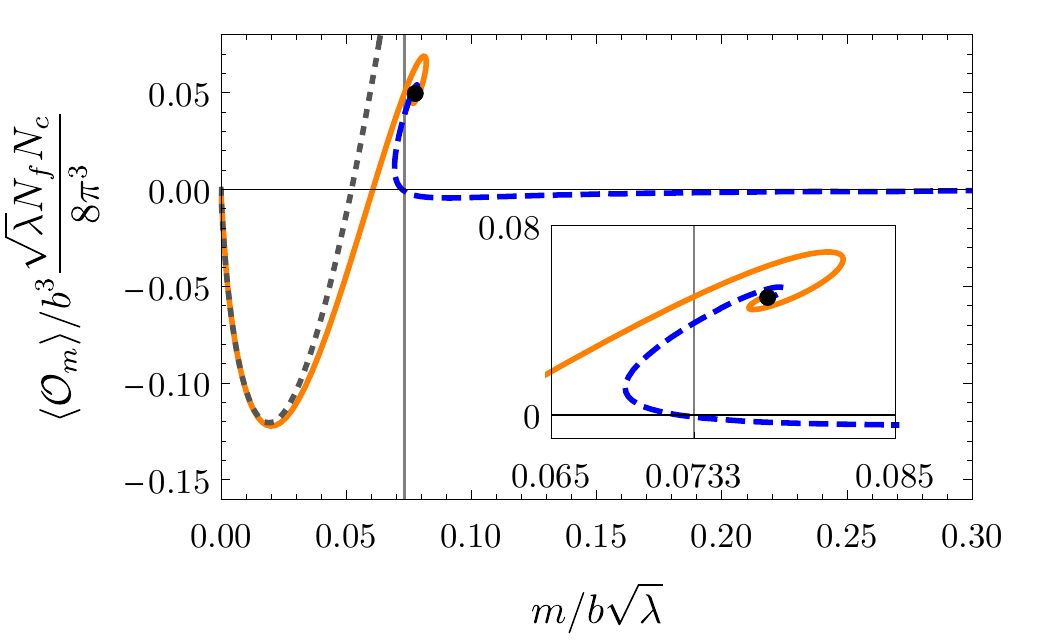}
    \caption{$\Omv$ at $T/b=0$}
    \label{fig:T0_condensate}
\end{subfigure}
\end{center}
\caption{
    \textbf{(a)} Examples of our numerical solutions for $R(r)/(L^2b)$ as functions of $r/(L^2 b)$, at $T/b=0$. The dashed blue, solid orange, and dot-dashed black lines are solutions obeying the $r \to 0$ boundary conditions in eqs.~\eqref{eq:T0_boundary_condition_analytic} (Minkowski),~\eqref{eq:T0_boundary_condition_non_analytic} (black-hole-like), and~\eqref{eq:T0_boundary_condition_critical} (critical), respectively.
    \textbf{(b)} Our numerical result for the free energy density $f/(b^4\frac{\lambda N_fN_c}{16\pi^4})$, as a function of $m/(b \sqrt{\lambda})$, at $T/b=0$, with the same colour coding as (a).  The dotted gray line is the small mass approximation in eq.~\eqref{eq:small_m_free_energy_approximate}. The black dot indicates the critical solution, and the inset is a close-up near the critical solution, showing the ``swallow tail'' shape characteristic of a first-order transition. The first order transition at $m/(b \sqrt{\l}) \approx 0.0733 $ is indicated by the vertical grey line. The horizontal grey line shows the large \(m/(b\sqrt{\l})\) limit of the free energy, \(f \approx - b^4 \l N_f N_c / (512 \pi^4)\).
    \textbf{(c)} The expectation value $\Omv/(b^3\frac{\sqrt{\lambda} N_fN_c}{8\pi^3})$ as a function of $m/(b\sqrt{\lambda})$, at $T/b=0$, with the same colour coding as (a) and (b). The dotted gray line is the small mass approximation in eq.~\eqref{eq:small_m_condensate_approximate}. The inset is a close-up of the spiral behaviour near the phase transition.
}
\label{fig:T0_plots}
\end{figure}

Figure~\ref{fig:T0_solutions} shows examples of our $T/b=0$ numerical solutions for $R(r)/(L^2 b)$. The dashed blue, solid orange, and dot-dashed black lines correspond to the $r=0$ boundary conditions in eq.~\eqref{eq:T0_boundary_condition_analytic} ($R_0\neq0$, Minkowski), eq.~\eqref{eq:T0_boundary_condition_non_analytic} (exponential, black-hole-like), and~\eqref{eq:T0_boundary_condition_critical} (critical), respectively. The limiting value that each solution approaches on the right-hand side of figure~\ref{fig:T0_solutions} determines $m$ as $\lim_{r \to \infty} R(r) = M L^2 = (2\pi\alpha') m$. Figure~\ref{fig:T0_solutions} shows that, broadly speaking, the dashed blue Minkowski embeddings only exist for large enough $m$, i.e. they describe large mass, while the solid orange black-hole-like embeddings only exist for small enough $m$, i.e. they describe small mass. Figure~\ref{fig:T0_solutions} also shows that both classes of embeddings produce the same values of $m$ for a range of $m$ near the critical solution, which will be crucially important when we consider $f$ below.

The holographic coordinate $\rho$ encodes the field theory energy scale, with the UV near the $AdS_5$ boundary $\rho\to \infty$ and the IR near $\rho = 0$. Given a value of the UV parameter $m/(b \sqrt{\lambda})$, the solution $R(r)$ encodes the corresponding renormalisation group (RG) flow, where the $r\to 0$ behaviours in eq.~\eqref{eq:T0_boundary_conditions} encode the IR degrees of freedom.

For example, the Minkowski embeddings obey eq.~\eqref{eq:T0_boundary_condition_analytic} and hence the $S^3 \in S^5$ collapses at $\rho = R_0\neq0$, as described above. As a result, the D7-branes are absent for $\rho < R_0$. The holographically dual statement is that sufficiently heavy hypermultiplets decouple at sufficiently low energy, and so disappear from the IR. Indeed, for Minkowski embeddings we expect the spectrum of linearised worldvolume excitations to be gapped and discrete~\cite{Hoyos:2006gb}.

In contrast, the black-hole-like embeddings have the exponential decay of eq.~\eqref{eq:T0_boundary_condition_non_analytic}, so that $\lim_{r\to0}R(r)=0$. The D7-branes thus reach the $AdS_5$ Poincar\'e horizon at $\rho=0$, as described above. The holographically dual statement is that for sufficiently light hypermultiplets the RG flow is to a gapless IR. Indeed, for black-hole-like embeddings we expect the spectrum of linearised worldvolume excitations to be gapless and continuous~\cite{Hoyos:2006gb}.

In fact, for the black-hole-like embeddings we can say more: a straightforward exercise shows that as $r\to0$ the D7-branes' worldvolume metric approaches that of $AdS_5 \times S^3$, with the same radius of curvature as that of the $r \to \infty$ region. The holographically dual statement is that the RG flow leads to an emergent conformal symmetry in the IR, and in fact the IR CFT is simply \textit{massless} probe hypermultiplets coupled to $\N=4$ SYM at large $N_c$ and large coupling. In particular, the $AdS_5$ isometry maps to the conformal symmetry while the $S^3$ isometry maps to the $SO(4) \simeq SU(2) \times SU(2)_R$ global symmetry.

For the critical solution, which has the linear in $r$ behaviour near $r =0$ of eq.~\eqref{eq:T0_boundary_condition_critical}, as $r \to 0$ the D7-branes' worldvolume metric approaches
\beq
\label{eq:critmetric}
P[G]_{ab} d\xi^a d\xi^b \approx L^2 \frac{dr^2}{r^2}+\frac{4}{3}\frac{r^2}{L^2} \left (-dt^2 + dx^2 + dy^2 \right) + \frac{1}{4} \, b^2 \, L^2 \, dz^2 + \frac{3}{4} \, L^2 \, ds^2_{S^3},
\eeq
which we recognise as that of $AdS_4$ with coordinates $(r,t,x,y)$ and radius $L$, times $\mathbb{R}$ with coordinate $z$, times $S^3$ with radius $L\sqrt{\frac{3}{4}}$. The holographically dual statement is that the RG flow leads to an emergent $(2+1)$-dimensional conformal symmetry in the IR, dual to the $AdS_4$ isometry, with a non-compact $U(1)$ symmetry, dual to translations in $z$, plus an $SO(4)$ symmetry, dual to the $S^3$ isometry. In other words, the critical RG flow leads to an emergent $(2+1)$-dimensional CFT. The fully holographic, bottom-up models of refs.~\cite{Landsteiner:2015lsa,Landsteiner:2015pdh,Copetti:2016ewq,Landsteiner:2019kxb,Juricic:2020sgg} have a similar critical solution, but with IR Lifshitz symmetry in which $z$ scales with a different power from $(t,x,y)$. In both our model and those models, the choice $\phi=b\,z$ breaks $SO(3)$ rotational symmetry of $(x,y,z)$ down to $SO(2)$ rotational symmetry of $(x,y)$, allowing for a lower-dimensional CFT or Lifshitz scaling in the IR.

Since only black-hole-like embeddings exist for sufficiently small $m/(b\sqrt{\lambda})$, and only Minkowski embeddings exist for sufficiently large $m/(b\sqrt{\lambda})$, as we increase $m/(b\sqrt{\lambda})$ a transition from black-hole-like to Minkowski embeddings must necessarily occur. A key question is the nature of that transition, including in particular its order.

Figure~\ref{fig:T0_free_energy} shows our numerical results for the free energy density $f$, in units of $b$ and normalised by $\frac{\lambda N_fN_c}{16\pi^4}$, as a function of $m/(b \sqrt{\lambda})$. The colour coding is the same as in figure~\ref{fig:T0_solutions}, while the black dot represents the critical solution and the dotted grey line is the small-$m$ approximation in eq.~\eqref{eq:small_m_free_energy_approximate}, showing excellent agreement with our numerics when $m/(b\sqrt{\lambda})\ll 1$. The horizontal grey line in the figure shows the analytic approximation for the free energy in the large-\(m\) limit, \(f \approx - b^4 \l N_f N_c /(512 \pi^4)\), which agrees well with our numerics when \(m/(b\sqrt{\l}) \gg 1\).

The inset in figure~\ref{fig:T0_free_energy} shows $f$ near the critical solution, which clearly exhibits the ``swallow tail'' shape characteristic of a first-order transition. Specifically, for a range of $m/(b\sqrt{\lambda})$ near the critical solution, $f$ is multi-valued, with both black-hole-like and Minkowski embeddings available to the system. The thermodynamically preferred solution is that with the lowest $f$. As we increase $m/(b\sqrt{\lambda})$ black-hole-like embeddings are preferred until $m/(b\sqrt{\lambda})\approx 0.0733$, denoted by the vertical line in figure~\ref{fig:T0_free_energy}, after which Minkowski embeddings are preferred. The first derivative $\partial f/\partial m$ is discontinuous at the transition, thus the transition is first order. The critical solution is never thermodynamically preferred.

Since we work in the strong coupling limit \(\lambda\gg1\), the phase transition occurs at \(m/b \sim \sqrt{\lambda} \gg 1\). This contrasts with the free model described in section~\ref{sec:intro}, for which the phase transition occurs at \(m/b=1/2\). Of course, there is no contradiction since the theory at \(\lambda\gg1\) is very far from being free. In general, dimensional analysis implies that for given values of \(N_f\) and \(N_c\), the phase transition occurs at \(m/b = \mathcal{F}(\l)\) for some dimensionless function \(\mathcal{F}(\lambda)\). The free model has \(\mathcal{F}(0)=1/2\), while we have found \(\mathcal{F}(\lambda \to \infty) \sim \sqrt{\l}\) at \(N_c \gg N_f \gg 1\). The same \(\sqrt{\lambda}\) scaling occurs in the meson-melting phase transition at \(b=0\) and non-zero \(T\): the mesons have binding energies \(\Delta E \sim m/\sqrt{\l}\), so the natural scale for the temperature of the transition is \(T \sim \Delta E \sim m/\sqrt{\l}\)~\cite{Kruczenski:2003be,Mateos:2006nu,Karch:2006bv,Mateos:2007vn}. We expect a similar interpretation to hold in our case, with \(T\) replaced by \(b\). We note that the factor of \(\sqrt{\lambda}\) arises naturally from string theory: the D3- and D7-branes are separated by a distance \(\Delta X\) that is of order one in \(AdS\) units, so strings stretched between them have masses \(m = \Delta X/2\pi \alpha' \sim \sqrt{\l}\).

Figure~\ref{fig:T0_condensate} shows some of our numerical results for $\Omv = \partial f/\partial m$, in units of $b$ and normalised by $\frac{\sqrt{\lambda} N_fN_c}{8\pi^3}$, with the same colour coding as figs.~\ref{fig:T0_solutions} and~\ref{fig:T0_free_energy}. At large $m$ clearly $\Omv \to 0$, as discussed below eq.~\eqref{eq:largemC}, and the  small-$m$ approximation of eq.~\eqref{eq:small_m_condensate_approximate} appears as the dotted grey line. The transition point $m/(b\sqrt{\lambda})\approx 0.0733$ is denoted by the vertical line. As expected, near the critical solution $\Omv$ is multi-valued, and as $m/(b\sqrt{\lambda})$ increases, at the transition point $\Omv$ jumps discontinuously from black-hole-like to Minkowski embeddings.

In figure~\ref{fig:T0_condensate} the inset is a close-up showing that $\Omv$ executes a spiral when approaching the critical solution, $R_{\mathrm{c}}(r)$. Such behaviour is familiar for probe branes, and arises from a discrete scale invariance of near-critical solutions, producing self-similarity~\cite{Frolov:1998td,Mateos:2006nu,Frolov:2006tc,Mateos:2007vn,Karch:2009ph,BitaghsirFadafan:2018iqr}. This discrete scale invariance explains why the transition is first order: discrete scale invariance of near-critical solutions implies that $\Omv$ executes a spiral and hence $f$ is multi-valued near the critical solution, which then guarantees that the transition is first order. We will not use the scaling symmetry here, so we will just sketch the derivation of the scaling exponents and self-similarity, leaving the details to refs.~\cite{Frolov:1998td,Mateos:2006nu,Frolov:2006tc,Mateos:2007vn,Karch:2009ph,BitaghsirFadafan:2018iqr}. The discrete scaling symmetry is manifest when we linearise the equation of motion eq.~\eqref{eq:T0_EOM} in $R(r)$ about the critical solution, $R_{\mathrm{c}}(r)$, which at small $r$ gives
\beq
\label{eq:nearcritsol}
R(r) \approx R_\mathrm{c}(r) + B \, r^{\n_+} + B^* \, r^{\n_-},
\eeq
with constant $B$ and $\n_\pm = - \frac{1}{2} \pm \frac{i}{2} \sqrt{23}$. The near-critical solutions have the scaling symmetry $R(r) \to \zeta R(r)$ and $r \to \zeta r$ with real, positive $\zeta$, under which $B \to \zeta^{1-\nu_+} B$. Since we have linearised the equation of motion, the map from the $r \to 0$ coefficients $B$ and $B^*$ to the $r \to \infty$ coefficients $M$ and $C$ is linear, which implies
\beq
\label{eq:spiral_parametric}
    M - M_\mathrm{c} \approx \kappa_1 \, \zeta^{1 - \n_+} + \kappa_1^* \, \zeta^{1 - \n_-},
    \qquad
    C - C_\mathrm{c} \approx \kappa_2 \, \zeta^{1 - \n_+} + \kappa_2^* \, \zeta^{1 - \n_-},
\eeq
with constants $\kappa_1$ and $\kappa_2$. Inserting eq.~\eqref{eq:spiral_parametric} into eq.~\eqref{eq:condensate_formula} we obtain a curve for $\Omv$ as a function of $m$, parametrised by $\zeta$. Re-writing
\beq
\zeta^{1\pm \nu_{\pm}} = e^{(1\pm\mathrm{Re}\,\nu_{\pm})\ln\zeta} \left[ \cos\left( \mathrm{Im}\,\nu_{\pm}\ln\zeta\right) + i \sin\left(\mathrm{Im}\,\nu_{\pm}\ln\zeta\right)\right],
\eeq
then shows that if we approach the critical solution by sending $\zeta \to 0$, then $\Omv$ as a function of $m$ will trace a spiral of decaying amplitude, with period $2\pi/\mathrm{Im}\,\nu_{\pm}$. In contrast, real-valued exponents $\nu_{\pm}$ would lead to single-valued $\Omv$ as a function of $m$ and hence a transition of second order or higher~\cite{Karch:2009ph}.

\subsection{Phase Transition at Non-Zero Temperature}
\label{sec:finite_T_solutions}

When $T>0$ the field theory has two free parameters, $m/(b\sqrt{\lambda})$ and $T/b$. The equation of motion for $R(r)$ derived from eq.~\eqref{eq:d7_action_LT} when $T/b>0$ and $p_{\f}=0$ is cumbersome and unilluminating, so we will not write it here.

As discussed in section~\ref{sec:holographic_model}, when $T >0$ two classes of solutions for $R(r)$ are possible. The first is black hole embeddings, which intersect the horizon at some $r_H$ such that $\rho_H^2 = r_H^2 + R(r_H)^2$. For black hole embeddings to be static, the D7-branes must intersect the horizon perpendicularly, $R'(r_H)=0$. The second class of solutions is Minkowski embeddings, which reach $r=0$ with $R_0>\rho_H$, so that the $S^3 \in S^5$ collapses outside of the horizon $\rho_H$. For Minkowski embeddings to be static, and in particular to avoid a conical singularity when the $S^3$ collapses~\cite{Karch:2006bv}, the D7-branes must hit $r=0$ perpendicularly, $R'(r=0)=0$, as in eq.~\eqref{eq:T0_boundary_condition_analytic}. These two classes are separated by a critical solution in which the $S^3 \in S^5$ collapses exactly at $\rho_H$.

When $T/b>0$ we know one exact solution for $R(r)$, namely the trivial solution $R(r)=0$, which is a black hole embedding, intersecting the horizon at $r=\rho_H$. The trivial solution describes $m=0$ and $\Omv=0$, but has non-zero free energy density eq.~\eqref{eq:free_energy_formula},
\beq
f = -\frac{\l N_f N_c}{16 \pi^4} \frac{\pi^4}{8} T^4.
\eeq
Plugging this $f$ into eqs.~\eqref{eq:entropy} and~\eqref{eq:soundspeed}, we obtain the corresponding entropy density, heat capacity density, and correction to the sound speed squared, respectively,
\beq
\label{eq:masslessthermo}
s = \frac{\l N_f N_c}{16 \pi^4}\frac{\pi^4}{2}T^3, \qquad c_V = 3\, s, \qquad \delta v^2 = 0.
\eeq
These results are the same as for hypermultiplets with $m=0$ and $b=0$. That is no surprise: the solution $R(r)=0$, and thus the Legendre-transformed action eq.~\eqref{eq:d7_action_LT} evaluated on $R(r)=0$, is independent of both $m$ and $b$. As a result, for the trivial solution all physical quantities are proportional to a power of $T$ dictated by dimensional analysis. Moreover, the sound speed squared $v^2$ must take the value required by $(d+1)$-dimensional scale invariance, $v^2 = 1/\sqrt{d}$, explaning why the $\mathcal{O}\left(N_f/N_c\right)$ correction vanishes, $\delta v^2 = 0$.

\begin{figure}
    \begin{subfigure}{0.5\textwidth}
        \includegraphics[width=\textwidth]{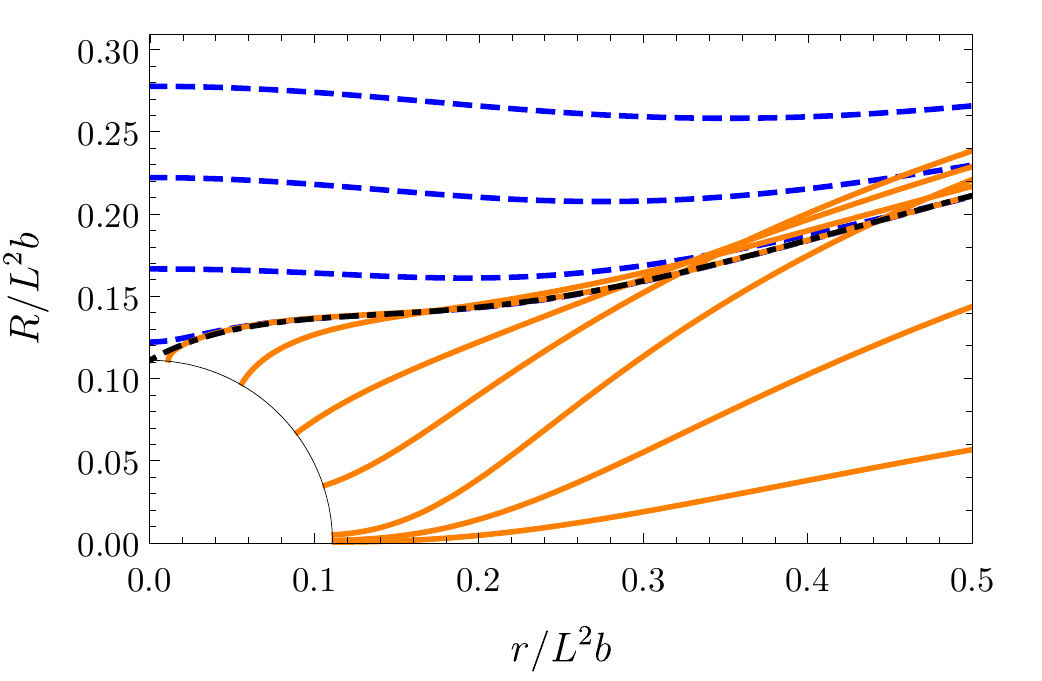}
        \caption{Embeddings at $T/b=0.05$}
        \label{fig:lowT_solutions1}
    \end{subfigure}
    \begin{subfigure}{0.5\textwidth}
        \includegraphics[width=\textwidth]{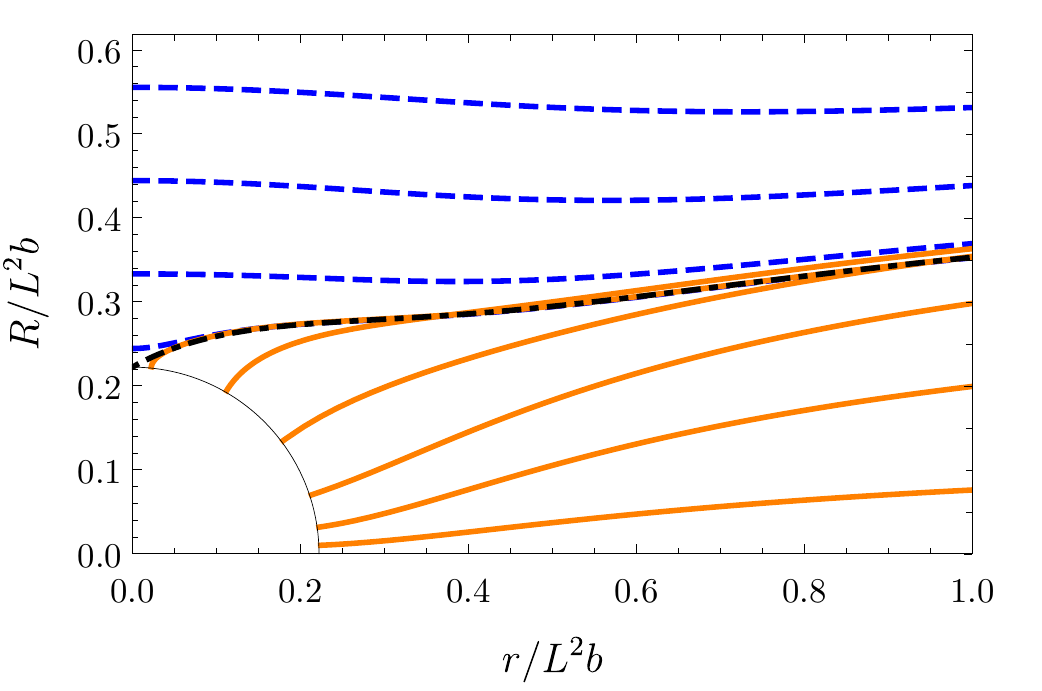}
        \caption{Embeddings at $T/b=0.1$}
         \label{fig:lowT_solutions2}
    \end{subfigure}
    \begin{subfigure}{0.5\textwidth}
        \includegraphics[width=\textwidth]{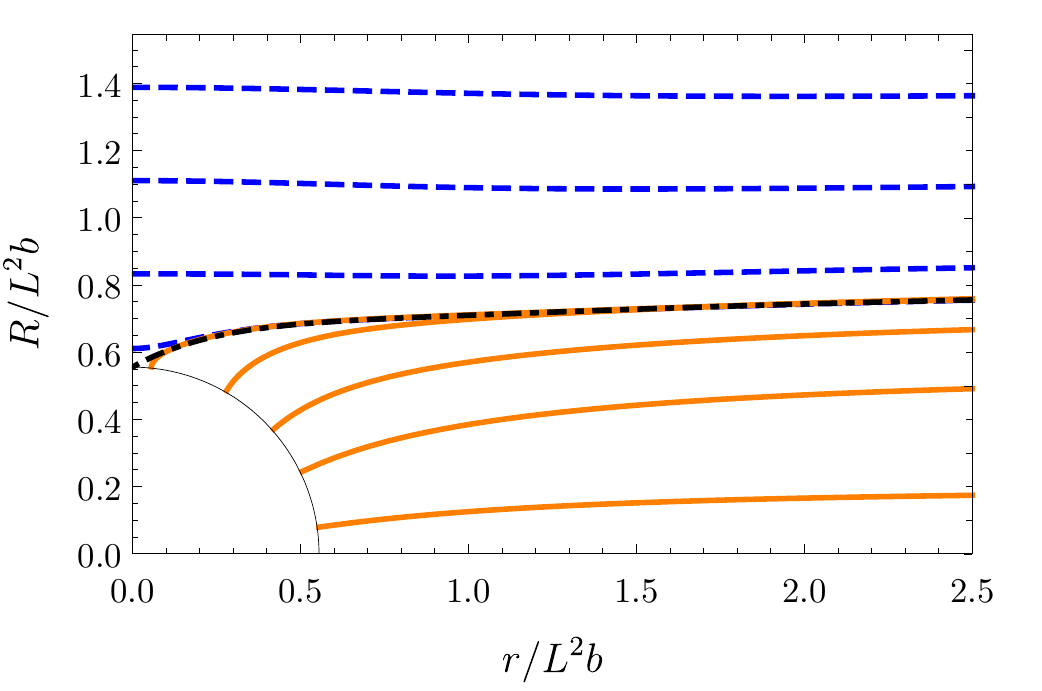}
        \caption{Embeddings at $T/b=0.25$}
    \end{subfigure}
    \begin{subfigure}{0.5\textwidth}
        \includegraphics[width=\textwidth]{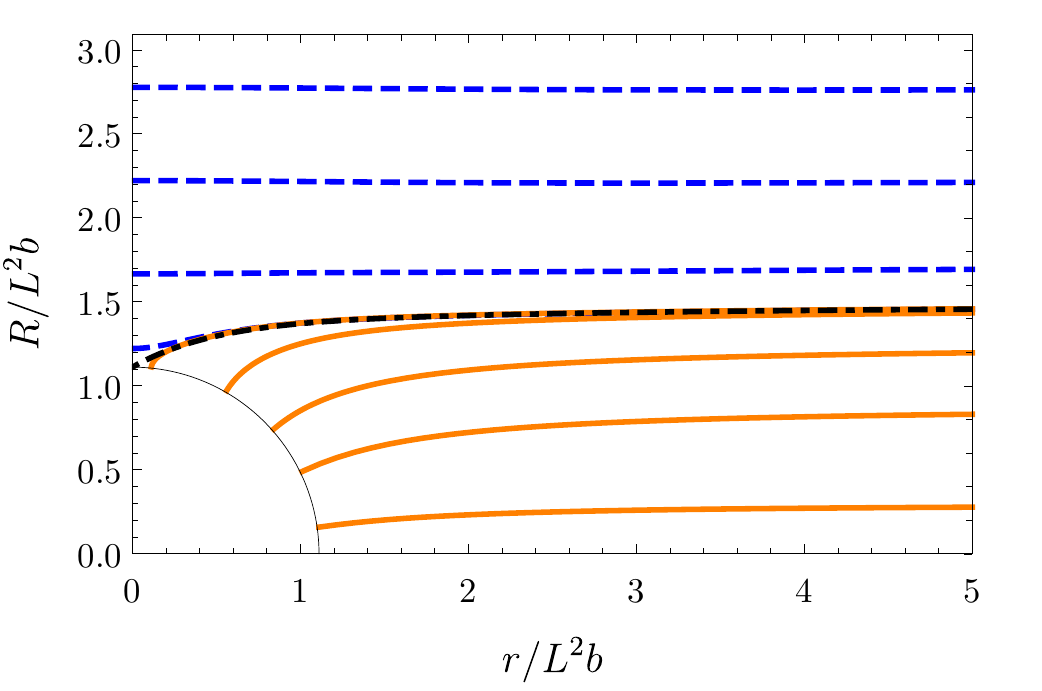}
        \caption{Embeddings at $T/b=0.5$}
    \end{subfigure}
    \caption{
        Examples of our numerical solutions for $R(r)/(L^2 b)$ as functions of $r/(L^2 b)$ when $T/b>0$. In each plot, the solid black quarter-circle is the horizon $\rho_H$, given by $\rho_H^2=r^2 + R^2 = \pi^2 L^4 T^2/2$. The dashed blue, solid orange, and dot-dashed black lines are Minkowski, black hole, and the critical embeddings, respectively, for \textbf{(a)} $T/b=0.05$, \textbf{(b)} $T/b=0.1$, \textbf{(c)} $T/b=0.25$, and \textbf{(d)} $T/b=0.5$.
    }
    \label{fig:temperature_solutions}
\end{figure}

We obtain solutions for $R(r)$ describing non-zero $m$ numerically, in a fashion similar to the $T=0$ case of section~\ref{sec:T0_solutions}. For black hole embeddings we shoot from $r_H$ with boundary conditions $R(r_H)=\sqrt{\rho_H^2-r_H^2}$ and $R'(r_H)=\sqrt{(\r_H/r_H)^2-1}$, where the latter condition is imposed by  \(R(r)\)'s equation of motion. For Minkowski solutions we shoot from $r=0$ with the boundary conditions $R(r=0)=R_0>\rho_H$ and $R'(r=0)=0$. Figure~\ref{fig:temperature_solutions} shows examples of our $T/b>0$ numerical solutions for $R(r)/(L^2b)$, where the dashed blue, solid orange, and dot-dashed black lines correspond to Minkowski, black hole, and the critical embeddings, respectively. Similar to the $T/b=0$ case of section~\ref{sec:T0_solutions}, Minkowski embeddings only exist for large enough $m$ while black hole embeddings only exist for small enough $m$, and both classes of embeddings exist for a range of $m$ near the critical embedding.

Among the black hole embeddings we find solutions whose boundary conditions approach the exponential behaviour of eq.~\eqref{eq:T0_boundary_condition_non_analytic} as $T/b \to 0$. Examples of these appear in figure~\ref{fig:lowT_solutions1} for $T/b=0.05$, as the three lowest solid orange lines, corresponding to the three lowest values of $m$, or in figure~\ref{fig:lowT_solutions2} for $T/b=0.1$ as the lowest few solid orange lines. These solutions describe RG flows to the IR CFT with non-zero temperature, i.e. massless hypermultiplets with $T/b>0$. In other words, some of the solutions with exponential boundary conditions at $T/b=0$, which describe RG flows to massless hypermultiplets, survive at sufficiently small non-zero $T/b$. However, as $T/b$ increases the horizon eventually ``hides'' any exponential behaviour. In CFT terms, once $T$ is sufficiently large compared to $m$, the IR CFT is ``washed out'' in the plasma.

For a given solution $R(r)$ we perform a numerical fit to the large-$r$ asymptotics in eq.~\eqref{eq:R_near_boundary}, extract $M$ and $C$, and plug these into eq.~\eqref{eq:condensate_formula} to obtain $\Omv$. We then calculate $f$ by performing the integral in eq.~\eqref{eq:free_energy_formula} numerically.

\begin{figure}
    \begin{subfigure}{0.5\textwidth}
        \includegraphics[width=\textwidth]{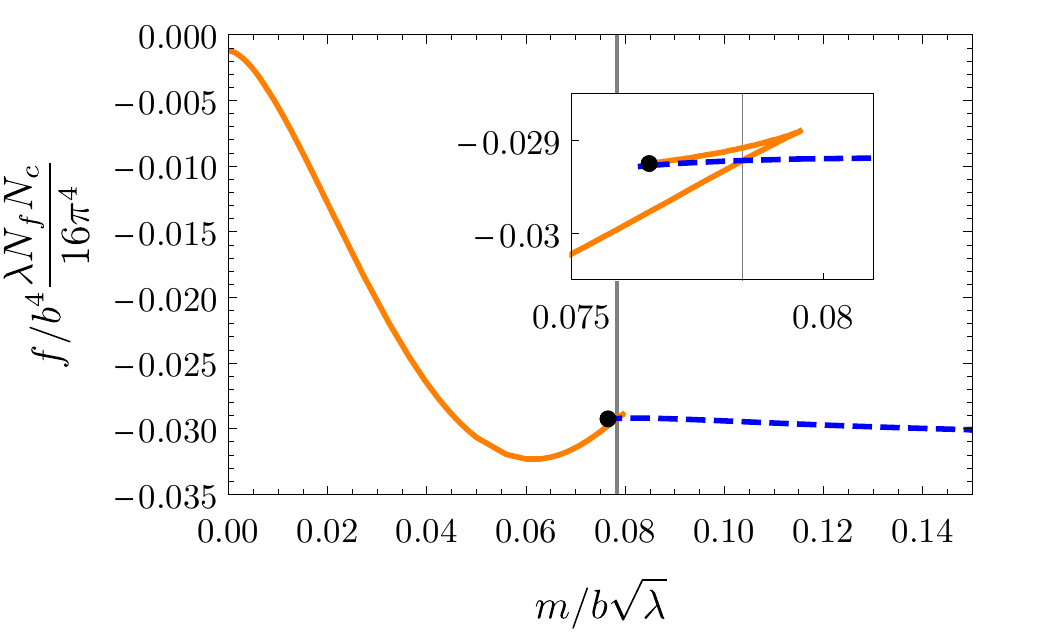}
        \caption{Free energy density at $T/b = 0.1$.}
    \end{subfigure}
    \begin{subfigure}{0.5\textwidth}
        \includegraphics[width=\textwidth]{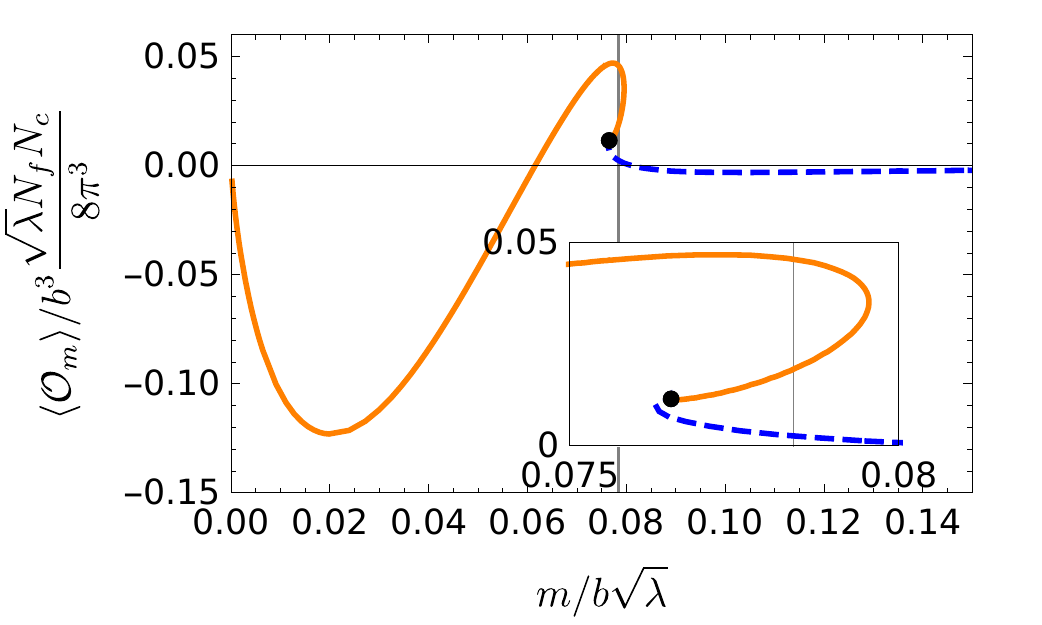}
        \caption{$\Omv$ at $T/b = 0.1$.}
    \end{subfigure}
    \begin{subfigure}{0.5\textwidth}
        \includegraphics[width=\textwidth]{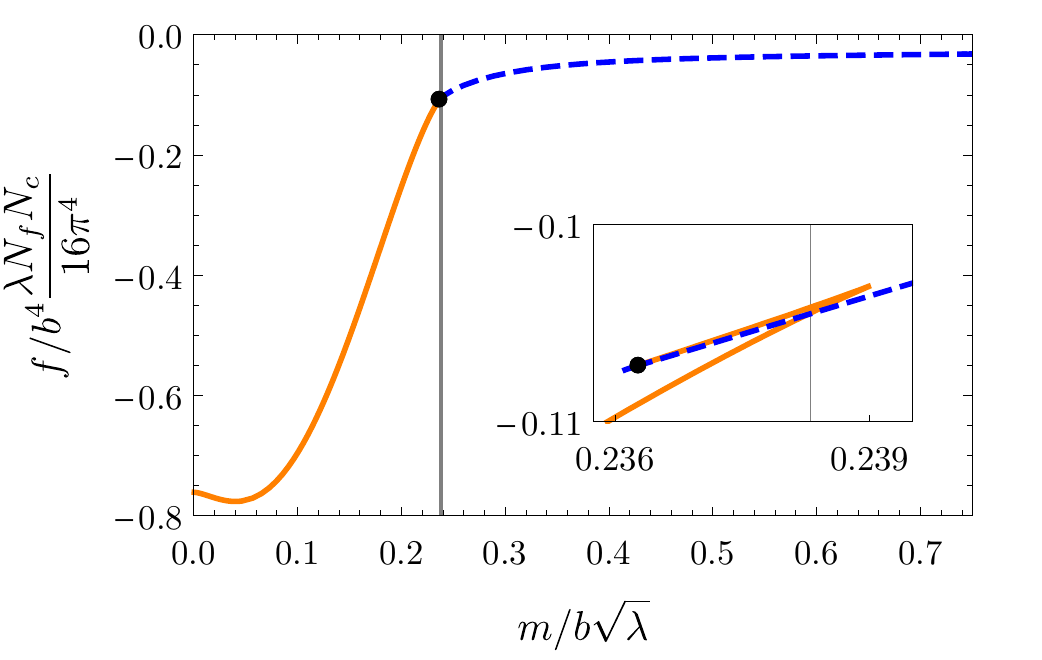}
        \caption{Free energy density at $T/b = 0.5$.}
    \end{subfigure}
    \begin{subfigure}{0.5\textwidth}
        \includegraphics[width=\textwidth]{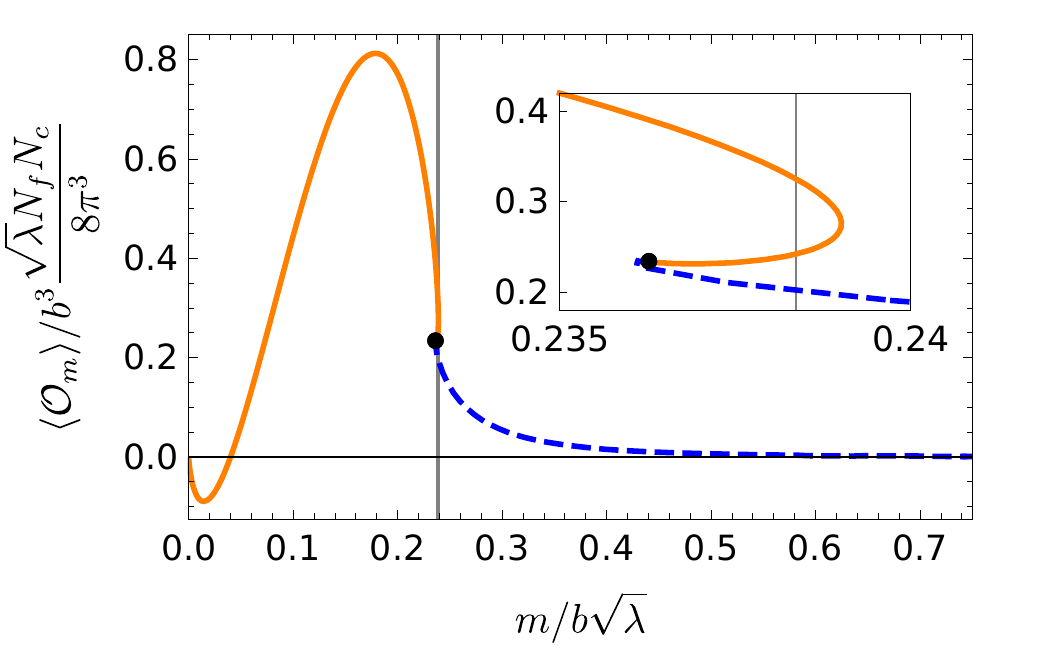}
        \caption{$\Omv$ at $T/b = 0.5$.}
    \end{subfigure}
    \begin{subfigure}{0.5\textwidth}
        \includegraphics[width=\textwidth]{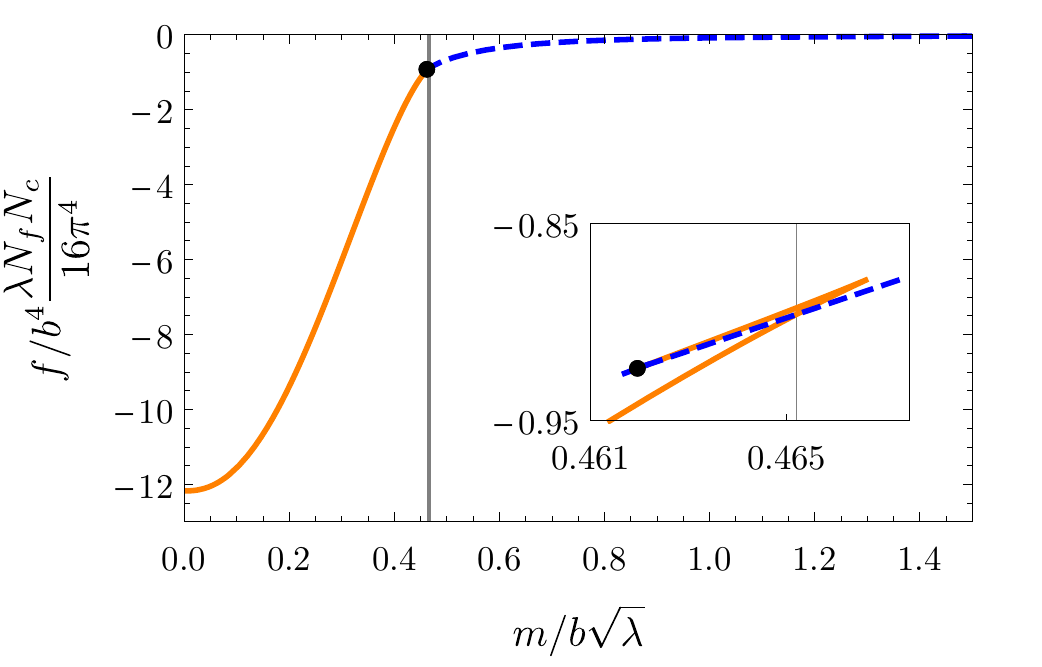}
        \caption{Free energy density at $T/b = 1$.}
    \end{subfigure}
    \begin{subfigure}{0.5\textwidth}
        \includegraphics[width=\textwidth]{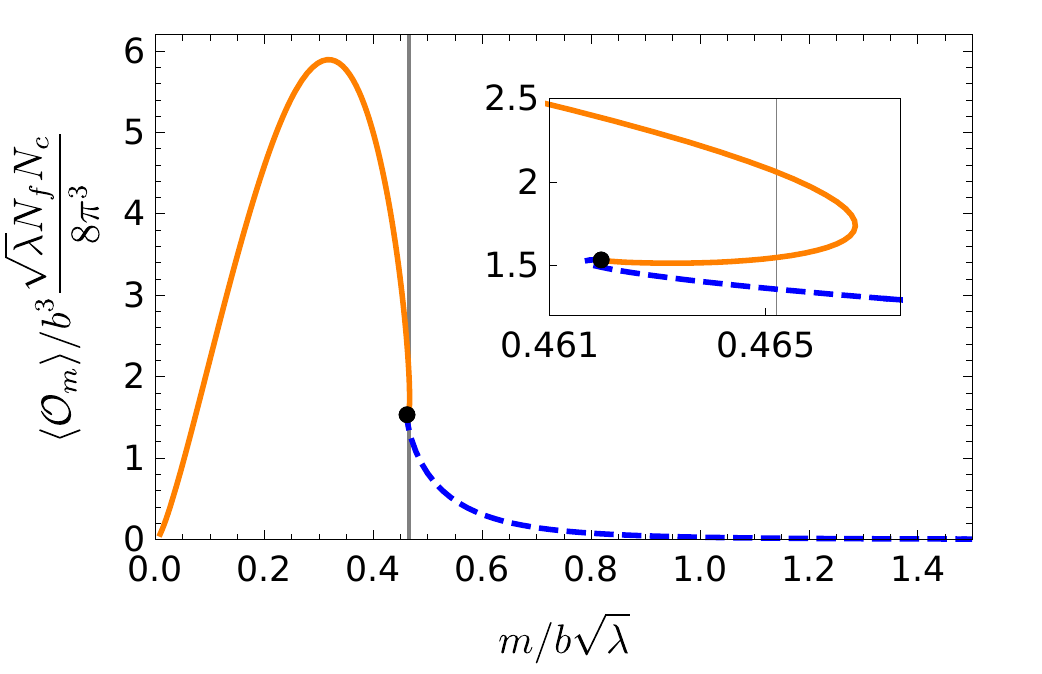}
        \caption{$\Omv$ at $T/b = 1$.}
    \end{subfigure}
    \caption{Our numerical results for the free energy density $f/(b^4\frac{\lambda N_fN_c}{16\pi^4})$ and for $\Omv/(b^3\frac{\sqrt{\lambda} N_fN_c}{8\pi^3})$ as functions of $m/(b \sqrt{\lambda})$, for \textbf{(a)} and \textbf{(b)} $T/b=0.1$, \textbf{(c)} and \textbf{(d)} $T/b=0.5$, and \textbf{(e)}, and \textbf{(f)} $T/b=1$. As in figs.~\ref{fig:T0_plots} and~\ref{fig:temperature_solutions}, the solid orange and dashed blue lines correspond to black hole and Minkowski embeddings, respectively, and the black dots denote critical embeddings. The insets are close-ups near critical embeddings, showing $f$'s ``swallow tail'' and $\Omv$'s spiral, both characteristic of first-order transitions. The vertical gray lines indicate the values of $m/(b\sqrt{\l})$ at the first-order transitions.}
    \label{fig:finite_T_free_energy_condensate}
\end{figure}

Figure~\ref{fig:finite_T_free_energy_condensate} shows some of our numerical results for $f$, normalised by $\frac{\lambda N_fN_c}{16\pi^4}$, and for $\Omv$, normalised by $\frac{\sqrt{\lambda} N_fN_c}{8\pi^3}$, both in units of $b$, as functions of $m/(b \sqrt{\lambda})$. Figure~\ref{fig:finite_T_free_energy_condensate} has the same colour coding as figure~\ref{fig:temperature_solutions}, and the black dot denotes the critical solution. Our results show clearly that the first-order transition we found at $T/b=0$ persists to $T/b>0$, with the same qualitative characteristics. In particular, for some range of $m/(b\sqrt{\lambda})$ near the critical solution both $f$ and $\Omv$ are multi-valued, and the insets in figure~\ref{fig:finite_T_free_energy_condensate} are close-ups near the critical solution showing that $f$ exhibits a ``swallow tail'' shape and $\Omv$ exhibits a spiral shape, similar to the $T/b=0$ case in figure~\ref{fig:T0_plots}. Clearly, as we increase $m/(b\sqrt{\lambda})$ a transition from a black hole to a Minkowski embedding occurs in which $f$ is continuous but its first derivative $\Omv = \partial f/\partial m$ is not. In figure~\ref{fig:finite_T_free_energy_condensate} we denote the transition point with a vertical line. The critical embedding is never thermodynamically preferred.

\begin{figure}
\begin{center}
    \includegraphics[width=0.6\textwidth]{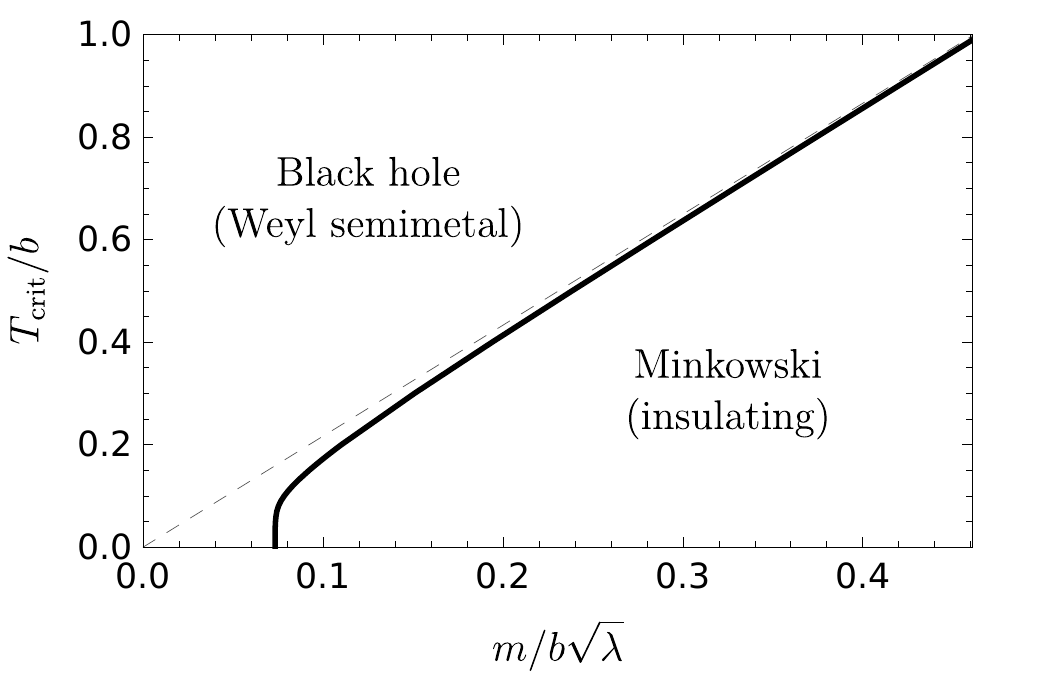}
\end{center}
\caption{The phase diagram of our holographic model. The solid black line is the critical temperature $T_\mathrm{crit}$, in units of $b$, where we find a first-order transition from black hole to Minkowski embeddings, as a function of $m/(b\sqrt{\l})$. To the left of this line, black hole embeddings are thermodynamically preferred, whereas to the right of this line, Minkowski embeddings are thermodynamically preferred. In section~\ref{sec:conductivities} we show the black hole embeddings are dual to WSM states, while Minkowski embeddings are dual to trivially insulating states. The dashed gray line shows the $b=0$ result $T_\mathrm{crit} \approx 2.166 \,m /\sqrt{\l}$ of ref.~\cite{Mateos:2007vn}, which our transition line approaches as $m/(b\sqrt{\lambda})\to\infty$.}
\label{fig:phase_diagram}
\end{figure}

We find that the first-order transition persists to all $T/b$. Figure~\ref{fig:phase_diagram} is the phase diagram of our model, showing our numerical result for the critical temperature of the first-order transition, $T_\mathrm{crit}$, in units of $b$, as a function of $m/(b \sqrt{\l})$. As we found in section~\ref{sec:T0_solutions}, at $T/b=0$ the transition occurs at $m/(b\sqrt{\lambda})\approx 0.0733$, and as $m/(b\sqrt{\lambda})$ increases, $T_{\mathrm{crit}}/b$ increases. As both $m$ and $T_{\mathrm{crit}}$ grow, we expect the influence of $b$ to fade, and the first-order transition to approach that at $b=0$~\cite{Mateos:2006nu,Mateos:2007vn}. Figure~\ref{fig:phase_diagram} confirms that expectation: the dashed grey line denotes the $b=0$ transition at $T_\mathrm{crit}/(m/\sqrt{\lambda}) \approx 2.166$~\cite{Mateos:2007vn}\footnote{The definition of the 't Hooft coupling $\lambda$ in ref.~\cite{Mateos:2007vn} is smaller than ours by a factor of $2$.}, which our $T_{\mathrm{crit}}$ indeed approaches as $m \to \infty$.

Similar to the $T/b=0$ case in eq.~\eqref{eq:nearcritsol}, the behaviours of $f$ and $\Omv$ in figure~\ref{fig:finite_T_free_energy_condensate} arise from a discrete scale invariance of solutions near the critical solution, characterised by a set of complex exponents. One difference between the transitions at $T/b=0$ and $T/b>0$ is the value of these exponents. When $T/b=0$ the exponents in eq.~\eqref{eq:nearcritsol} were $\n_\pm = - \frac{1}{2} \pm \frac{i}{2} \sqrt{23}$. When $T/b>0$ the analogous exponents are defined in an expansion of $R(r)$ around the critical solution in powers of $r$, and take the values $- \frac{3}{2} \pm \frac{i}{2} \sqrt{7}$, the same as for the first-order transition when $b=0$~\cite{Mateos:2006nu,Karch:2009ph}. These exponents are different because the critical solutions have different topology when $T/b=0$ or $T/b>0$. The difference is clearest in Euclidean signature, where when $T/b=0$ the Euclidean time direction is non-compact, but when $T/b>0$ the Euclidean time is an $S^1$ that collapses to zero size at the horizon. In Euclidean signature, when $T/b=0$ the critical solution approaches Euclidean $AdS_4 \times \mathbb{R} \times S^3$ deep in the bulk, as in eq.~\eqref{eq:critmetric}, whereas when $T/b>0$ the critical solution has both a collapsing $S^1$ and a collapsing $S^3$.

\begin{figure}
	\begin{subfigure}{0.5\textwidth}
		\includegraphics[width=\textwidth]{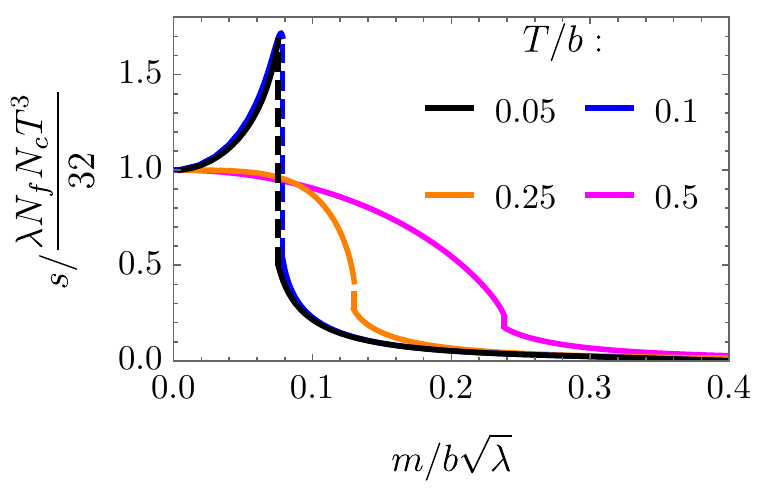}
		\caption{Entropy density}
		\label{fig:numentropy}
	\end{subfigure}
	\begin{subfigure}{0.5\textwidth}
		\includegraphics[width=\textwidth]{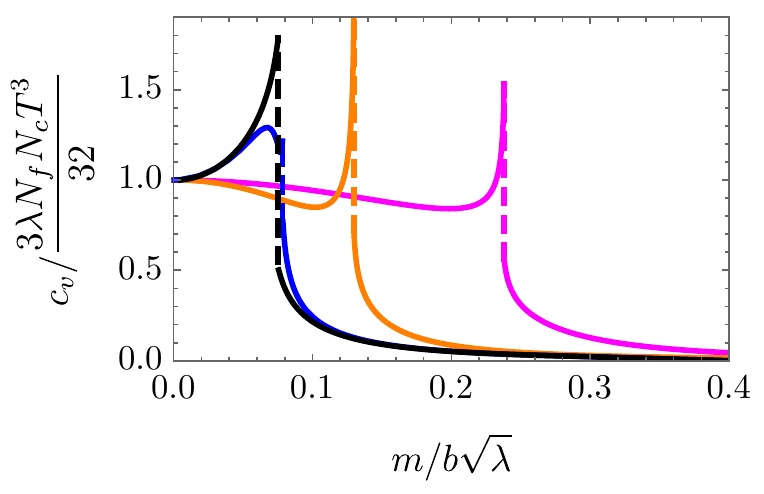}
		\caption{Heat capacity density}
	\end{subfigure}
	\begin{center}
		\begin{subfigure}{0.5\textwidth}
			\includegraphics[width=\textwidth]{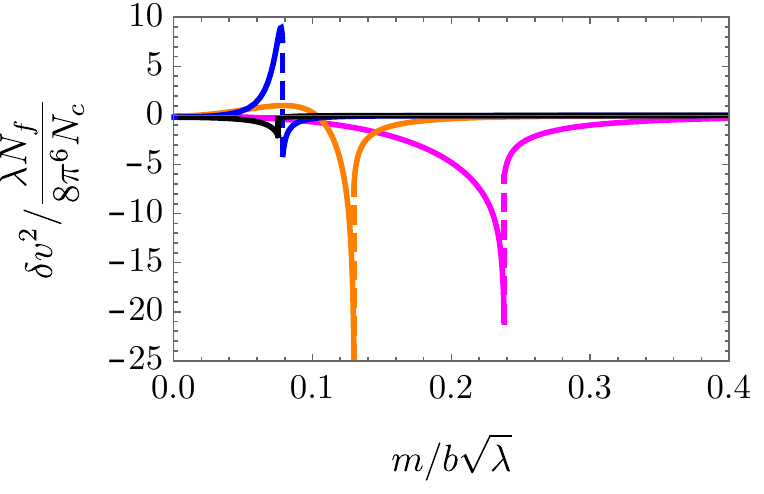}
			\caption{Correction to the sound speed squared, \(\delta v^2\)}
			\label{fig:speedcorr}
		\end{subfigure}
	\end{center}
	\caption{(a) Our numerical results for the entropy density, $s$, normalised by its $m=0$ value in eq.~\eqref{eq:masslessthermo}, $\lambda N_f N_c T^3/32$, as a function of $m/(b \sqrt{\lambda})$, for $T/b=0.05$ (solid black), $0.1$ (solid blue), $0.25$ (solid orange), and $0.5$ (solid pink). (b) Our numerical results for the heat capacity density, $c_V$, normalised by its $m=0$ value in eq.~\eqref{eq:masslessthermo}, $3\lambda N_f N_c T^3/32$, as a function of $m/(b \sqrt{\lambda})$, for the same values of $T/b$ as in (a). (c) Our numerical results for the $\mathcal{O}\left(N_f/N_c\right)$ correction to the sound speed squared, $\delta v^2$ in eq.~\eqref{eq:soundcorr}, as a function of $m/(b \sqrt{\lambda})$, for the same values of $T/b$ as in (a) and (b). In each case the dashed vertical lines denote a first-order phase transition.
	}
	\label{fig:thermodynamics_solutions}
\end{figure}

In appendix~\ref{sec:thermodynamics} we write the entropy density $s$ in a form suited for our numerics,
\beq
\label{eq:entropynumerics}
s = - \frac{\l N_f N_c}{16 \pi^4 L^8} \biggl[   
\int_{r_0}^{r_c} dr \, r^3\sqrt{1 + R'^2} \frac{\partial }{\partial T} \bigg( g\, h \sqrt{1 + \frac{L^4 b^2 R^2}{h\,\left(r^2+R^2\right)^2}}  \bigg)\biggr].
\eeq
From $s$ we then compute heat capacity density $c_V$ numerically using finite differences, and from $s$ and $c_V$ we compute the $\mathcal{O}\left(N_f/N_c\right)$ correction to the sound speed squared, $\delta v^2$ via eq.~\eqref{eq:soundcorr}. Figure~\ref{fig:thermodynamics_solutions} shows some of our numerical results for $s$, $c_V$, and $\delta v^2$ as functions of $m/(b \sqrt{\lambda})$ for several values of $T/b$. Both $s$ and $c_V$ take forms characteristic of a first-order transition, for example $c_V$ grows rapidly when approaching the transition.

Exceptional behaviour appears in $s$, which at low $T/b$ exhibits a dramatic increase as $m/(b \sqrt{\lambda})$ approaches the transition from below: see figure~\ref{fig:numentropy} with $T/b=0.05$ (solid black) and $0.1$ (solid blue). These results arise from solutions with exponential behaviour at low $T/b$, namely those we discussed in figures~\ref{fig:lowT_solutions1} and~\ref{fig:lowT_solutions2}. In CFT terms these are cases where the IR is the massless hypermultiplet CFT. Recalling that $s$ counts thermodynamic degrees of freedom, the rise in $s$ at $T/b=0.05$ or $0.1$ presumably comes from these additional massless degrees of freedom. More generally, such increases or spikes in $s$ may serve as signals of emergent massless degrees of freedom.

Figure~\ref{fig:speedcorr} shows that in the conformal limits $m/(b \sqrt{\lambda})\to 0$ or $\to \infty$ we find $\delta v^2 \to 0$, as expected. In most cases, for $m/(b \sqrt{\lambda})$ near the transition $\delta v^2$ spikes down to negative values. However, for some $T/b$, near the transition $\delta v^2$ becomes positive: in figure~\ref{fig:speedcorr} see $T/b= 0.1$ (solid blue) and $0.25$ (solid orange). In those cases the sound speed squared in eq.~\eqref{eq:soundspeed} is greater than the conformal value, $v^2 > 1/3$, thus violating the bound conjectured for $v^2$ in ref.~\cite{,Cherman:2009tw}. The significance of such behaviour, if any, we leave for future research.

\section{Conductivity}
\label{sec:conductivities}

In this section we compute our system's $U(1)_V$ DC longitudinal and Hall conductivities, \(\s_{xx}\) and \(\s_{xy}\), respectively. To do so, we use the method of refs.~\cite{Karch:2007pd,OBannon:2007cex}, wherein we introduce a non-dynamical, constant $U(1)_V$ electric field in the $x$ direction, $E$, compute the resulting expectation values of $U(1)_V$ currents, $\Jx$ and $\Jy$, and from these extract \(\s_{xx}\) and \(\s_{xy}\). As a check, in appendix~\ref{sec:kubo} we also compute $\s_{xx}$ and $\s_{xy}$ at $T=0$ using Kubo formulas, finding perfect agreement with the results in this section.

As before, we parameterize the D7-branes' worldvolume coordinates as \(\xi^a = (t,x,y,z,r)\) plus the \(S^3\) coordinates. Our ansatz for the worldvolume fields again includes $R(r)$ and $\f(z,r)=b z + \Phi(r)$. As mentioned in section~\ref{sec:holographic_model}, the $U(1)_V$ current is dual to the D7-brane's $U(1)$ worldvolume gauge field, so now our ansatz also includes two gauge field components. The first is $A_x = E t + a_x(r)$, where the first term introduces the electric field \(E\) in the \(x\) direction, while the second term, $a_x(r)$, allows for a non-zero $\Jx$. We use rotational symmetry in the \(xy\)-plane to set \(E > 0\). Second is $A_y(r)$, which allows for a non-zero $\Jy$.  The D7-brane action~\eqref{eq:d7_action} evaluated on this ansatz is
\begin{subequations}
\beq
\label{eq:d7_action_ansatz_E_field}
    S_\mathrm{D7} = - \cN \int d r \le(
        \sqrt{w_1(r) \le(h + h R'^2 + A_y'^2 \ri) + w_2(r) \f'^2+w_3(r) A_x'^2} - w_4(r) A_y'
    \ri),
\eeq
\begin{align}
 \label{eq:action_coefficients}
    w_1 &\equiv r^6 \le(h + \frac{L^4 b^2 R^2}{\left(r^2+R^2\right)^2} \ri) \le( g^2-  \frac{L^4 E^2}{\left(r^2+R^2\right)^2} \ri) , \\
    w_2 &\equiv r^6 h^2 R^2 \le(g^2- \frac{L^4 E^2}{\left(r^2+R^2\right)^2} \ri), \\
    w_3 &\equiv r^6 g^2 \le(h + \frac{L^4 b^2 R^2}{\left(r^2+R^2\right)^2} \ri),\\
    w_4 &\equiv \frac{L^4 r^4}{\left(r^2+R^2\right)^2} \, b \, E,
\end{align}
\end{subequations}
where $R' \equiv \partial R/\partial r$, and similarly for the other fields. The action in eq.~\eqref{eq:d7_action_ansatz_E_field} depends on \(\f'\), \(A_x'\), and \(A_y'\) and not on $\f$, $A_x$, or $A_y$. As a result, the equations of motion imply that the corresponding canonical momenta are independent of $r$. To be explicit, the canonical momenta conjugate to $\f$, $A_x$, and $A_y$ are, respectively,
\begin{subequations}
    \label{eq:canonical_momenta}
\begin{align}
    P_\f &\equiv \frac{\d S_\mathrm{D7}}{\d \f'} = - \cN \frac{w_3 \f'}{\sqrt{w_1(r) \le(h + h R'^2 + A_y'^2 \ri) + w_2(r) \f'^2+w_3(r) A_x'^2}},
    \\
    P_x &\equiv \frac{\d S_\mathrm{D7}}{\d A_x'} = - \cN \frac{w_2 A_x'}{\sqrt{w_1(r) \le(h + h R'^2 + A_y'^2 \ri) + w_2(r) \f'^2+w_3(r) A_x'^2}},
    \\
    P_y &\equiv \frac{\d S_\mathrm{D7}}{\d A_y'} = - \cN \frac{w_1 A_y'}{\sqrt{w_1(r) \le(h + h R'^2 + A_y'^2 \ri) + w_2(r) \f'^2+w_3(r) A_x'^2}} + \cN w_4,
\end{align}
\end{subequations}
and the corresponding Euler-Lagrange equations are, respectively, $\partial_r P_\f=0$, $\partial_r P_x=0$, and $\partial_r P_y=0$. We therefore write the canonical momenta as \(P_\f = \cN p_\f\), \(P_x = \cN j_x\) and \(P_y = \cN j_y\) with constants \(p_\f\), \(j_x\) and \(j_y\). In appendix~\ref{sec:holo_rg} we show that these constants determine the one-point functions of the dual operators: we again have \(\langle \cO_\f \rangle = \cN p_\f\), while
\beq 
\label{eq:current_vevs}
    \langle J_x \rangle = - \left(2 \pi \a'\right) \cN j_x,
    \qquad
    \langle J_y \rangle = - \left(2 \pi \a' \right)\cN j_y.
\eeq

To obtain an action for \(R(r)\) alone, we use a similar strategy to that in section~\ref{sec:holographic_model}, eliminating $\phi$, $A_x$, and $A_y$ in favour of $P_\f$, $P_x$, and $P_y$ by a Legendre transform. To be explicit, we insert \(P_\f = \cN p_\f\), \(P_x = \cN j_x\) and \(P_y = \cN j_y\) into eq.~\eqref{eq:canonical_momenta}, solve for $\f'$, $A_x'$, and $A_y'$, plug these solutions into the action eq.~\eqref{eq:d7_action_ansatz_E_field}, and then Legendre transform with respect to $\f$, $A_x$, and $A_y$. The result is
\begin{align}
    \tilde{S}_\mathrm{D7}  &\equiv S_\mathrm{D7} - \int d r \le(\f' P_\f + A_x' P_x + A_y' P_y \ri)
    \nonumber \\
    &= - \cN \int d r \,  \sqrt{h} \sqrt{1 + R'^2} \sqrt{
        w_1(r) - \frac{w_1(r)}{w_2(r)} p_\f^2 - \frac{w_1(r)}{w_3(r)} j_x^2 - \le[j_y - w_4(r) \ri]^2
    },
    \label{eq:LT_action_final}
\end{align}

Similar to $\tilde{S}_\mathrm{D7}$ in eq.~\eqref{eq:d7_action_LT} , the integrand of $\tilde{S}_\mathrm{D7}$ in eq.~\eqref{eq:LT_action_final} includes a product of three square roots. The first two of these, $\sqrt{h}$ and $\sqrt{1 + R'^2}$, are manifestly real for all $r$. However the third square root is not necessarily real for all $r$. To see why in detail, we re-write the factors under the third square root,
\begin{subequations}
\beq
    \tilde{S}_\mathrm{D7}  = - \cN \int d r \,  \sqrt{h} \sqrt{1 + R'^2} \sqrt{\alpha(r) \beta(r) - \gamma(r)
    },
    \label{eq:LT_action_final2}
\eeq
\begin{align}
    \a(r) &\equiv  g^2 - \frac{L^4 E^2}{\left(r^2+R^2\right)^2},\label{eq:adef}\\
    \b(r) &\equiv r^6 \le(h + \frac{L^4 b^2 R^2}{\left(r^2+R^2\right)^2}\ri) - \frac{ j_x^2 }{g^2},\label{eq:bdef}\\
    \g(r) &\equiv \frac{p_\f^2}{h^2} \le( \frac{h}{R^2} + \frac{L^4 b^2}{\left(r^2+R^2\right)^2} \ri) + \le(j_y - \frac{L^4 r^4 b E}{\left(r^2+R^2\right)^2}\ri)^2.\label{eq:cdef}
\end{align}
\end{subequations}
Clearly $\gamma(r) \geq 0$ for all $r$. However, $\alpha(r)$ and $\beta(r)$ can change sign. For example, if $T>0$ then each of $\alpha(r)$ and $\beta(r)$ is positive at the $AdS_5$ boundary, $r \to \infty$, and negative at the horizon, where $g=0$. Each must therefore change sign at some $r$ in between. If one of $\alpha(r)$ or $\beta(r)$ changes sign and the other does not, then $\alpha(r) \beta(r) - \gamma(r) <0$ for some range of $r$ (until the other also changes sign). In that case, $\tilde{S}_\mathrm{D7}$ acquires a non-zero imaginary part, signaling a tachyonic instability, as mentioned in section~\ref{sec:holographic_model}. The method of refs.~\cite{Karch:2007pd,OBannon:2007cex} is to adjust $p_\f$, $j_x$, and $j_y$ such that $\alpha(r)$ and $\beta(r)$ change sign at the same value of $r$, such that $\alpha(r) \beta(r) - \gamma(r) \geq 0$ for all $r$, thus avoiding the instability. With $\Jx = -(2\pi\alpha')\N j_x$ and $\Jy = -(2\pi\alpha')\N j_y$ thus fixed, we extract the DC conductivities via
\beq
\label{eq:condlimit}
\sigma_{xx} = \lim_{E \to 0} (2\pi\alpha') \Jx /E, \qquad \sigma_{xy} = - \sigma_{yx} = -\lim_{E \to 0} (2\pi\alpha') \Jy /E,
\eeq
where the factors of $(2\pi\alpha')$ come from our normalisation of the D7-brane's worldvolume gauge field, described below eq.~\eqref{eq:d7_action}.

We will thus impose the conditions $\a(r_*)=0$, $\b(r_*)=0$, and $\g(r_*)=0$ at some $r_*$. The location $r_*$ is in fact a horizon of the open string metric on the D7-brane worldvolume~\cite{Kundu:2018sof,Kundu:2019ull}. This horizon has an associated Hawking temperature,\footnote{Whether any entropy can be associated with this horizon is an open question~\cite{Sonner:2013mba,OBannon:2016exv,Kundu:2018sof,Kundu:2019ull}.} in general larger than the background $AdS_5$ black hole's Hawking temperature $T$. The difference in temperatures signals that these solutions do not describe thermal equilibrium states, since heat will flow from the D7-brane to the background black hole. In fact, stationary solutions with $E>0$ describe non-equilibrium steady states. For a review of their physics, see refs.~\cite{Kundu:2018sof,Kundu:2019ull} and references therein. We will ultimately take $E\to0$, as in eq.~\eqref{eq:condlimit}, so we only use the $E>0$ solutions in intermediate steps. However, the effect of $b>0$ on solutions with $E>0$ is worth studying in future research, as we discuss in section~\ref{sec:summary}.

Not all D7-brane embeddings have a worldvolume horizon when $E>0$. In particular, in some Minkowski embeddings the D7-brane ends before either $\a(r)$ or $\b(r)$ changes sign. Whether a worldvolume horizon appears thus depends on the boundary conditions, which in turn depend on $T$. We consider $T=0$ first, in section~\ref{sec:T0_conductivities}, and then $T>0$ in section~\ref{sec:T_conductivities}.

\subsection{Conductivity at Zero Temperature}
\label{sec:T0_conductivities}

If $T=0$ then $g=1$ and $h=1$. We first consider D7-brane embeddings with a worldvolume horizon at some $r_*$, the value of which is fixed by $\a(r_*)=0$. With the notation $R_* \equiv R(r_*)$ we have from eq.~\eqref{eq:adef}
\beq
\label{eq:T0_wvh_alpha_condition}
1 - \frac{L^4 E^2}{\left(r_*^2+R_*^2\right)^2} =0 \qquad \Rightarrow \qquad r_*^2 + R_*^2 = L^2 E.
\eeq
In the $(r,R)$ plane the worldvolume horizon is thus a circle of radius $L\sqrt{E}$. The conditions $\b(r_*)=0$ and $\g(r_*)=0$ then give, respectively,
\begin{subequations}
\beq
    r_*^6 \le(1 + \frac{L^4 b^2 R_*^2}{\left(r_*^2+R_*^2\right)^2} \ri) - j_x^2 = 0,
    \label{eq:T0_wvh_beta_condition}
\eeq
\beq
    p_\f^2 \le( \frac{1}{R_*^2} + \frac{L^4 b^2}{\left(r_*^2+R_*^2\right)^2} \ri) + \le(j_y - \frac{L^4 r_*^4 b E}{\left(r_*^2+R_*^2\right)^2}\ri)^2 = 0,
    \label{eq:T0_wvh_gamma_condition}
\eeq
\end{subequations}
In eq.~\eqref{eq:T0_wvh_gamma_condition} the left-hand side is a sum of squares, which vanishes if and only if each term in the sum vanishes independently. The first term vanishes only if $p_\f=0$, which implies $\Ophiv=0$. We thus take $p_\f=0$ henceforth. In that case eqs.~\eqref{eq:T0_wvh_beta_condition} and~\eqref{eq:T0_wvh_gamma_condition} determine $j_x$ and $j_y$, and thus $\Jx$ and $\Jy$ via eq.~\eqref{eq:current_vevs}, giving
\begin{subequations}
\label{eq:T0_wvh_currents}
\begin{eqnarray}
 \Jx & = & (2\pi\alpha')\N \, r_*^3\, \sqrt{1 +  \frac{L^4b^2R_*^2}{\left(r_*^2+R_*^2\right)^2}},\\
\Jy &=&-(2\pi\alpha') \N \, L^4 \,b\,E \frac{r_*^4}{\left(r_*^2+R_*^2\right)^2}.
\end{eqnarray}
\end{subequations}

We now need $r_*$ and $R_*$, at least in the limit $E \to 0$. If $E=0$ then clearly eq.~\eqref{eq:T0_wvh_alpha_condition} implies $r_*=0$ and $R_*=0$, or in other words $R_0\equiv R(r=0)=0$. We thus learn that when $E \to 0$ the solutions with worldvolume horizon reduce to the black-hole like embeddings with the exponential boundary condition in eq.~\eqref{eq:T0_boundary_condition_non_analytic}. As discussed in section~\ref{sec:T0_solutions}, these embeddings describe RG flows to an IR CFT, the massless probe hypermultiplet CFT.

To determine $r_*$ and $R_*$ at small $E$ we expand \(R(r) = R^{(0)}(r) + R^{(1)}(r) + \dots\), where \(R^{(0)}(r)\) is the solution at \(E=0\), \(R^{(1)}(r)\) is the first correction at small non-zero \(E\), and so on. All we will need to know about the terms in this expansion is that \(R^{(0)}\) obeys the boundary condition of eq.~\eqref{eq:T0_boundary_condition_non_analytic}, and in particular \(R^{(0)}(r) \propto \exp\le(-L^2 b/r\ri)\) at small \(r\). At leading order in small \(E\), eq.~\eqref{eq:T0_wvh_beta_condition} becomes
\beq
\label{eq:rstar_small_E}
    r_*^2 + R^{(0)}(r_*)^2 \approx L^2 E.
\eeq
When $E\to0$ we have $r_* \to 0$, but \(R^{(0)}(r_*) \propto \exp\le(-L^2 b/r_*\ri)\to0\) more quickly, so to leading approximation eq.~\eqref{eq:rstar_small_E} gives $r_* \approx L \sqrt{E}$ and $R^{(0)}(r_*)\approx0$, and hence $R_*\approx 0$. Plugging these into eq.~\eqref{eq:T0_wvh_currents} gives
\begin{subequations}
\label{eq:T0_wvh_currents_smallE}
\begin{eqnarray}
 \Jx & = & (2\pi\alpha')\N \, L^3 \, E^{3/2}\,\\
\Jy &=&-(2\pi\alpha') \N \, L^4 \, b \, E.
\end{eqnarray}
\end{subequations}
Given that our IR is a CFT, the powers of $E$ in eq.~\eqref{eq:T0_wvh_currents_smallE} are dictated by dimensional analysis and the fact that $\Jy$ must be proportional to the $\mathcal{T}$-breaking parameter $b$. Plugging eq.~\eqref{eq:T0_wvh_currents_smallE} into eq.~\eqref{eq:condlimit} we obtain
\beq
\label{sec:T0_conductivities_low_mass}
\s_{xx} = 0, \qquad \s_{xy} = \frac{N_f N_c}{4 \pi^2}\, b.
\eeq
As mentioned above, in appendix~\ref{sec:kubo} we also computed $\s_{xx}$ and $\s_{xy}$ at $T=0$ using Kubo formulas, and found perfect agreement with eq.~\eqref{sec:T0_conductivities_low_mass}.

We thus find that the black-hole like embeddings with the exponential boundary condition in eq.~\eqref{eq:T0_boundary_condition_non_analytic} describe RG flows from small values of \(m/(b\sqrt{\l})\) in the UV to massless hypermultiplets in the IR, with vanishing DC longitudinal conductivity, $\sigma_{xx}=0$, and an anomalous Hall conductivity, $\sigma_{xy} \neq 0$. The latter indicates that $\mathcal{T}$ is broken in the IR.

Remarkably, our $\sigma_{xy}$ in eq.~\eqref{sec:T0_conductivities_low_mass} is \textit{independent} of the hypermultiplet mass $m$. In fact, it takes the value determined by the $U(1)_A$ anomaly when $m=0$, but now extended to cases with small \(m/(b\sqrt{\l})\), described by our black-hole like embeddings. In contrast, $\sigma_{xy}$ of the free Dirac fermion in eq.~\eqref{eq:wsm_toy_model_hall_current} and of previous holographic models~\cite{Landsteiner:2015lsa,Landsteiner:2015pdh,Copetti:2016ewq,Liu:2018spp,Landsteiner:2019kxb,Juricic:2020sgg} depended on $m$, and in particular decreased as $m$ increased, reaching $\sigma_{xy}=0$ at a quantum critical point. The reason for this difference is clear from a holographic perspective. In previous holographic models, $\sigma_{xy}$ was proportional to the product of the Chern-Simons coefficient and the value of the $U(1)_A$ gauge field at the horizon. Our result in eq.~\eqref{sec:T0_conductivities_low_mass} has the same form, but in our case the $U(1)_A$ gauge field is $\partial_{\mu} \phi/2$, as mentioned below eq.~\eqref{eq:axialpot}. Our ansatz is $\phi(z,r) = bz + \Phi(r)$, and our solution includes $p_\f=0$, which implies $\phi(z,r)$ is actually independent of $r$ and hence $\Phi(r)=0$. As a result, our $U(1)_A$ gauge field is simply $A_z^5 = \partial_z \phi/2 = b/2$, leading to our $m$-independent result for $\sigma_{xy}$ in eq.~\eqref{sec:T0_conductivities_low_mass}. Crucially, this $m$-independence is not required by any symmetry, and is not determined by the $U(1)_A$ anomaly alone, but comes from dynamics, and specifically from the fact that we had to take $p_\f=0$ to avoid a tachyonic instability, as explained above.

It is possible that the \(m\)-independence of \(\s_{xy}\) is an artefact of the probe limit, holding only in the limit \(N_f/N_c \to 0\). Away from this limit one should solve the full, coupled equations of motion for both the bulk fields (such as the metric) and the worldvolume fields of the D7-brane. These equations may admit solutions with \(A_z^5 = \p_z \f/2\) depending non-trivially on \(r\), leading to a different form for \(\s_{xy}\) as a function of \(m/(b\sqrt{\lambda})\). Further, even within the probe limit, if the solutions that minimise the free energy are not captured by our ansatz for \(\f(z,r)\) then \(\s_{xy}\) would most likely take a different form.

We now consider the case where the D7-brane has no worldvolume horizon. These embeddings are necessarily Minkowski, and in particular the D7-brane should reach $r=0$ outside of the worldvolume horizon described by the semicircle in eq.~\eqref{eq:T0_wvh_alpha_condition}. Indeed, demanding $\a(r)\geq0$ for all $r$ gives $r^2+R(r)^2 \geq L^2 E$ for all $r$. Evaluating this at $r=0$ gives $R_0 > L \sqrt{E}$. Similarly we demand $\beta(r)\geq0$ for all $r$. Evaluating this at $r=0$ gives $\beta(0)=-j_x^2\geq 0$, which implies $j_x=0$, so that in fact $\beta(0)=0$. Finally we demand $\alpha(r) \beta(r) - \gamma(r) \geq 0$ for all $r$. Evaluating this at $r=0$ gives $-\gamma(0)\geq0$, where
\beq
\g(0) = \frac{p_\f^2}{R_0^2} \le(1 + \frac{L^4 b^2}{R_0^2} \ri) + j_y^2.
\eeq
Clearly $-\gamma(0) \geq 0$ is possible if and only if $p_\f=0$ and $j_y=0$, so that in fact $\gamma(0)=0$. We thus find that Minkowski embeddings with $R_0 \geq L \sqrt{E}$ have $p_\f=0$, $j_x=0$, and $j_y=0$, implying $\Ophiv=0$, $\Jx=0$, and $\Jy=0$, respectively. As a result, $\sigma_{xx}=0$ and $\sigma_{xy}=0$. Minkowski embeddings typically have discrete spectra~\cite{Karch:2002sh,Kruczenski:2003be,Hoyos:2006gb}, so given that \(\s_{ij}=0\) we identify these embeddings as describing trivially insulating states.

If we now take $E \to 0$, then $R_0 \geq L\sqrt{E}$ becomes simply $R_0\geq0$. The \(E \to 0\) limit of embeddings with no worldvolume horizon thus correspond to the Minkowski embeddings with the $r\to 0$ boundary condition in eq.~\eqref{eq:T0_boundary_condition_analytic}, describing large values of \(m/(b\sqrt{\l})\). We have therefore learned that the large \(m/(b\sqrt{\l})\) phase exhibits no current flow in response to an applied electric field, as both the longitudinal and Hall conductivities vanish, \(\s_{xx} =0\) and \(\s_{xy} = 0\), respectively. The latter indicates that $\mathcal{T}$ is preserved in the IR.

To summarise, when $T=0$ and $E=0$ we find that $\Ophiv=0$ and \(\s_{xx} = 0\) for all \(m/(b\sqrt{\l})\), while \(\s_{xy}\) takes the non-zero value in eq.~\eqref{sec:T0_conductivities_low_mass} at small \(m/(b\sqrt{\l})\), dual to black-hole-like embeddings, but vanishes at large \(m/(b\sqrt{\l})\), dual to Minkowski embeddings. In section~\ref{sec:T0_solutions} we found a first-order transition from black hole-like embeddings to Minkowski embeddings at  \(m/(b\sqrt{\l}) \approx  0.0733 \), so the Hall conductivity in our model is
\beq
\label{eq:T0_Hall}
    \s_{xy} = \begin{cases}
            \dfrac{N_f N_c}{4 \pi^2} \,b, \quad & m/(b \sqrt{\l}) \lesssim 0.0733,
            \\[1em]
            0,  \quad & m/(b \sqrt{\l}) \gtrsim 0.0733,
        \end{cases}
\eeq
and correspondingly in the IR $\mathcal{T}$ is broken when $\sigma_{xy}\neq0$ and is preserved when $\sigma_{xy}=0$. Remarkably, $\sigma_{xy}$ when $m/(b \sqrt{\l}) \lesssim 0.0733$ is independent of $m$, in contrast to the free Dirac fermion in eq.~\eqref{eq:wsm_toy_model_hall_current} and previous holographic models~\cite{Landsteiner:2015lsa,Landsteiner:2015pdh,Copetti:2016ewq,Liu:2018spp,Landsteiner:2019kxb,Juricic:2020sgg}. This $\sigma_{xy}$ is precisely the value dictated by the $U(1)_A$ anomaly when $m=0$, but now extended to $m/(b \sqrt{\l}) \lesssim 0.0733$. We thus identify the $m/(b \sqrt{\l}) \lesssim 0.0733$ phase as a WSM and the $m/(b \sqrt{\l}) \gtrsim 0.0733$ phase as a trivial insulator, as indicated in our phase diagram, figure~\ref{fig:phase_diagram}.

\subsection{Conductivity at Non-Zero Temperature}
\label{sec:T_conductivities}

At $T>0$ we first consider embeddings with a worldvolume horizon at $r_*$, determined by $\a(r_*)=0$. Denoting the values of $R$ and $g$ at $r_*$ as $R_*$ and $g_*$, respectively, and using the definition of $\a(r)$ in eq.~\eqref{eq:adef} and $g$ in eq.~\eqref{eq:background_solution}, we find
\beq
\label{eq:finite_T_wvh_location}
g_*^2 - \frac{L^4 \, E^2}{\left(r_*^2+R_*^2\right)^2}=0, \qquad \Rightarrow \qquad r_*^2 + R_*^2 =  L^2 \, E/2 + \sqrt{L^4 \, E^2/4 + \rho_H^4}.
\eeq
In the $(r,R)$ plane the worldvolume horizon is thus again a semicircle, which at $E=0$ coincides with the $AdS_5$-Schwarzschild horizon $\rho_H$, and as $E$ grows, monotonically moves to larger $r$. In particular, $r_*^2 + R_*^2 \geq \rho_H^2$ for all $r$, i.e. the worldvolume horizon is always coincident with or outside the $AdS_5$-Schwarzschild horizon. In general, when $T>0$ and $E>0$, embeddings with a worldvolume horizon fall into two categories. The first are black hole embeddings. The second are Minkowski embeddings in which the D7-brane has a worldvolume horizon but ends before reaching the black hole horizon.

Denoting the value of $h$ at $r_*$ as $h_*$, $\b(r_*)=0$ and $\g(r_*)=0$ give, respectively,
\begin{subequations}
\beq
\label{eq:finite_T_wvh_beta_condition}
r_*^6 \le(h_* + \frac{L^4 b^2 R_*^2}{\left(r_*^2+R_*^2\right)^2}\right) - \frac{ j_x^2 }{g^2_*} = 0,
\eeq
\beq
\label{eq:finite_T_wvh_gamma_condition}
\frac{p_\f^2}{h^2_*} \le( \frac{h_*}{R_*^2} + \frac{L^4 b^2}{\left(r_*^2+R_*^2\right)^2} \ri) + \le(j_y - \frac{L^4 r_*^4 \, b \, E}{\left(r_*^2+R_*^2\right)^2}\ri)^2 = 0.
\eeq
\end{subequations}
Similar to the $T=0$ case, in eq.~\eqref{eq:finite_T_wvh_gamma_condition} the left-hand side is a sum of squares, which vanishes if and only if each term in the sum vanishes independently. The first term vanishes only if $p_\f=0$, so we take $p_\f=0$ henceforth, and thus $\Ophiv=0$. In that case eqs.~\eqref{eq:finite_T_wvh_beta_condition} and~\eqref{eq:finite_T_wvh_gamma_condition} determine $j_x$ and $j_y$, and thus $\Jx$ and $\Jy$,
\begin{subequations}
\label{eq:finite_T_wvh_currents}
\begin{eqnarray}
 \Jx & = & (2\pi\alpha')\N \, r_*^3\,\frac{L^2 E}{r_*^2+R_*^2} \,\sqrt{h_* + \frac{L^4 b^2 R_*^2}{\left(r_*^2+R_*^2\right)^2}},\\
\Jy &=&-(2\pi\alpha') \N \, L^4 \, b\, E \frac{ r_*^4}{\left(r_*^2+R_*^2\right)^2}.
\end{eqnarray}
\end{subequations}

We are interested in the limit $E\to0$. As mentioned above, when $E=0$ the worldvolume horizon coincides with the black hole horizon, or in other words, only black hole embeddings have a worldvolume horizon, which is at $r_H$. When $E \to 0$ we thus have $r_* \to r_H$ and $r_*^2 + R_*^2 \to r_H^2+R(r_H)^2 = \rho_H^2$. Expanding $\Jx$ and $\Jy$ in eq.~\eqref{eq:finite_T_wvh_currents} in $E$, and using \(\r_H = \pi L^2 T/\sqrt{2}\) from eq.~\eqref{eq:hawkingT}, we thus find
\begin{subequations}
\beq
\Jx = (2\pi\alpha')\N \, r_H^3 \, \frac{4 E}{\pi^4 L^4 T^4} \sqrt{\frac{1}{2}\pi^4 L^4 T^4+b^2 \, R(r_H)^2} + \cO(E^2),
\eeq
\beq
\Jy = - (2\pi\alpha')\N\,L^4\,b\,E \frac{4 r_H^4}{\pi^4 L^8 T^4} + \cO(E^2).
\eeq
\end{subequations}
Using eq.~\eqref{eq:condlimit} we then find
\begin{subequations}
\label{eq:finite_T_conductivities}
\beq
\s_{xx} = \frac{N_f N_c}{4\pi^2}\,r_H^3\, \frac{4}{\pi^4 L^8 T^4} \sqrt{\frac{1}{2}\pi^4 L^4 T^4+b^2 R(r_H)^2},
\eeq
\beq
\s_{xy} = \frac{N_f N_c}{4\pi^2} \,b\,\frac{4 \,r_H^4}{\pi^4 L^8 T^4}.
\eeq
\end{subequations}
In general, we must determine $\sigma_{xx}$ and $\sigma_{xy}$ as functions of $m/(b\sqrt{\l})$ numerically. In particular, as explained at the beginning of section~\ref{sec:finite_T_solutions}, for black hole embeddings we choose $r_H$, which determines $R(r_H)=\sqrt{\rho_H^2-r_H^2}$, and impose $R'(r_H)=0$, numerically solve the equation of motion, and then from the large-$r$ asymptotics we extract $m/(b\sqrt{\l})$.

However, we know one black hole embedding exactly, namely the trivial solution $R(r)=0$, corresponding to $m=0$. The trivial solution has $r_H=\rho_H$, so from eq.~\eqref{eq:finite_T_conductivities} we find
\beq
\label{eq:conductivities_trivial}
\s_{xx} = \frac{N_f N_c}{4\pi^2} \, \pi T,
\qquad
\s_{xy} = \frac{N_f N_c}{4\pi^2} \, b,
\qquad
m=0.
\eeq
Clearly when $m=0$ all $b$ dependence disappears from $\sigma_{xx}$, which takes the $b=0$ and $m=0$ value of ref.~\cite{Karch:2007pd}, and all $T$ dependence disappears from $\sigma_{xy}$, which takes the value we found in eq.~\eqref{sec:T0_conductivities_low_mass} at $T=0$ and small $m/(b\sqrt{\l})$.

We now consider the case where the D7-brane has no worldvolume horizon. Our arguments here are very similar to the $T=0$ case, so we will be brief. These embeddings are necessarily Minkowski, with the D7-brane ending at $r=0$ outside of the worldvolume horizon in eq.~\eqref{eq:finite_T_wvh_location}. We demand $\a(r) \geq 0$ and $\b(r) \geq 0$ for all $r\in [0,\infty)$. The condition $\b(0)\geq0$ is satisfied if and only if $j_x=0$, so that in fact $\b(0)=0$. We also demand $\a(r)\beta(r)-\g(r)\geq0$ for all $r\in [0,\infty)$, which when evaluated at $r=0$ becomes $-\gamma(0)\geq0$, which is satisfied if and only if $p_\f=0$ and $j_y=0$. As a result, embeddings without a worldvolume horizon describe states with $\Ophiv=0$, $\Jx=0$, and $\Jy=0$, and hence $\sigma_{xx}=0$ and $\sigma_{xy}=0$, i.e. trivially insulating states.

\begin{figure}
    \begin{subfigure}{0.5\textwidth}
        \includegraphics[width=\textwidth]{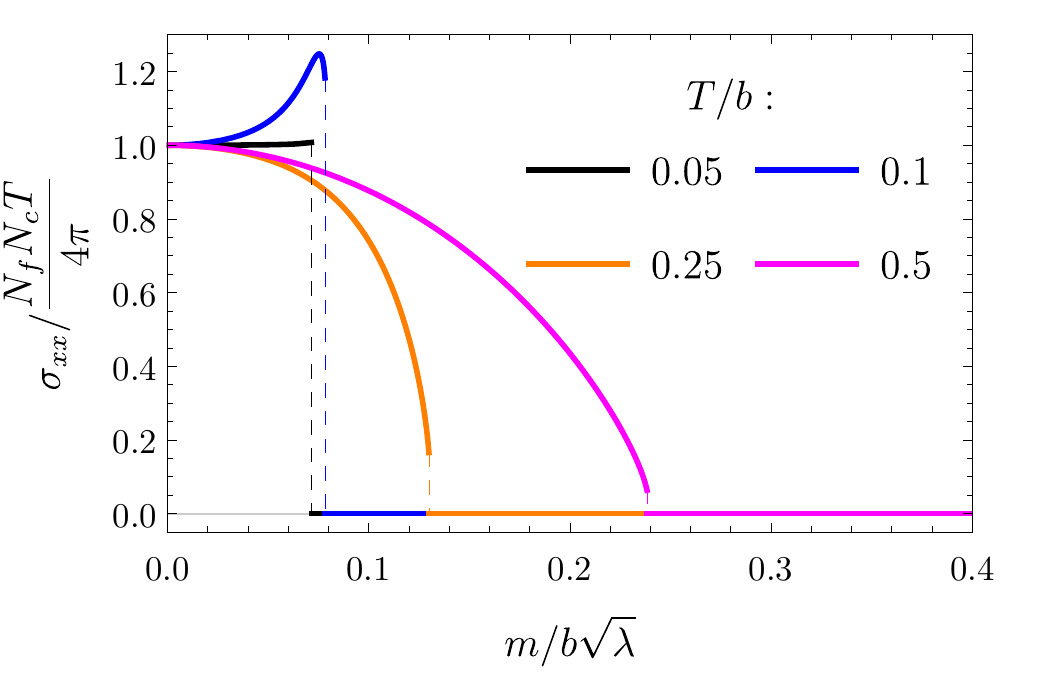}
        \caption{Longitudinal conductivity \(\s_{xx}\).}
        \label{fig:longitudinal_conductivity}
    \end{subfigure}
    \begin{subfigure}{0.5\textwidth}
        \includegraphics[width=\textwidth]{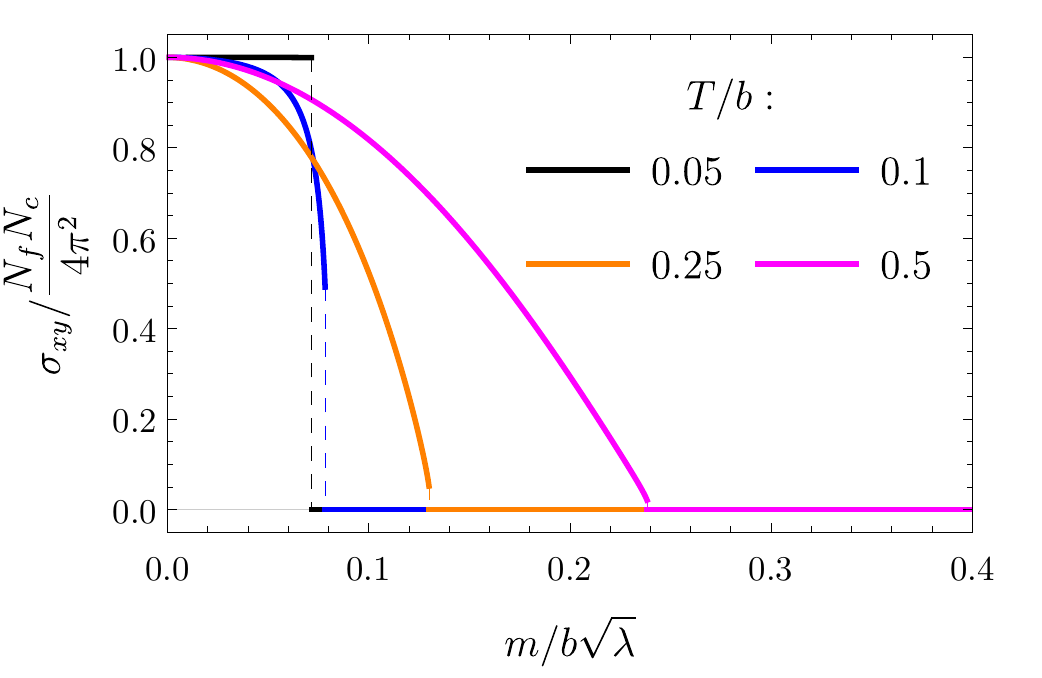}
        \caption{Hall conductivity \(\s_{xy}\).}
        \label{fig:hall_conductivity}
    \end{subfigure}
    \begin{subfigure}{0.5\textwidth}
        \includegraphics[width=\textwidth]{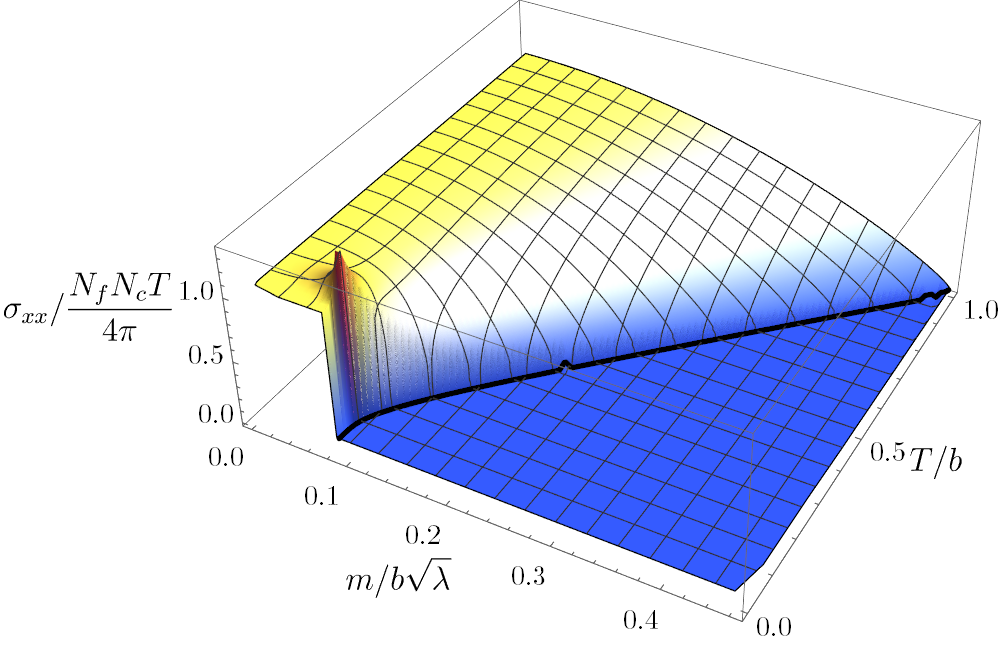}
        \caption{Longitudinal conductivity \(\s_{xx}\).}
        \label{fig:longitudinal_conductivity_3d}
    \end{subfigure}
    \begin{subfigure}{0.5\textwidth}
        \includegraphics[width=\textwidth]{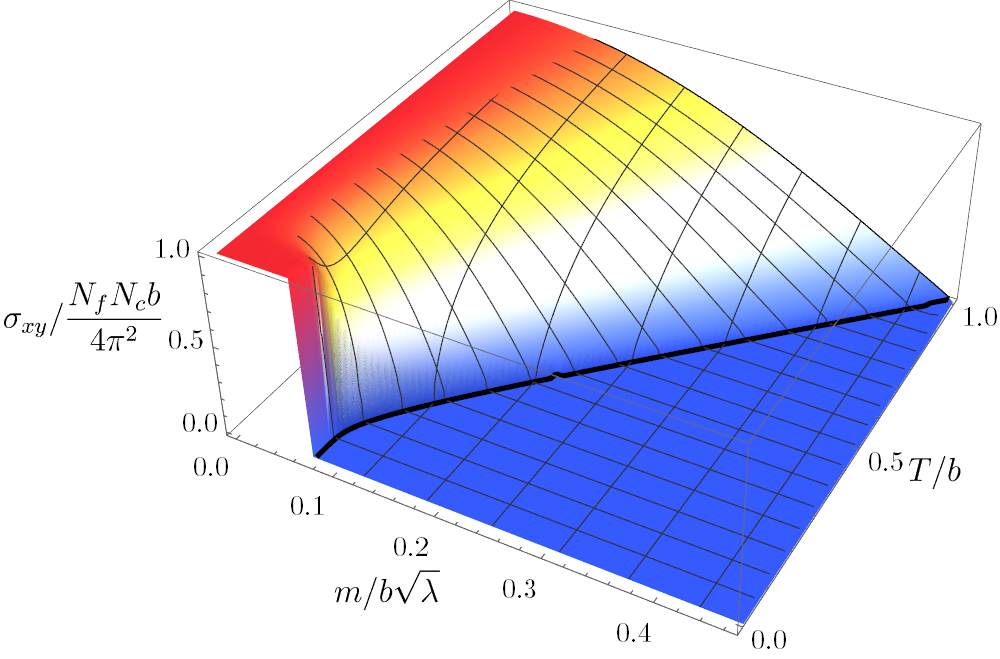}
        \caption{Hall conductivity \(\s_{xy}\).}
        \label{fig:hall_conductivity_3d}
    \end{subfigure}
    \caption{
        Our numerical results for \textbf{(a)} the DC longitudinal conductivity, \(\s_{xx}/(N_fN_cT/4\pi)\), and \textbf{(b)} the DC Hall conductivity, \(\s_{xy}/(N_fN_c/(4\pi^2))\), as functions of \(m/(b\sqrt{\l})\), for \(T/b=0.05\) (black) $0.1$ (blue), $0.25$ (orange), and $0.5$ (purple). In each case the vertical dashed line indicates the first order phase transition of figure~\ref{fig:phase_diagram}. As $m/(b\sqrt{\l})$ approaches the transition from below, $\sigma_{xy}$ decreases monotonically for all $T/b$, reaching $\sigma_{xy}=0$ at the transition. In contrast, $\sigma_{xx}$ decreases monotonically for larger $T/b$, but exhibits a maximum near the transition at small $T/b$. For $m/(b\sqrt{\l})$ above the transition, $\sigma_{xx}=0$ and $\sigma_{xy}=0$. \textbf{(c)} and~\textbf{(d)} 3D plots of our numerical results for \(\s_{xx}/(N_fN_cT/4\pi)\) and \(\s_{xy}/(N_fN_c b/(4\pi^2))\), respectively, as functions of \(m/(b\sqrt{\l})\) and \(T/b\).
    }
    \label{fig:finite_T_conductivities}
\end{figure}

Figure~\ref{fig:finite_T_conductivities} shows some of our numerical results for \(\s_{xx}\) and \(\s_{xy}\), normalised by their $m=0$ values in eq.~\eqref{eq:conductivities_trivial}. Figures~\ref{fig:longitudinal_conductivity} and~\ref{fig:hall_conductivity} show \(\s_{xx}/(N_fN_cT/4\pi)\) and \(\s_{xy}/(N_fN_c/(4\pi^2))\) as functions of \(m/(b \sqrt{\l})\), for sample values of \(T/b\). In both figures the vertical dashed lines indicate the first order phase transition of figure~\ref{fig:phase_diagram}. For small $T/b$, such as for example $T/b=0.1$ (blue in figure~\ref{fig:longitudinal_conductivity}), as we increase $m/(b\sqrt{\l})$ we find $\sigma_{xx}$ exhibits a maximum just below the transition. Such behaviour likely indicates a pole in $J_x$'s retarded two-point function in Fourier space near the origin of the complex frequency plane, and may be related to the IR CFT, similar to what we discussed for the entropy density in figure~\ref{fig:thermodynamics_solutions}. At larger $T/b$ however, $\sigma_{xx}$ decreases monotonically, reaching $\sigma_{xx}=0$ at the transition. In contrast, for all $T/b>0$ we find $\sigma_{xy}$ decreases monotonically as $m/(b\sqrt{\l})$ increases, reaching $\sigma_{xy}=0$ at the transition. For $m/(b\sqrt{\l})$ above the transition, $\sigma_{xx}=0$ and $\sigma_{xy}=0$.

Figures~\ref{fig:longitudinal_conductivity_3d} and~\ref{fig:hall_conductivity_3d} show 3D plots of \(\s_{xx}\) and \(\s_{xy}\), normalised by their $m=0$ values in eq.~\eqref{eq:conductivities_trivial}, as functions of \(m/(b\sqrt{\l})\) and \(T/b\). These plots summarise all of our main results. For example, the phase diagram of figure~\ref{fig:phase_diagram} is apparent in the plane of \(m/(b\sqrt{\l})\) and \(T/b\), the step-function in $\sigma_{xy}$ of eq.~\eqref{eq:T0_Hall} is obvious in figure~\ref{fig:hall_conductivity_3d} at $T=0$, and so on. Figure~\ref{fig:longitudinal_conductivity_3d} also shows a spike in $\sigma_{xx}$ at low $T/b$ and $m/(b\sqrt{\l})$ near the transition, consistent with the maximum in figure~\ref{fig:longitudinal_conductivity}. As mentioned above, this spike likely comes from a pole in $J_x$'s retarded two-point function near the origin of the complex frequency plane.

\section{Summary and Outlook}
\label{sec:summary}

In this paper we studied a top-down holographic model of a WSM, namely probe D7-branes in the $AdS_5 \times S^5$ background of type IIB supergravity, dual to probe hypermultiplets in $\N=4$ SYM at large $N_c$ and large coupling $\lambda$, with worldvolume fields describing non-zero hypermultiplet mass $m$ and background spatial $U(1)_A$ gauge field $b$. The latter explicitly breaks time reversal symmetry, $\mathcal{T}$.

At zero temperature, $T=0$, we found that sufficiently small values of $m/(b\sqrt{\lambda})$ in the UV renormalise to zero mass in the IR, so that the IR is a CFT, namely $\N=4$ SYM coupled to massless probe hypermultiplets. As we increased $m/(b\sqrt{\lambda})$ we found a first-order quantum phase transition, at $m/(b\sqrt{\lambda})\approx 0.0733$. When $m/(b\sqrt{\lambda})\lesssim 0.0733$, we found a WSM with $\sigma_{xx}=0$ but non-zero anomalous Hall conductivity $\sigma_{xy}$, and hence broken $\mathcal{T}$ in the IR. Remarkably, this $\sigma_{xy}$ at $T=0$ was independent of $m/(b\sqrt{\lambda})$, retaining its $m=0$ value, determined by the $U(1)_A$ anomaly, for all $m/(b\sqrt{\lambda})\lesssim 0.0733$. When $m/(b\sqrt{\lambda})\gtrsim 0.0733$ we found a trivial insulator with $\sigma_{xx}=0$ and $\sigma_{xy}=0$, and hence restored $\mathcal{T}$ in the IR. The first order transition survived for all $T/b>0$, as summarised in our phase diagram, figure~\ref{fig:phase_diagram}. The biggest effect of $T/b>0$ was the fact that both $\sigma_{xx}$ and $\sigma_{xy}$ acquired non-trivial dependence on $m/(b\sqrt{\lambda})$ and $T/b$ in the WSM phase, though both still vanished in the trivial insulator phase, as summarised in figure~\ref{fig:finite_T_conductivities}. We also studied our model's thermodynamics, finding among other things a rise in the entropy density at low $T/b$ for $m/(b\sqrt{\lambda})$ just below the transition, presumably coming from the emergent IR CFT degrees of freedom.

Our model had several non-trivial features distinct from previous models, such as the free Dirac fermion model and previous holographic models that we discussed in section~\ref{sec:intro}. Chief among these differences was our first order transition for all $T/b$, including $T/b=0$, in contrast to the second-order (quantum) phase transitions of most previous models, as well as the fact that our $T=0$ anomalous Hall conductivity was independent of $m$.

These results raise many crucial questions for future research on this model. For example, what is the spectrum of excitations of our model? In particular, how do the retarded Green's functions in the probe sector depend on $m/(b\sqrt{\lambda})$ and $T/b$? Where are their poles, representing the excitations of the system? How do the corresponding spectral functions behave? How does the IR CFT affect these? Does a pole in the retarded two-point function of the $U(1)_V$ current produce the maximum we saw in $\sigma_{xx}$ at sufficiently small $m/(b\sqrt{\lambda})$ and $T/b$ in figure~\ref{fig:finite_T_conductivities}? How does the spectrum of excitations differ between modes propagating parallel and perpendicular to the axial gauge field? More generally, the spectrum of excitations could reveal whether our model has perturbative instabilities, long-lived propagating modes, the expected Fermi surfaces and associated topological invariants, and more.

Particularly important excitations characterising the WSM phase are of course Fermi arcs. Does our model support Fermi arcs? These may be ``washed out'' at strong coupling, nevertheless boundary currents required by the $U(1)_A$ anomaly, and hence topologically protected, should still appear~\cite{Ammon:2016mwa}. Does our model support such boundary currents?

Due to the anisotropy introduced by the axial gauge field, the hydrodynamic description of WSMs contains a richer set of transport coefficients than rotationally invariant systems, for example there are two shear viscosities governing dissipation of momentum in different directions~\cite{Landsteiner:2019kxb}. Holographic models have also revealed anomalous Hall viscosities in WSMs~\cite{Landsteiner:2016stv}. Does our model support an anomalous Hall viscosity? From the holographic perspective this phenomenon arises from a mixed $U(1)_A$-gravitational Chern-Simons term~\cite{Landsteiner:2016stv}. The D7-brane WZ terms indeed include a term of the correct form~\cite{Johnson:2000ch}, which however comes with an additional factor of \(\a'^2 = L^4/\l\) compared to the WZ term we included in eq.~\eqref{eq:d7_action}, and hence is suppressed when $\l \gg 1$. We therefore expect that our model indeed exhibits anomalous Hall viscosities, albeit vanishing as \(1/\l\) at strong coupling.

Holographic probe brane models exhibit several special phenomena, especially in transport. For example, in our model a non-zero electric field, $E>0$, can induce negative differential conductivity, in which the longitudinal conductivity $\sigma_{xx}$ is a \textit{decreasing} function of the electric field~\cite{Nakamura:2010zd}. In contrast, in typical metals an increasing electric field produces a larger current. For a WSM in parallel electric and magnetic fields, the $U(1)_A$ anomaly can induce negative magneto-resistance, in which $\sigma_{xx}$ is an \textit{increasing} function of the magnetic field, in contrast to typical metals~\cite{Son_2013}. Remarkably, our model exhibits negative magneto-resistance already when $b=0$~\cite{Ammon:2009jt,Baumgartner:2017kme}. How does non-zero $b$ affect these phenomena? Could this model suggest any unusual transport in real strongly-coupled WSMs?

More generally, this paper opens the way for top-down holographic probe brane models of many semi-metal phenomena, such as type II WSMs, nodal line semi-metals, nodal loop semi-metals, and more. We intend to pursue many of these in future research, using this paper as a foundation.

\section*{Acknowledgments}

We would like to thank Henk Stoof for useful discussions. We acknowledge support from STFC through Consolidated Grant ST/P000711/1. A. O'B. is a Royal Society University Research Fellow. The work of R. R. was supported by the D-ITP consortium, a program of the Netherlands Organisation for Scientific Research (NWO) that is funded by the Dutch Ministry of Education, Culture and Science (OCW).

\appendix
\section{Holographic Renormalization}
\label{sec:holo_rg}

\subsection*{On-shell Action}

For the purposes of holographic renormalization, it will be convenient to replace \(r\) and \(R\) with new coordinates \(u = L^2/\sqrt{r^2 + R^2}\) and \(\q = \arctan (R/r)\). The inverse transformation is \(r= L^2 u^{-1} \cos\q\), \(R= L^2 u^{-1} \sin\q\).  It will also be convenient to use units in which the \(AdS\) radius is \(L\equiv1\), restoring factors of \(L\) by dimensional analysis at the end. In these coordinates, the asymptotically \(AdS_5 \times S^5\) black brane background in eq.~\eqref{eq:background_solution} becomes
\begin{align}
    ds^2 &= \frac{1}{u^2} \le[ -\frac{g^2(u)}{h(u)} dt^2 + h(u) d\vec{x}^2 + du^2\ri] + d\q^2 + \sin^2 \q \, d \f^2 + \cos^2 \q \, ds_{S^3}^2
    \nonumber \\
    C_4 &= \frac{1}{u^4} h^2(u) d t \wedge d x \wedge d y \wedge d z - \cos^4 \q \,d \f \wedge d s_{S^3},
    \label{eq:background_solution_fg_gauge}
\end{align}
where in a slight abuse of notation we have defined \(g(u) \equiv 1 - \r_0^4 u^4\) and \(h(u) \equiv 1 + \r_0^4 u^4\). The boundary is at \(u=0\), while the horizon of the black brane is at \(u = 1/\r_0\).

In this coordinate system, our ansatz for the D7-brane embedding becomes
\beq \label{eq:holo_rg_E0_ansatz}
    \q = \q(u), \quad \f = b z + \Phi(u).
\eeq
From the near-boundary expansion of \(R(r)\) in eq.~\eqref{eq:R_near_boundary} we can find the near-boundary expansion of \(\q(u)\),
\beq \label{eq:theta_small_u}
    \q(u) = M \le(u + \frac{b^2}{2} u^3\log u \ri) + \le(C + \frac{M^3}{6}\ri) u^3 + \dots \; .
\eeq
Solving eq.~\eqref{eq:phi_momentum} for \(\phi'\) in terms of \(p_\f = P_\f/\cN\) and \(R\), replacing \((r,R)\) with \((u,\q)\), and using $\theta(u)$'s near-boundary expansion in eq.~\eqref{eq:theta_small_u}, we also find $\f$'s near-boundary expansion,
\beq \label{eq:phi_small_u}
    \f = b z - \frac{p_\f}{2} u^2 + \dots \; .
\eeq

The D7-brane action evaluated on the ansatz in eq.~\eqref{eq:holo_rg_E0_ansatz} is
\beq \label{eq:holo_rg_bulk_action}
    S_\mathrm{D7} = - \cN \int du \frac{\cos^3 \q}{u^5} g(u) h(u) \sqrt{
        \le(1 + \frac{b^2 u^2 \sin^2\q}{h(u)} \ri) \le(1 + u^2 \q'^2 \ri) + u^2 \sin^2 \q \, \f'^2
    }.
\eeq
If we plug in the near-boundary expansions in eqs.~\eqref{eq:theta_small_u} and~\eqref{eq:phi_small_u}, then we find that the integrand diverges near \(u=0\). We regularize this divergence by introducing a small-\(u\) cutoff at \(u=\e\). We then find that the on-shell action is
\beq \label{eq:on_shell_action_divergences}
    S^\star_\mathrm{D7} = \cN \le[-\frac{1}{4\e^4} + \frac{M^2}{2 \e^2} + b^2 M^2 \log \e + \cO(\e^0) \ri],
\eeq
where the \(\cO(\e^0)\) term cannot be determined from the near-boundary analysis.

To have a well-defined variational principle we need to remove the small-\(\e\) divergences in eq.~\eqref{eq:on_shell_action_divergences} by the addition of counterterms to the action. The full D7-brane action is then \(S = S_\mathrm{D7} + S_\mathrm{ct}\) where the counterterms are given by \(S_\mathrm{ct} = \sum_i  S_{\mathrm{ct},i}\) with~\cite{Karch:2005ms,Karch:2006bv,Hoyos:2011us,Hoyos:2011zz}
\begin{gather}
    S_{\mathrm{ct},1} = \cN \frac{1}{4} \sqrt{-\g},
    \qquad
    S_{\mathrm{ct},2} = - \cN \frac{1}{2} \sqrt{-\g} |\Theta|^2,
    \qquad
    S_{\mathrm{ct},3} = \cN \frac{5}{12} \sqrt{-\g} |\Theta|^4,
    \nonumber \\[0.5em]
    S_{\mathrm{ct},4} = \cN \frac{1}{2} \sqrt{-\g} \Theta^* \Box_\g \Theta \log |\Theta| ,
    \qquad
    S_{\mathrm{ct},5} = \cN \frac{1}{4} \sqrt{-\g} \Theta^* \Box_\g \Theta.
    \label{eq:counterterms}
\end{gather}
where \(\Theta \equiv \q e^{i\f}\), \(\g_{\m\n} = \e^{-2} \h_{\m\n}\) is the induced metric on the intersection of the brane with the cutoff surface at \(u = \epsilon\), and \(\Box_\g \Theta = \frac{1}{\sqrt{-\g}} \p_\m \le(\sqrt{-\g} \, \g^{\m\n} \p_\n \Theta \ri)\) . Evaluating these counterterms using the small-\(u\) expansions in eqs.~\eqref{eq:theta_small_u} and~\eqref{eq:phi_small_u}, we find
\beq \label{eq:counterterms_expanded}
    S_\mathrm{ct} = \cN \le[\frac{1}{4 \e^4} - \frac{M^2}{2 \e^2} - b^2M^2 \log \e - M C +\frac{M^2}{4} \le(M^2 - b^2 \ri) - \frac{1}{2} b^2 M^2 \log M \ri],
\eeq
where we have suppressed terms that vanish when \(\e \to 0\). The counterterms cancel the divergences in the bulk action eq.~\eqref{eq:on_shell_action_divergences}, as expected, and also provide a finite contribution to the action.

The contribution of the D7-branes to free energy density in the dual field theory is given by its on-shell action density in Euclidean signature. Since our solutions are time-independent, this is just minus the on-shell action density in Lorentzian signature, \(f = - S^\star = - (S_\mathrm{D7}^\star + S_\mathrm{ct}^\star)\). The bulk D7-brane action's contribution may be found from eq.~\eqref{eq:d7_action_on_ansatz}, by substituting our numerical solution for \(R(r)\) up to some large-\(r\) cutoff \(r_c\),\footnote{Since all of our solutions have \(p_\f=0\), they also have \(\f' = 0\) by eq.~\eqref{eq:phi_momentum}.}
\beq \label{eq:d7_on_shell_action}
    S_\mathrm{D7}^\star = - \cN \int_{r_0}^{r_c} d r \, r^3 g(\r) h(\r)
    \sqrt{1 + \frac{b^2 R^2}{\r^4 h(\r)}} \sqrt{1 + R'^2},
\eeq
where the lower limit of integration is $r=0$ for Minkowski embeddings, and is $r=r_H$ for black hole embeddings.

The contribution from the counterterms may be obtained by exchanging the small-\(u\) cutoff \(\e\) in eq.~\eqref{eq:counterterms_expanded} for the large-\(r\) cutoff \(r_c\). To do so, we expand the relation \(\e = 1/\sqrt{r_c^2 + R(r_c)^2}\) for \(r_c\), making use of \(R\)'s near boundary expansion in eq.~\eqref{eq:R_near_boundary} to find
\beq
    \frac{1}{\e} = r_c + \frac{M^2}{2 r_c} - \frac{b^2 M^2}{2 r_c^3} \log r_c + \frac{M}{r_c^3} \le(C - \frac{M^3}{8} \ri) + \dots \; .
\eeq
Substituting this into eq.~\eqref{eq:counterterms_expanded} we find that the counterterm contribution is
\beq \label{eq:counterterms_rc}
    S_\mathrm{ct}^\star = \cN \le[\frac{r_c^4}{4} + \frac{b^2 M^2}{2} \log r_c - \frac{b^2 M^2}{4} \le(1 + 2 \log M\ri) \ri].
\eeq
Combining this with eq.~\eqref{eq:d7_on_shell_action} we find an expression for the free energy density
\begin{align}
    f = \cN \biggl[&
        \int^{r_c} dr \, r^3 g(\r) h(\r) \sqrt{1 + \frac{b^2 R^2}{\r^4 h(\r)}} \sqrt{1 + R'^2}
    \nonumber \\  &\hspace{2cm}
        - \frac{r_c^4}{4} - \frac{b^2 M^2}{2} \log r_c + \frac{b^2 M^2}{4} \le(1 + 2 \log M\ri)
    \biggr].
\end{align}
Using dimensional analysis to restore factors of \(L\) and taking \(\cN = \l N_c N_f/(16 \pi^4 L^8)\) from eq.~\eqref{eq:cNdef} then yields eq.~\eqref{eq:free_energy_formula}.

\subsection*{Scalar One-point Functions}

The one-point functions of the operators \(\cO_m\) and \(\cO_\f\) defined in eq.~\eqref{eq:field_theory_operators} are proportional to the functional derivatives of the on-shell action with respect to the boundary values of \(\q\) and \(\f\),
\beq \label{eq:Om_Ophi_function_derivatives}
    \langle \cO_m \rangle = - 2\pi\a'\lim_{\e \to 0} \e \frac{\d S}{\d \q(\e)},
    \quad
    \langle \cO_\f \rangle = - \lim_{\e \to 0} \frac{\d S}{\d \f(\e)}.
\eeq

In order to compute \(\langle \cO_m \rangle\), let us consider a small variation \(\q(u) \to \q(u) + \d \q(u)\). Writing \(S_\mathrm{D7} = \int d u \, \cL\), the resulting variation in the action is
\beq
    \d S_\mathrm{D7} = \int_\e du \le[ \frac{\p \cL}{\p \q'(u)} \d \q'(u) - \frac{\p \cL}{\p \q(u)} \d\q(u)\ri]
    = - \le. \frac{\p \cL}{\p \q'(u)} \d \q(u) \ri|_{u = \e}.
\eeq
The second equality is obtained using integration by parts and the Euler-Lagrange equation for \(\q\). The derivative \(\p \cL/\p \q'(u)\) may be computed from the action in eq.~\eqref{eq:holo_rg_bulk_action}. Inserting the near-boundary expansions in eqs.~\eqref{eq:theta_small_u} and~\eqref{eq:phi_small_u} gives a result that diverges as \(\e^{-3}\). The \(\e^{-3}\) and an \(\e^{-1} \log \e\) divergence are cancelled by the variation of the counterterms in eq.~\eqref{eq:counterterms}, so we obtain a finite result from eq.~\eqref{eq:Om_Ophi_function_derivatives},
\beq
    \langle \cO_m \rangle = 2 \pi \a' \cN L^6 \le[- 2 C + \frac{b^2 M}{2} + b^2 M \log \le(M L\ri)\ri],
\eeq
where we used dimensional analysis to restore factors of \(L\) on the right-hand side. Using \(\cN = \l N_c N_f/(16 \pi^4 L^8)\) from eq.~\eqref{eq:cNdef} and \(L^4 = \a'^2 \l \) then yields eq.~\eqref{eq:condensate_formula}.

Similarly under a small variation \(\f \to \f + \d \f\) we find
\beq \label{eq:dS_phi}
    \d S_\mathrm{D7} = - \le. \frac{\p \cL}{\p \f'} \d\f \ri|_{u=\e} = - \cN p_\f \, \d\f(\e),
\eeq
where we have used \(\p \cL/\p \f' \equiv P_\f = \cN p_\f\). In this case the counterterms do not contribute to the one-point function, and we can read off \(\langle \cO_\f \rangle\) from eq.~\eqref{eq:dS_phi} using eq.~\eqref{eq:Om_Ophi_function_derivatives},
\beq
    \langle \cO_\f \rangle = \cN p_\f.
\eeq
Since \(p_\f=0\) in all the embeddings we consider, all phases of our model have \(\langle \cO_\f \rangle = 0\).

\subsection*{Current One-point Functions}

In order to compute the one-point functions of the current, we need to allow for non-zero \(A_x\) and \(A_y\),
\beq \label{eq:holo_rg_finite_E_ansatz}
    \q = \q(u), \quad \f = b z + \Phi(u),
    \quad
    A_x = E t + a(u),
    \quad
    A_y = A_y(u).
\eeq
Non-zero \(A\) changes the coefficient of the \(\log \e\) divergence of the D7-brane action. This change is cancelled by an additional  counterterm~\cite{Karch:2005ms,Karch:2006bv,Hoyos:2011us,Hoyos:2011zz},
\beq
    S_\mathrm{ct,6} = - \cN \frac{1}{4} \sqrt{-\g} F^{\m\n} F_{\m\n} \log \e.
\eeq

The one-point functions of the currents are given by
\beq
    \langle J^\m \rangle = 2\pi\a' \lim_{\e \to 0} \frac{\d S^\star}{\d A_\m(\e)}.
\eeq
The calculation proceeds similarly to that of \(\langle \cO_\f \rangle\). We find that the small variation in the bulk D7-brane action resulting from \(A_\m \to A_\m + \d A_\m\) is
\beq \label{eq:dS_A}
    \d S_\mathrm{D7} = - \le. \frac{\p \cL}{\p A_\m'(u)} \ri|_{u =\e} \d A_\m(\e).
\eeq
The variation of the counterterms vanishes at leading order in \(\d A_\m\), so eq.~\eqref{eq:dS_A} is the only contribution to the one-point function. Using \(\p \cL / \p A_\m' = \cN j_\m\), we then find
\beq
    \langle J_x \rangle = - 2 \pi \a' \cN j_x,
    \quad
    \langle J_y \rangle = - 2 \pi \a' \cN j_y.
\eeq

\section{Details of Thermodynamics}
\label{sec:thermodynamics}

The free energy given in eq.~\eqref{eq:free_energy_formula} and calculated in appendix A is
\begin{align}
f = \frac{\l N_c N_f}{16 \pi^4 L^8} \biggl[&
\int_{r_0}^{r_c} dr \, r^3 g(\r) h(\r) \sqrt{1 + \frac{L^4 b^2 R^2}{\r^4 h(\r)}} \sqrt{1 + R'^2}
\nonumber
\\  &\hspace{1.5cm}
- \frac{r_c^4}{4} - \frac{L^8 b^2 M^2}{2} \log \le( r_c / L\ri) + \frac{L^8 b^2 M^2}{4} \le(1 + 2 \log \le(M L \ri)\ri)
\biggr],
\end{align}
where $r_0=0$ for Minkowski embeddings and $r_0=r_H$ for black hole embeddings.

To calculate the entropy density, $s$ we have to differentiate the free energy with respect to temperature. It is easiest to do this in steps. First, we have the contribution coming from the endpoints of integration, and second from the integrand itself. The upper bound for both the Minkowski and black hole embeddings has no temperature dependence, nor does the lower bound for the Minkowski solution. This leaves the contribution from the lower bound for the black hole embedding, $r_0 = \frac{\pi L^2}{\sqrt{2}} T\,\sin \theta$,
\beq
s_{r_0} = - \frac{\l N_c N_f}{16 \pi^4 L^8}  \biggl[
 r_0^3 g(\r_0) h(\r_0) \sqrt{1 + \frac{L^4 b^2 R^2}{\r_0^4 h(\r_0)}} \sqrt{1 + R'^2}\biggr].
\eeq 
As $g(\r_0) h(\r_0) = 0$ this contribution is zero, $s_{r_0} = 0$. As a result, only the contribution from the integrand itself contributes:

\beq
s = - \frac{\l N_c N_f}{16 \pi^4 L^8} \biggl[
\int_{r_0}^{r_c} dr \,  \frac{\partial }{\partial T} \bigg(r^3 g(\r) h(\r) \sqrt{1 + \frac{L^4 b^2 R^2}{\r^4 h(\r)}} \sqrt{1 + R'^2} \bigg) \biggr].
\eeq
All three of $\r_H$, $R$ and $R'$ have $T$ dependence. We can therefore split the differentiation into two parts, one where we vary $\r_H$ while holding $R, R'$ constant, and another where we vary $R, R'$ while holding $\r_H$ constant. The former contribution is
\beq
s_{i} = - \frac{\l N_c N_f}{16 \pi^4 L^8} \biggl[
\int_{r_0}^{r_c} dr \, r^3\sqrt{1 + R'^2} \frac{\partial }{\partial T} \bigg( g(\r) h(\r) \sqrt{1 + \frac{L^4 b^2 R^2}{\r^4 h(\r)}}  \bigg) \bigg|_{R,R'} \biggr],
\label{eq:entropy_first}
\eeq
where the differentiation can be explicitly performed. The latter contribution is
\beq
s_{ii} = - \frac{\l N_c N_f}{16 \pi^4 L^8} \biggl[
\int_{r_0}^{r_c} dr \,  \frac{\partial }{\partial T} \bigg( r^3\sqrt{1 + R'^2} g(\r) h(\r) \sqrt{1 + \frac{L^4 b^2 R^2}{\r^4 h(\r)}}  \bigg) \bigg|_{\r_0} \biggr].
\eeq
Following ref.~\cite{Mateos:2007vn} this term can be simplified by noticing that the differentiation with temperature can be viewed as a variation $\delta R$. Given that the entropy is evaluated on-shell this means we only have to calculate a boundary term,
\beq
s_{ii} = -\frac{\l N_c N_f}{16 \pi^4 L^8} \frac{r^3}{\sqrt{1 + R'^2}} g(\r) h(\r)\sqrt{1 + \frac{L^4 b^2 R^2}{\r^4 h(\r)}} R' \frac{\partial R}{\partial T} \bigg|^{r_c}_{r_0},
\eeq
where we have three cases to consider. First is the case where the lower bound is $r_0 = 0$, i.e the Minkowksi embedding. We then have the boundary condition $R(0) = R_0$ where $R_0$ is a constant, hence $R' = 0$ and the contribution to the entropy density is zero. Next we consider the lower bound for the black hole embeddings, where $r_0 = \frac{\pi L^2}{\sqrt{2}} T \,\sin \theta$. However, as above this implies $g(\r_0) h(\r_0) = 0$, so the contribution to the entropy density is zero. Finally we have the upper bound $r_c$. As this is at the asymptotically $AdS_5$ boundary we can use the embedding's asymptotic expansion in eq.~\eqref{eq:R_near_boundary},
\beq 
R = L^2 M \le(1 - L^4 b^2 \frac{\log (r/L)}{2 r^2} \ri) + \frac{L^6 C}{r^2} + \dots .
\eeq
This term has no temperature dependence and so $\frac{\partial R}{\partial T} = 0$, and again the contribution to the entropy density is zero. Ultimately, then, the total entropy density is given by eq.~\eqref{eq:entropy_first},
\beq
\label{eq:entropynumericsapp}
s = - \frac{\l N_c N_f}{16 \pi^4 L^8} \biggl[
\int_{r_0}^{r_c} dr \, r^3\sqrt{1 + R'^2} \frac{\partial }{\partial T} \bigg( g(\r) h(\r) \sqrt{1 + \frac{L^4 b^2 R^2}{\r^4 h(\r)}}  \bigg) \bigg|_{R,R'} \biggr],
\eeq
as stated in eq.~\eqref{eq:entropynumerics}. From eq.~\eqref{eq:entropynumericsapp} we calculate the heat capacity $c_V$ numerically using finite differences.

\section{Conductivity from Kubo Formula}
\label{sec:kubo}

We can reproduce our results for the $T=0$ conductivity in eq.~\eqref{eq:T0_Hall} from the Kubo formulas,
\beq \label{eq:conductivity_kubo}
    \s_{xx} \equiv \lim_{\w \to 0} \frac{1}{i\w} G_{J_x J_x}(\w, k=0),
    \quad
    \s_{xy} \equiv \lim_{\w \to 0} \frac{1}{i\w} G_{J_x J_y}(\w, k=0),
\eeq
where \(G_{J_\m J_\n}\) is the retarded two-point function of \(J_\m\) with \(J_\n\) in Fourier space, $\omega$ is frequency, and $k$ is momentum. To compute these we introduce small time-dependent fluctuations of the gauge field, \(A_x(t,r)\) and \(A_y(t,r)\). Expanding the D7-brane action eq.~\eqref{eq:d7_action} in powers of these fluctuations, we find that the quadratic term is
\begin{multline}
    S_\mathrm{D7}^{(2)} = \frac{\cN}{2} \int d t \, d r \frac{r^3}{\r^6} \sqrt{
        \frac{\r^4 + L^4 b^2 R^2}{1 + R'^2}
    } \le[
        L^4 \le( 1 + R'^2 \ri) \le(\dot{A}_x^2 + \dot{A}_y^2 \ri) - \r^4 \le(A_x'^2 + A_y'^2 \ri)
    \ri]
    \\
    + \cN L^4 b \int dt \, dr \frac{r^4}{\r^4} \le( \dot{A}_y A_x' - \dot{A}_x A_y' \ri),
    \label{eq:conductivity_action}
\end{multline}
where $\dot{A}_x\equiv \frac{\partial}{\partial t} A_x$ and $A'_x\equiv \frac{\partial}{\partial r} A_x$, and similarly for $A_y$. The first line in eq.~\eqref{eq:conductivity_action} comes from the DBI term in the action, while the second line, which couples \(A_x\) and \(A_y\), comes from the WZ term.

The equations of motion for \(A_x\) and \(A_y\) are, respectively,
\begin{subequations}
\label{eq:conductivity_fluctuation_equation}
\begin{align}
    \le(\frac{r^3}{\r^2} \sqrt{ \frac{\r^4 + L^4 b^2 R^2}{1 + R'^2} } A_x' \ri)'     
    + \w^2 \frac{L^4 r^3}{\r^6} \sqrt{ \le(\r^4 + L^4 b^2 R^2\ri)\le( 1 + R'^2 \ri) } A_x
    &= - i \w L^4 b \le(\frac{r^4}{\r^4}\ri)' A_y,
    \\
    \le(\frac{r^3}{\r^2} \sqrt{ \frac{\r^4 + L^4 b^2 R^2}{1 + R'^2} } A_y' \ri)'     
    + \w^2  \frac{L^4 r^3}{\r^6} \sqrt{ \le(\r^4 + L^4 b^2 R^2\ri)\le( 1 + R'^2 \ri) }  A_y
    &= i \w L^4 b \le(\frac{r^4}{\r^4}\ri)' A_x,
\end{align}
\end{subequations}
where we have Fourier transformed with respect to time, \(A_x (t,r) = \int \frac{d\w}{2\pi} e^{-i \w  t} A_x(\w,r)\), and similarly for $A_y$. The near-boundary expansions are
\begin{subequations}
\begin{align}
    A_x(\w,r) = A_x^{(0)}(\w) \le[1 + \frac{L^4 \w^2}{2r^2} \log(r/L) \ri] + \frac{A_x^{(2)}(\w)}{r^2} + \dots \;
    \\
    A_y(\w,r) = A_y^{(0)}(\w) \le[1 + \frac{L^4 \w^2}{2r^2} \log(r/L) \ri] + \frac{A_y^{(2)}(\w)}{r^2} + \dots \;
\end{align}
\end{subequations}

When the equations of motion are satisfied, the action reduces to a boundary term,
\beq
    S_\mathrm{D7}^{(2)\star} = - \frac{\cN}{2} \int dt \le[
        \frac{r^3}{\r^2} \sqrt{
            \frac{\r^4 + L^4 b^2 R^2}{1 + R'^2}
        } \le( A_x A_x' + A_y A_y' \ri)
        - \frac{b r^4}{\r^4} \le(A_x \dot{A_y} - \dot{A}_x A_y \ri)
    \ri]_{r = r_c}
\eeq
Substituting the near-boundary expansions of \(R\), \(A_x\) and \(A_y\) and Fourier transforming with respect to time, we find
\begin{align}
    S_\mathrm{D7}^{(2)\star} =
    - \cN \int \frac{d \w}{2\pi} \Bigl[&
        A_x^{(0)}(-\w) A_x^{(2)}(\w)
        + A_y^{(0)}(-\w) A_y^{(2)}(\w)
        + \w^2 L^4 A_x^{(0)}(-\w) A_x^{(0)}(\w)
    \nonumber \\ &
        + \w^2 L^4 A_y^{(0)}(-\w) A_y^{(0)}(\w)
        - i \w L^4 b A_x^{(0)}(-\w) A_y^{(0)}(\w)
    \Bigr]_{r=r_c}.
\end{align}
Applying the Minkowski correlator prescription of refs.~\cite{Son:2002sd,Herzog:2002pc} we can read off expressions for the two-point functions. Substituting these into the Kubo formulas eq.~\eqref{eq:conductivity_kubo} we find
\beq
    \s_{xx} = (2 \pi \a')^2 \cN \lim_{\w \to 0} \frac{2}{i\w} \le. \frac{A_x^{(2)}(\w)}{A_x^{(0)} (\w)} \ri|_{A_y^{(0)} = 0},
    \quad
    \s_{xy} =  \frac{N_f N_c}{4 \pi^2} b + (2 \pi \a')^2  \cN \lim_{\w \to 0} \frac{2}{i\w} \le. \frac{A_x^{(2)}(\w)}{A_y^{(0)} (\w)} \ri|_{A_x^{(0)} = 0}.
    \label{eq:conductivity_greens_functions}
\eeq

To evaluate these expressions we use the membrane paradigm approach of ref.~\cite{Iqbal:2008by}. We Fourier transform the action in eq.~\eqref{eq:conductivity_action}, and write it as
\begin{align}
    S_\mathrm{D7}^{(2)} = \int \frac{d \w}{2\pi} \int d r \Bigl\{&
        \frac{1}{2} \cF_1(r) \le[A_x'(-\w) A_x'(\w) + A_y'(-\w) A_y'(\w) \ri]
        \nonumber \\ &
        + \frac{1}{2} \w^2 \cF_2(r) \le[ A_x(-\w) A_x(\w) + A_y(-\w) A_y(\w) \ri]
        \label{eq:conductivity_action_abc} \\ &
        + i \w \cF_3(r) \le[ A_x'(-\w) A_y(\w) + A_x(-\w) A_y'(\w) \ri]
    \Bigr\},
    \nonumber
\end{align}
where we suppressed the \(r\)-dependence of \(A_x\) and $A_y$ for notational simplicity, and defined
\begin{subequations}
\label{eq:Fdefs}
\begin{align}
    \cF_1(r) &\equiv - \cN \frac{r^3}{\r^2} \sqrt{
        \frac{\r^4 + L^4 b^2 R^2}{1 + R'^2}
    },
     \\
    \cF_2(r) &\equiv \cN \frac{r^3 L^4}{\r^6} \sqrt{
        \le(\r^4 + L^4 b^2 R^2\ri)\le(1 + R'^2\ri)
    },
    \\
    \cF_3(r) &\equiv \cN L^4 b \frac{r^4}{\r^4}.
    \end{align}
\end{subequations}

From eq.~\eqref{eq:conductivity_action_abc} we find the canonical momenta
\begin{subequations}
\begin{align}
    P_x(\w) \equiv \frac{\d S}{\d A_x'(-\w)} = \cF_1(r) A_x'(\w) + i \w \cF_3(r) A_y(\w),
     \\
    P_y(\w) \equiv \frac{\d S}{\d A_y'(-\w)} = \cF_1(r) A_y'(\w) - i \w \cF_3(r) A_x(\w).
    \label{eq:conductivity_momenta}
\end{align}
\end{subequations}
The $\omega \to 0$ limits of \(P_x/i\w A_x\) and \(P_x/i\w A_y\) yield the longitudinal and transverse conductivities. To see this, we expand the canonical momenta in eq.~\eqref{eq:conductivity_momenta} at large \(r\), finding
\begin{subequations}
\begin{align}
    \lim_{\w \to 0} \lim_{r \to \infty} \frac{P_x}{i \w A_x}
    &= \cN \lim_{\w \to 0} \lim_{r \to \infty} \le( \frac{2}{i \w} \frac{A_x^{(2)}}{A_x^{(0)}}
    + L^4 b \frac{A_y^{(0)}}{A_x^{(0)}}\ri),
    \label{eq:longitudinal_conductivity_momentum}
    \\
    \lim_{\w \to 0} \lim_{r \to \infty} \frac{P_x}{i \w A_y}
    &=  \cN \lim_{\w \to 0} \lim_{r \to \infty} \le( \frac{2}{i \w} \frac{A_x^{(2)}}{A_y^{(0)}}
    + L^4 b \ri).
    \label{eq:transverse_conductivity_momentum}
\end{align}
\end{subequations}
Comparing to eq.~\eqref{eq:conductivity_greens_functions}, we see that if we use boundary conditions such that \(A_y^{(0)} = 0\), then the right-hand side of eq.~\eqref{eq:longitudinal_conductivity_momentum} is \(\s_{xx}/(2\pi\a')^2\). Similarly, if we use boundary conditions such that \(A_x^{(0)} = 0\) then the right-hand side of eq.~\eqref{eq:transverse_conductivity_momentum} is \(\s_{xy}/(2\pi\a')^2\).

We will now apply the method of ref.~\cite{Iqbal:2008by} to show that the ratios \(P_x/i\w A_x\) and \(P_x/i\w A_y\) are independent of \(r\) in the low frequency limit. This means we do not need take the \(r\to\infty\) limit in eq.~\eqref{eq:longitudinal_conductivity_momentum} and eq.~\eqref{eq:transverse_conductivity_momentum}, as instead we can compute the ratios at \(r=0\). A key ingredient will be the Hamiltonian form of the equations of motion. The Hamiltonian is
\begin{align}
    H &\equiv \int \frac{d \w}{2\pi} \int d r \le[P_x(-\w) A_x'(\w) + P_y(-\w) A_y'(\w) \ri] - S_\mathrm{D7}^{(2)}
    \nonumber \\
    &= \int \frac{d \w}{2\pi} \int d r \biggl\{
        \frac{1}{2 \cF_1(r)} \le[P_x(-\w) P_x(\w) + P_y(-\w) P_y(\w) \ri]
        \nonumber \\ &\phantom{\int \frac{d \w}{2\pi} \int d r \biggl\{} 
        - \frac{i \w \cF_3(r)}{\cF_1(r)} \le[A_y(-\w) P_x(\w) - A_x(-\w) P_y (\w) \ri]
        \nonumber \\  &\phantom{\int \frac{d \w}{2\pi} \int d r \biggl\{} 
        - \frac{1}{2}\w^2 \le[\cF_2(r) - \frac{\cF_3(r)^2}{\cF_1(r)} \ri] \le[A_x(-\w) A_x(\w) + A_y(-\w) A_y(\w) \ri]
    \biggr\}.
\end{align}
From this we can read off Hamilton's equations,
\begin{subequations}
\begin{align}
    A_x'(\w) &= \frac{1}{\cF_1(r)} P_x(\w) - \frac{i \w \cF_3(r)}{\cF_1(r)} A_y(\w),
    \label{eq:conductivity_HE_AxPrime}
    \\
    A_y'(\w) &= \frac{1}{\cF_1(r)} P_y(\w) + \frac{i \w \cF_3(r)}{\cF_1(r)} A_x(\w),
    \label{eq:conductivity_HE_AyPrime}
    \\
    P_x'(\w) &= - \frac{i \w \cF_3(r)}{\cF_1(r)} P_y(\w) + \w^2 \le[\cF_2(r) - \frac{\cF_3(r)^2}{\cF_1(r)} \ri] A_x(\w),
    \label{eq:conductivity_HE_PixPrime}
    \\
    P_y'(\w) &= \frac{i \w \cF_3(r)}{\cF_1(r)} P_x(\w) + \w^2 \le[\cF_2(r) - \frac{\cF_3(r)^2}{\cF_1(r)} \ri] A_y(\w).
    \label{eq:conductivity_HE_PiyPrime}
\end{align}
\end{subequations}

Now consider the formula for the transverse conductivity,
\beq
    \s_{xy} = (2\pi\a')^2 \lim_{\w \to 0} \lim_{r \to \infty} \frac{P_x}{i \w A_y}.
\eeq
We are supposed to take the limit of zero frequency, keeping \(P_x\) and \(i \w A_y\) fixed. Eq.~\eqref{eq:conductivity_HE_AyPrime} implies \(\p_r \le(i \w A_y\ri) \sim \cO(\w)\), and eq.~\eqref{eq:conductivity_HE_PixPrime} implies \(\p_r P \sim \cO(\w)\), so to leading order at small \(\w\) the ratio \(P_x/i \w A_y\) is independent of \(r\). We can therefore evaluate it at \(r=0\) instead of \(r \to \infty\). Using eq.~\eqref{eq:conductivity_momenta} and \(\cF_1(r)\) and \(\cF_3(r)\) from eq.~\eqref{eq:Fdefs}, we may then write
\beq
    \s_{xy} = \frac{N_f N_c}{4 \pi^2} \lim_{\w \to 0} \lim_{r \to 0} \le(
    b \frac{r^4}{(r^2 + R^2)^2} + \frac{A_x'}{i \w A_y}\frac{r^3}{L^4 (r^2+R^2)} \sqrt{\frac{(r^2 + R^2)^2 + L^4 b^2 R^2}{1+R'^2}} \ri),
    \label{eq:transverse_conductivity_horizon}
\eeq
where we used \((2\pi\a')^2 \cN L^4 = N_f N_c/4\pi^2\).

We need to understand the \(r \to 0\) limit of the two terms in eq.~\eqref{eq:transverse_conductivity_horizon} for the different types of solution we have found. In the low \(M/b\) phase we have \(R \to 0\) exponentially quickly as \(r \to 0\), as given in eq.~\eqref{eq:T0_boundary_condition_non_analytic}. The first term in the brackets therefore just gives \(b\) while the second term vanishes.\footnote{From the equations of motion eq.~\eqref{eq:conductivity_fluctuation_equation} one finds that in the low \(M/b\) phase, \(A_x'/A_y\) diverges as \(1/r^2\) at small \(r\), while the factor multiplying \(A_x'/A_y\) in eq.~\eqref{eq:transverse_conductivity_horizon} vanishes as \(r^3\) at small \(r\). Hence, the second term in eq.~\eqref{eq:conductivity_fluctuation_equation} vanishes in the limit \(r \to 0\).} For the critical solution, we instead have \(R \sim r/\sqrt{3}\) as \(r \to 0\), as given in eq.~\eqref{eq:T0_boundary_condition_critical}. In that case, the second term in eq.~\eqref{eq:transverse_conductivity_horizon} still vanishes, but the first term in the brackets now evaluates to \(9 b/16\). Finally, in the large \(M/b\) phase \(R\) remains finite as \(r \to 0\), as given in eq.~\eqref{eq:T0_boundary_condition_analytic}. In that case, both terms in eq.~\eqref{eq:transverse_conductivity_horizon} vanish. To summarise:
\beq
    \s_{xy} = \frac{N_f N_c}{4\pi^2} \, b \,\times \begin{cases}
        1, \quad & \textrm{(black hole-like embeddings)},
        \\
        \dfrac{9}{16}, \quad  &\textrm{(critical embedding)},
        \\
        0, \quad & \textrm{(Minkowski embeddings)}.
    \end{cases}
    \label{eq:T0_Hall_conductivity}
\eeq
As we argued in section~\ref{sec:T0_solutions}, when $T=0$ and $m/(b\sqrt{\l})$ increases, the critical solution is never the minimum of the free energy because a first order transition occurs from a black hole-like to a Minkowski embedding. As a result, our $\sigma_{xy}$ behaves as the Heaviside step function in eq.~\eqref{eq:T0_Hall}.

For the longitudinal conductivity we have
\beq
    \s_{xx} = \frac{N_f N_c}{4 \pi^2} \lim_{\w \to 0} \lim_{r \to 0} \le(
        \frac{r^4}{(r^2 + R^2)^2} \frac{A_y}{A_x} b
        + \frac{A_x'}{i \w A_x}\frac{r^3}{L^4 (r^2+R^2)} \sqrt{\frac{(r^2 + R^2)^2 + L^4 b^2 R^2}{1+R'^2}} \ri).
\eeq
The second term in the brackets vanishes for all three kinds of solution, similar to \(\s_{xy}\). The first term looks like the term that gave us $\sigma_{xy}$, except multiplied by \(A_y/A_x\). This ratio is \(\cO(\w)\), and so vanishes when we take \(\w \to 0\). To see this, recall that when computing \(\s_{xx}\) we have the boundary condition \(A_y^{(0)} = 0\), so any non-zero \(A_y\) must be sourced by \(A_x\). Since we take $\omega \to 0$ with \(\w A_x\) fixed, from the equations of motion in eq.~\eqref{eq:conductivity_fluctuation_equation} we expect that the non-zero \(A_y\) sourced by \(A_x\) will be of order \(\cO(\w^0)\). Hence \(A_y/A_x \sim \cO(\w)\), and therefore the ratio vanishes as \(\w \to 0\). The conclusion is that the longitudinal conductivity vanishes in all phases at \(T=0\), consistent with the analysis in section~\ref{sec:T0_conductivities}.

\bibliographystyle{JHEP}
\bibliography{d3d7_wsm}

\end{document}